\documentclass[10pt]{article}

\usepackage[english]{babel}
\usepackage{latexsym,amsmath,amssymb,amsbsy,amstext,amscd,amsfonts}
\usepackage{graphics,graphicx}
\usepackage{color}
\usepackage{tabularx}
\usepackage{multirow}
\usepackage{booktabs}
\usepackage{epstopdf}
\usepackage{natbib}
\usepackage{ulem}
\usepackage [latin1]{inputenc}
\newtheorem{remark}{Remark}
\date{}

\title{On the application of simplified  rheological models of fluid in the hydraulic fracture problems\footnote{Preprint submitted to \textit{International Journal of Engineering Science}}}
\author{Michal Wrobel
\\
{\it \!Department of Civil and Environmental Engineering, University of Cyprus,}
\\ {\it 75 Kallipoleos Street, 1678 Nicosia, Cyprus}
\\{\it wrobel.michal@ucy.ac.cy}}

\begin{document}

\maketitle

\begin{abstract}
In this paper we analyse a problem of a hydraulic fracture driven by a non-Newtonian shear-thinning fluid. For the PKN fracture geometry we consider three different rheological models of fluid: i) the Carreau fluid, ii) the truncated power-law fluid, iii) the power-law fluid. For each of these models a number of simulations are performed. The results are post-processed and compared with each other in order to find  decisive factors for similarities/dissimilarities. It is shown that under certain conditions even the basic power-law rheology can be a good substitute for the Carreau characteristics. Although for a particular fluid such a conclusion cannot be made a priori,  post-processing based on average values of the fluid shear rates is a very good  tool to verify credibility of the results obtained for simplified rheological models. The truncated power-law rheology is a good alternative for the Carreau model. It always produces results that are very similar to those obtained with the equivalent Carreau fluid and simultaneously provides a relative ease of numerical implementation.
\end{abstract}

\providecommand{\keywords}[1]
{
  \small	
  \textbf{\textit{Keywords:}} #1
}

%\end{frontmatter}

\section{Introduction}

The phenomenon of hydraulic fracture (HF) is encountered in many natural and man-made processes. One of its most prominent applications is  fracking technology used to stimulate hydrocarbon reservoirs. The highly multiphysical nature of the underlying physical mechanism necessitates careful analysis of the interactions between respective component physical fields in order to properly predict evolution of hydraulically induced fractures and optimally design the treatments. 

The hydraulic fracturing process is influenced essentially by rheological properties of the fracturing fluid. Depending on the geology of  formation, economic factors,  stage and overall scenario of the treatment, the fracturing fluids are engineered accordingly so as to achieve optimal combination of chemical and mechanical properties. There are many requirements that involve  physical behaviour of fracturing fluids. One can mention among them \citep{Barbati_2016}: i) viscosity sufficient to create desirable fracture width, ii) suspending properties that facilitate proppant transport under both dynamic and static conditions and mitigate the risk of bridging phenomenon \citep{Garagash_2019}, iii) low leak-off to formation, iv) short time of fracture closure after the influx shut-off to prevent proppant settling, and others. Moreover, the desired properties should be retained over specific temperature ranges and chemical environments. No wonder, all these needs can hardly be addressed by any Newtonian fluid. For this reason complex fluids have been widely employed in the oilfield industry. As the shear-thinning rheology improves suspending properties of the fluid, many fracturing fluids are intentionally made shear-thinning (e.g. by adding polymers \citep{Bao_2017}).

In fact, rheology of numerous  fracturing fluids yields shear-thinning behaviour only over some limited range of shear rates. At low shear rates a Newtonian plateau is observed for which the apparent viscosity achieves a maximum. It is only above a critical value of shear rate that the fluid shear thins. Similarly,  at very high shear rates the viscosity reaches another Newtonian plateau corresponding to that of the base solvent used \citep{Lecampion_2018}. It is still not well recognised how the viscosity plateaus and the shear-thinning amplitude affect the propagation of hydraulic fractures. 

Such a complex behaviour of a fracturing fluid can be well reproduced by four parameter rheological models e.g. Carreau or Cross \citep{Bird_1987}. Unfortunately, when using these models respective flow equations cannot be integrated analytically to obtain expressions for the fluid velocity and the fluid flow rate in the form used routinely for the hydraulic fracture problem in the framework of lubrication theory.  Instead a power-law rheology is usually employed \citep{Adachi_2002,Garagash_2006,Peck_2018_1,Peck_2018_2} which enables a derivation of the Poiseulle-type relation for the fluid flow rate. Moreover, in petroleum industry it is customary to sample only a limited viscosity data, in a narrow shear rate range (typically 25 $\frac{1}{\text{s}}$ to  100 $\frac{1}{\text{s}}$), in order to find fitting parameters for the averaged power-law characteristics \citep{Huang_2004}. Naturally, such an oversimplified model cannot correctly describe the beahviour of fracturing fluid in a broad range of shear rates. Considering a substantial gradation of the shear rate  values along  the fracture length it is evident that the pure power-law model does not reflect properly the near-tip high shear rate behaviour of the fluid and, depending on the process parameters, can largely overestimate the viscosity in the proximity of the crack mouth. Furthermore, when the hydraulic fracture model accounts for the hydraulically induced tangential tractions on the crack faces \citep{Wrobel_2017,Wrobel_2018}, the elasticity equation cannot be asymptotically balanced near the fracture tip for the power-law rheology.

A study on the near-tip behaviour of a hydraulic fracture driven by Carreau fluid was conducted in \citet{Lecampion_2018} where the authors analysed a problem of a semi-infinite plane strain crack propagating at a constant speed  in an impermeable material. Quantification of influence of the fracturing fluid rheology on the fluid lag was performed. Nevertheless, a problem of a finite hydraulic  fracture and its temporal evolution still needs to be addressed. A question whether the frequently used power-law rheology can be an acceptable substitute for the Carreau-like model is yet to be answered. Some indication on the significance of this issue can be found in \cite{Huang_2004} where the authors investigate the hydraulic fracture problem for a fluid with a single shear stress plateau assuming the PKN fracture geometry \citep{Nordgren}. A piecewise power-law model is introduced to describe the fluid rheology. The authors conclude that the conventional power-law model may be inadequate for accurate prediction of the fracture geometry.

In \cite{Lavrov_2015} a concept of truncated power-law  fluid was used to analyse the velocity profiles and fluid flow rates in a slit flow (thin flat channel). The truncated power-law rheology, being a four-parameter model, constitutes a simple regularisation of the power-law model, where cut-off viscosities are introduced for the high and low shear rates. In this way, the truncated power-law model can reproduce correctly the limiting behaviour of the Carreau or Cross fluid with the interim power-law approximation of the Carreu/Cross characteristics. The analysis presented in \cite{Lavrov_2015} shows that the truncated power-law rheology eliminates inherent  drawbacks of the power-law model, producing results that are much closer to those obtained for the Carreau fluid even in the low and high shear rate ranges. However, a question whether such an approximation is sufficient for the hydraulic fracturing problems still remains open. 

An efficient algorithm for numerical computation of the velocity profiles and fluid flow rates for a class of generalised Newtonian fluids \citep{Bird_1987} was introduced in \cite{Wrobel_Arxiv}. The computational scheme assumes piecewise approximation of the apparent viscosity with subsequent analytical integration of the resulting flow equations. Using the example of a slit flow the author showed that the algorithm can provide any desirable accuracy of solution at a computational cost that is only a fraction of those produced by other schemes available in the literature. As such, the new algorithm can be a numerical substitute of the Poiseulle-type relation in the hydraulic fracture problems. 

In this paper we address a  problem of a hydraulic fracture driven by a shear-thinning  fluid. For the analysis we assume the PKN fracture geometry. The algorithm from \cite{Wrobel_Arxiv} is adapted to compute the fluid flow rates in the case of fracture of elliptic cross section. This subroutine is integrated with the  hydraulic fracture solver developed in \cite{Wrobel_2015,Perkowska_2016}. A number of simulations are performed for three different rheological models of fluids: i) the Carreau model, ii) the truncated power-law model, iii) the power-law model. Based on the numerical results we verify whether and under what conditions  the simplified rheologies can be considered credible substitute for the Carreau law. 

The paper is structured as follows. In Section \ref{gen_rel} we introduce general relations for the hydraulic fracture problem of the PKN geometry. Section \ref{fluid_eq} includes  constitutive relations for respective rheological models of fluid together with corresponding expressions for the fluid flow rates. Computational relations for the Carreau variant are derived in Appendix \ref{ap_A}. In Section \ref{num_res} we perform a number of simulations for four different fracturing fluids. Each of these fluids is described by all three analysed rheologies.  A discussion on the numerical results is provided in Section \ref{disc}. Final conclusions are given in Section \ref{conc}.

\section{General relations}
\label{gen_rel}
\begin{figure}[htb!]
\begin{center}
\includegraphics[scale=0.45]{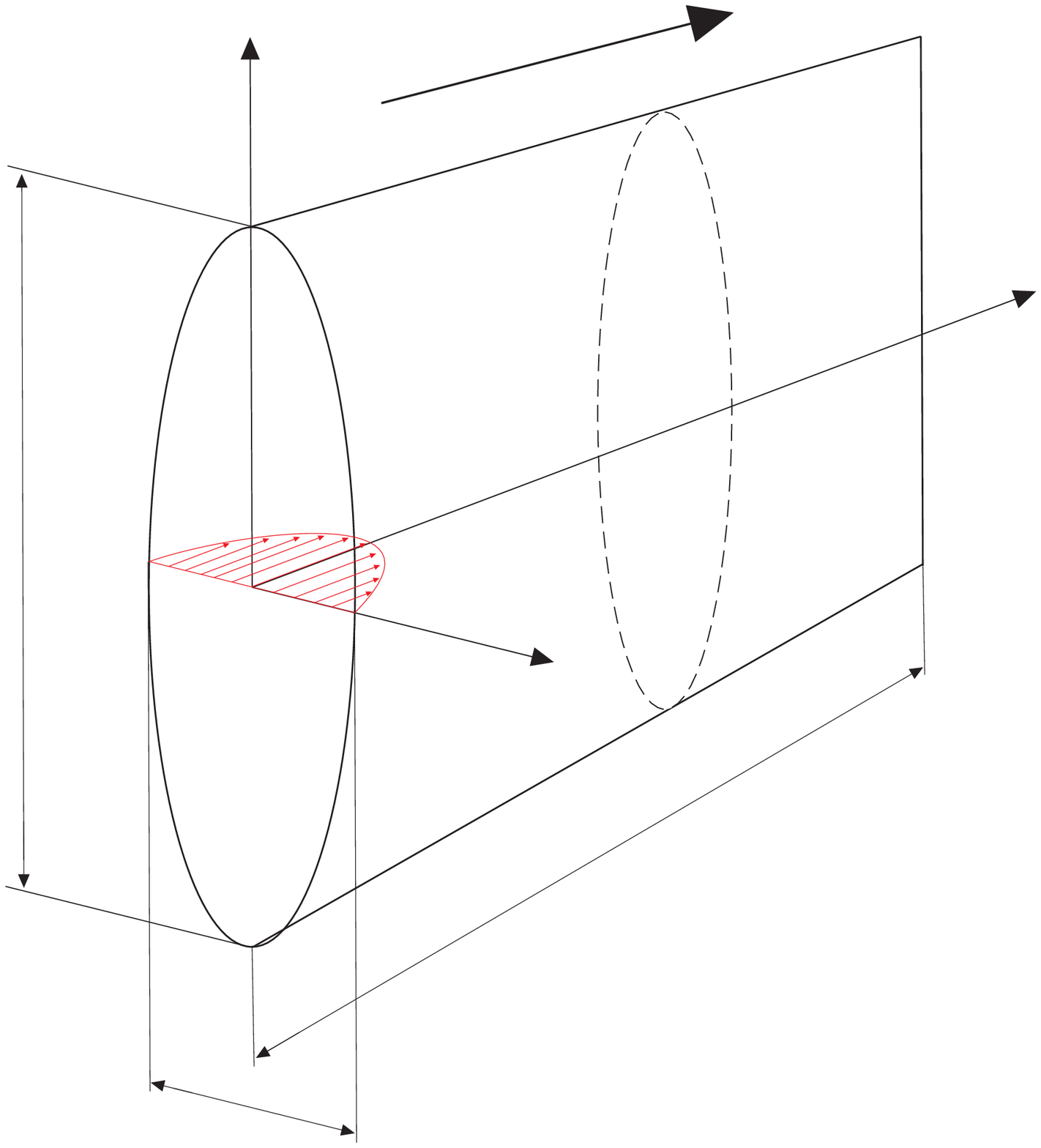}
\put(-140,227){\rotatebox{14}{\text{flow direction}}}
\put(-163,225){$z$}
\put(-13,183){$x$}
\put(-112,109){$y$}
\put(-183,15){\rotatebox{-15}{$w(x,t)$}}
\put(-225,120){\rotatebox{90}{$H$}}
\put(-100,65){\rotatebox{30}{$L(t)$}}
\caption{The PKN fracture geometry. Only one of the fracture wings is shown. Velocity profile in the cross section $z=0$ is marked by red arrows.}
\label{PKN_geom}
\end{center}
\end{figure}

Let us consider a hydraulic fracture whose geometry is defined by the classical PKN model \citep{Nordgren}. The symmetrical two-winged fracture of length $2L$ propagates in the plane $x \in[-L,L]$, where $L=L(t)$. In the following we analyse only one of the symmetrical parts, i.e.  $x \in[0,L]$, as shown in Fig. \ref{PKN_geom}. The fracture height, $H$, is assumed constant, while the fracture opening, $w(x,t)$, depends on the net fluid pressure, $p(x,t)$, and is an element of the solution. The relation between $p$ and $w$ is of the following form:
\begin{equation}
\label{elast}
p(x,t)=kw(x,t),
\end{equation}
where $k=\frac{E}{2(1-\nu^2)H}$, with $E$ and $\nu$ being the Young modulus and the Poisson's ratio, respectively. The mass conservation principle expressed by the continuity equation yields:
\begin{equation}
\label{cont}
\frac{\partial w}{\partial t}+\frac{\partial q}{\partial x}+q_\text{l}=0,
\end{equation}
where $q(x,t)$ is the normalised fluid flow rate through the fracture cross sections  and $q_\text{l}(x,t)$ stands for the normalised leak-off function (both quantities use a normalisation factor: $H\pi /4$ - compare e.g. \cite{Nordgren}). The fluid velocity averaged over the fracture cross section is defined as:
\begin{equation}
\label{v_gen}
v=\frac{q}{w}.
\end{equation}
We assume that there is no lag between the fluid front and the fracture tip and the leak-off is bounded at the crack apex, which implies:
\begin{equation}
\label{SE}
v(L,t)=\frac{\text{d}L}{\text{d}t}.
\end{equation}

Respective boundary conditions for the problem include:
\begin{itemize}
\item{two tip boundary conditions:
\begin{equation}
\label{BCs_tip}
w(L,t)=0, \quad q(L,t)=0,
\end{equation}}
\item{the influx boundary condition:
\begin{equation}
\label{BC_q}
q(0,t)=q_0(t).
\end{equation}}
\end{itemize}
Finally, the initial conditions define the initial crack length and the initial fracture aperture:
\begin{equation}
\label{init_cond}
L(0)=L_*, \quad w(x,0)=w_*(x).
\end{equation}

\section{Fluid flow equations}
\label{fluid_eq}

In the paper we compare results obtained for three rheological models of fluids: i) the power-law fluid, ii) the Carreau fluid, ii) the truncated power-law fluid, each of them complying with a definition of the generalised Newtonian fluid. Respective equations for the fluid flow rate, that supplement the problem formulation from the previous section, are given below.

\subsection{Power-law model}

The simplest model that can describe a non-Newtonian behaviour of a fluid is the power-law model \citep{Bird_1987,Gholipour_2018_1} for which the apparent viscosity is expressed as:
\begin{equation}
\label{PL_eta}
\eta_\text{a}= C |\dot \gamma|^{n-1},
\end{equation}
where $C$ is the consistency index,  $n$ stands for the fluid behaviour index, while $\dot \gamma$ denotes the shear rate. For $n<1$ it reflects the shear-thinning properties, while $n>1$ produces the shear-thickening characteristic. Unfortunately, this model yields unphysical results for  low and high shear rates. In the case of shear-thinning behaviour one obtains infinite viscosity for zero shear rate and zero viscosity as $\dot \gamma \to \infty$.  For the shear-thickening variant a reverse trend holds. 
The big advantage of the power-law model is that it enables analytical integration of the respective flow equations to obtain expression for the average fluid flow rate. For the elliptical channel in the PKN model the respective normalisedfluid flux is given by the formula (see Remark \ref{rem_q} in Appendix \ref{ap_A}):
\begin{equation}
\label{q_PL}
q=\frac{n}{1+3n}2^{-\frac{n+1}{n}}\left( -\frac{1}{C}\frac{\partial p}{\partial x}w^{2n+1}\right)^{1/n}.
\end{equation}

\subsection{Carreau model}

We employ the model of Carreau-Yasuda fluid \citep{Habibpour_2017} whose apparent viscosity can be described by the following relation:
\begin{equation}
\label{eta_carr}
\eta_\text{a}^{(c)}=\eta_\infty+(\eta_0-\eta_\infty)\left[1+|\lambda \dot \gamma |^a \right]^\frac{n-1}{a},
\end{equation}
$\eta_0$ is the viscosity at zero shear rate,  $\eta_\infty$ is the limiting viscosity for $\dot \gamma \to \infty$, while $\lambda$, $a$ and $n$ are fitting parameters. The Carreau-Yasuda model has been recognised to imitate well the physical behaviour of many fracturing fluids \citep{Lecampion_2018}. It eliminates the inherent deficiency of the classical power-law model described above. Unfortunately, expression \eqref{eta_carr} does not allow analytical integration of the respective fluid flow equations to obtain an average fluid flow rate even in the conduits of simple geometries.

In \cite{Wrobel_Arxiv} a numerical scheme was proposed that enables effective computation of the velocity and fluid flow rates for the generalised Newtonian fluids in conduits of simple geometries. It was shown that the procedure can be successfully used as a numerical substitute for the Poiseulle-type relation for $q$. 
The scheme assumes piecewise approximation of the apparent viscosity in the form:
\begin{equation}
\label{eta_ap}
\eta_\text{a}=
  \begin{cases}
		\eta_0       & \quad \text{for } \quad |\dot \gamma |<|\dot \gamma_1|,\\
    C_j |\dot \gamma|^{n_j-1}       & \quad \text{for} \quad |\dot \gamma_j|<|\dot \gamma|<|\dot \gamma_{j+1}| \quad j=1,...,N-1,\\
		
    \eta_\infty  & \quad \text{for } \quad |\dot \gamma|>|\dot \gamma_N|,
  \end{cases}
\end{equation}
where the values of  $\dot \gamma_j$, $C_j$ and $n_j$  are taken in a way to preserve continuity of $\eta_\text{a}$ and provide the best approximation of  the original rheological law for the chosen value of $N$. A method to construct approximation \eqref{eta_ap} is given in \cite{Wrobel_Arxiv}. 

In Appendix \ref{ap_A} we extend the algorithm from \cite{Wrobel_Arxiv}, originally proposed for the slit flow, to the case of an elliptic channel for the PKN geometry. 
The corresponding expressions for $q$ are  \eqref{q_sum} - \eqref{D_N}. Note that the fluid flow rate computed in this way takes into account  full velocity profile across the channel height. 

As explained in Appendix \ref{ap_A}, when analysing the flow in the elliptic cross section of the channel for the viscosity model \eqref{eta_ap} one can distinguish up to $N+1$ shear rate layers in each of the symmetrical parts of the conduit (see Fig. \ref{channel}). Among them there are: i) a Newtonian type layer of viscosity $\eta_0$ at the core of the flow - its thickness in the plane $z=0$, $\delta_1$, is given by formula \eqref{delta_1_ap}, ii) $N-1$ power-law layers of  thicknesses ($z=0$), $\delta_j$, defined by \eqref{delta_j_ap}, iii) a Newtonian layer with viscosity $\eta_\infty$ adjacent to the channel wall whose thickness ($z=0$), $\delta_{N+1}$, is described by \eqref{delta_N_ap}. Comparing the total thickness  of the latter Newtonian layer, $2\delta_{N+1}$, with the overall fracture opening, $w$, we obtain:
\begin{equation}
\label{del_N_comp}
\frac{2\delta_{N+1}}{w}=1-\sum_{i=1}^N\frac{2\delta_i}{w}.
\end{equation}
Now, let us recall that:
\begin{equation}
\label{del_i_est}
\delta_i \propto \left(-\frac{\partial p}{\partial x} \right)^{-1}, \quad i=1,...,N.
\end{equation}
When approaching the fracture tip the pressure gradient tends to $-\infty$. Thus, the thicknesses of respective layers tend to zero. For the standard estimation of the crack tip asymptotics:
\[
w \propto (L-x)^\alpha, \quad x \to L,
\]
we have that:
\begin{equation}
\label{del_j_est}
\frac{\delta_j}{w} \propto (L-x)^{1-2\alpha}, \quad x \to L, \quad i=1,...,N.
\end{equation}
In this way, for any permissible value of $\alpha$ for the shear-thinning fluids in the PKN model ($1/3 \leq \alpha<1/2$ - see e.g. \cite{Perkowska_2016}) the following estimations are satisfied:
\begin{equation}
\label{asym_est_1}
\frac{\delta_j}{w} \to 0,  \quad x \to L, \quad i=1,...,N,
\end{equation}
\begin{equation}
\label{asym_est_2}
\frac{2\delta_{N+1}}{w} \to 1, \quad x \to L.
\end{equation}

As can be seen, in the immediate vicinity of the fracture tip the high shear rate Newtonian layer of viscosity $\eta_\infty$ tends to occupy the whole width of the fracture. Thus, over at least some small distance behind the crack front the fracturing fluid behaves like a Newtonian fluid of viscosity $\eta_\infty$. 

Having the above feature in mind, we introduce in our analysis the following formulation of the fluid flow rate:
\begin{equation}
\label{q_car_def}
q=-\frac{1}{16\eta_\infty}w^3\frac{\partial p}{\partial x}F\left(x,t \right),
\end{equation}
where:
\begin{equation}
\label{f_def}
F\left(x,t \right)=-\frac{128\eta_\infty}{w^4}\left(\frac{\partial p}{\partial x} \right)^{-1}\int_0^{w/2}yV\text{d}y.
\end{equation}
The integral on the right hand side of \eqref{f_def} is computed according to formulae \eqref{q_int_1}-\eqref{D_N} for $V$ being the fluid velocity profile in the plane $z=0$ (see Appendix \ref{ap_A}). Note that for the purely Newtonian regime of flow with viscosity $\eta_\infty$ function $F$ assumes a unit value, while for the Newtonian flow at low shear rates with viscosity $\eta_0$ $F$ yields $\eta_\infty/\eta_0$. In particular, when taking into account \eqref{asym_est_2}, one has:
\begin{equation}
\label{f_tip}
F \to 1 , \quad x \to L(t).
\end{equation}
 Thus,  $F(x,t)$ informs us to what degree the solution in a certain spatial and temporal location deviates from the high shear rate Newtonian regime of flow. As such, this function can be very instructive in understanding the underlying flow phenomena and it will be used in our analysis.
 
 \subsection{Truncated power-law model}
 
 The truncated power-law model constitutes a simple regularization of the pure power-law model, where  low and high shear rate cut-off viscosities, $\eta_0$ and $\eta_\infty$, are introduced.  As such it can be considered a special case of approximation \eqref{eta_ap} for $N=2$. This time, up to three shear rate layers can appear within each of the channel symmetrical parts, with two of them being Newtonian-type layers of viscosities $\eta_0$ and $\eta_\infty$, respectively. Consequently, the estimations \eqref{asym_est_1}--\eqref{asym_est_2} hold, which means that in the immediate vicinity of the fracture tip the fluid behaves like a Newtonian fluid of viscosity $\eta_\infty$. For the fluid flow rate, $q$, in this case we keep representation \eqref{q_car_def}--\eqref{f_def} with the computational relations \eqref{q_int_1} - \eqref{D_N} employed. 
 
 As shown in \cite{Lavrov_2015}, the truncated power-law model produces much more reliable results than the pure power-law in terms of fluid velocity and fluid flow rate. However, the presented data suggests that over some ranges of pressure gradient this approximation may not be sufficient for practical applications. This issue will be verified in the paper by comparison with  results obtained for the Carreau model. 
 
\section{The numerical results}
\label{num_res}

In this section we will simulate numerically the process of hydraulic fracture propagation for four different types of fracturing fluids:
\begin{itemize}
\item{The fluid described in \cite{Lavrov_2015}, where the Carreau model parameters and the corresponding truncated power-law coefficients are provided. It will henceforth be called \textit{fluid 1}.}
\item{Hydroxypropylguar (HPG) fluid - the Carreau model parameters are given in \cite{Lecampion_2018}. }
\item{Solution of partially hydrolyzed polyacrylamide (HPAM) with the concentartion of 150 weight parts per million (wppm)  for which the Careau-Yasuda model parameters are available in \cite{Habibpour_2017}.}
\item{Xanthan gum (XG) solution of 600 wppm concentration. The Careau-Yasuda model parameters are taken from  \cite{Habibpour_2017}.}
\end{itemize}

The parameters used for the Carreau-Yasuda and the truncated power-law models are collected in Table \ref{tab_fluid}. Respective values for the truncated power-law model were taken in a way to minimise the maximal relative deviation from the Carreau-Yasuda variant (for \textit{fluid 1} $C$ and $n$ were adopted directly from \cite{Lavrov_2015}). For the pure power-law model we use the same parameters ($C$ and $n$) as for the truncated power-law case. 

\begin{table}[]
\label{tab_fluid}
\begin{center}
\begin{tabular}{|c||c|c|c|c|c||c|c|c|c|}
\hline
\multirow{2}{*}{fluid} & \multicolumn{5}{c||}{Carreau}                   & \multicolumn{4}{c|}{Truncated power-law}        \\ \cline{2-10} 
                         & $\eta_0$, Pa$\cdot$s & $\eta_\infty$, Pa$\cdot$s  & $\lambda$, s & $a$ & $n$ & $C$, Pa$\cdot$s$^n$ & $n$ & $|\dot \gamma_1|$, s$^{-1}$ & $|\dot \gamma_2|$, s$^{-1}$ \\ 
                         \hline
                         \hline
\textit{fluid 1}                & 0.5        & $10^{-3} $            & 600         & 2  & 0.25  & $5\cdot 10^{-3}$  & 0.3  & $1.39\cdot 10^{-3}$                & 9.97               \\ \hline
HPG                      & 0.44        & $10^{-3} $             & 0.303        & 2  & 0.46 & 0.464  & 0.567  & 1.128                & $1.45\cdot 10^{6}$                 \\ \hline
150 wppm HPAM            & 0.2668        & $4.1\cdot 10^{-3}$             & 5.46        & 3.15  & 0.26  & $7.27\cdot 10^{-2}$   & 0.476  & $8.37\cdot 10^{-2}$                 & $241$                 \\ \hline
600 wppm XG              & 0.2689        & $4\cdot 10^{-3}$             & 5.34        & 1.92  & 0.43  & $8.82\cdot 10^{-2}$   & 0.6  & $6.17\cdot 10^{-2}$                 & 2283               \\ \hline
\end{tabular}
\end{center}
\caption{Parameters of the Carreau and the truncated power-law models (for the Carreau model respective data was taken from \cite{Lavrov_2015}, \cite{Lecampion_2018} and \cite{Habibpour_2017}). The limiting viscosities are the same for both models.}
\end{table}

\begin{figure}[htb!]
\begin{center}
\includegraphics[scale=0.4]{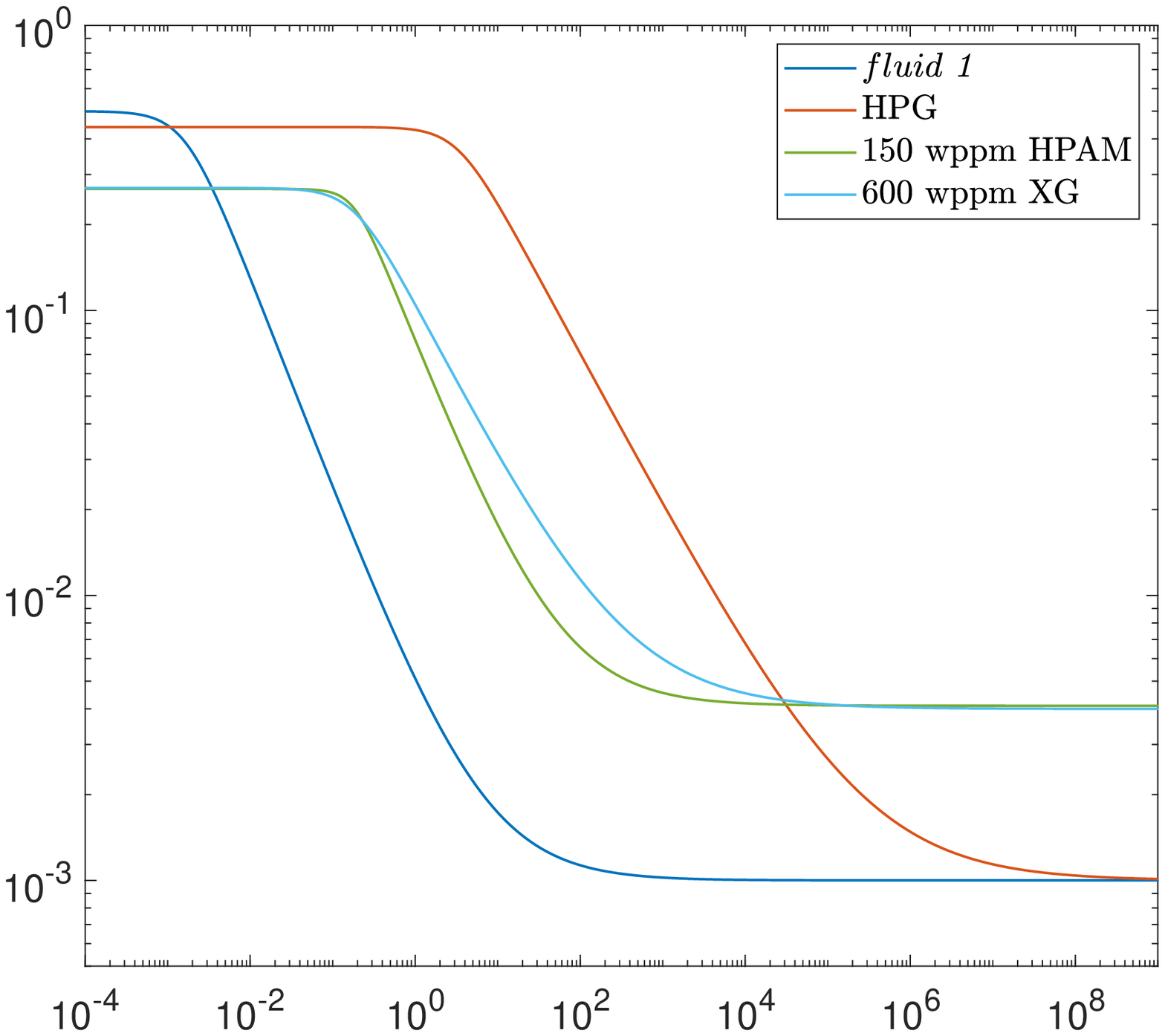}
\put(-110,-5){$|\dot \gamma|$}
\put(-240,90){$\eta_\text{a}$}
\caption{The apparent viscosities, $\eta_\text{a}$ [Pa$\cdot$s], according to the Carreau-Yasuda model for the analysed fluids. Respective parameters are collected in Table \ref{tab_fluid}.}
\label{eta_dif}
\end{center}
\end{figure}

In order to illustrate the influence of fluid rheology on the overall process we will make comparisons for two pairs of fluids: i) \textit{fluid 1} and HPG, ii) 150 wppm HPAM and 600 wppm XG. Note that in each of these pairs the limiting viscosities ($\eta_0$ and $\eta_\infty$) are virtually the same. As can be seen in Fig. \ref{eta_dif} however, the interim behaviours between the viscosity plateaus are different. For the first pair (\textit{fluid 1} and HPG) one can observe a very large (a few orders of magnitude in $|\dot \gamma|$) translation of the $\eta_\text{a}$ characteristics towards the high shear rate values for the HPG fluid. For the second pair similar trend is much less pronounced. 

Assuming some typical values of the HF parameters for the PKN model we will investigate to what degree the aforementioned rheological features affect the fracture evolution and whether the simplified viscosity models (power-law and truncated power-law) can be considered credible substitutes for the original Carreau rheology. Following e.g. \cite{Huang_2004,Wang_2018} we set: $E=10$ GPa, $\nu=0.2$, $H=20$ m, $q_0=1.5 \cdot 10^{-4}$ $ \frac{\text{m}^2}{\text{s}} $ (note that for the fluid influx the normalisation \eqref{q_norm} holds). The influx magnitude is increased from zero for $t=0$ s to the maximum  $q_0$ at $t_1=10$ s and then kept constant according to the following formula:
\begin{equation}
\label{TP_def}
\bar q_0(t)=
  \begin{cases}
		\left(\frac{3}{t_1^2}t^2-\frac{2}{t_1^3}t^3\right)q_0      & \quad t<t_1,\\
    q_0       & \quad \text{for} \quad t \geq t_1.
  \end{cases}
\end{equation}
This particular choice of $t_1$ enables observation of fracture evolution under gradual increase of $\bar q_0$. The leak-off to the rock formation is neglected ($q_l=0$) and the initial fracture length and velocity are assumed zero. The overall time of the process is set to $t_\text{end}=3600$ [s].

The computations are performed by the HF solver introduced in \cite{Wrobel_2015,Perkowska_2016} with some modifications to implement formula \eqref{q_sum} instead of the classical Poiseulle-type relation. The scheme is based on two blocks: i) subroutine computing the fluid velocity from the continuity equation \eqref{cont}, ii) subroutine for the fracture opening utilising elasticity operator \eqref{elast}. For the Carreau-Yasuda models we employ approximation \eqref{eta_ap} for $N=100$ which, according to the analysis conducted in \cite{Wrobel_Arxiv}, provides the accuracy of the order $10^{-5}$ for both, the apparent viscosity itself and the fluid flow rate.

\subsection{\textit{Fluid 1} and HPG fluid}

\begin{figure}[htb!]
\begin{center}
\includegraphics[scale=0.38]{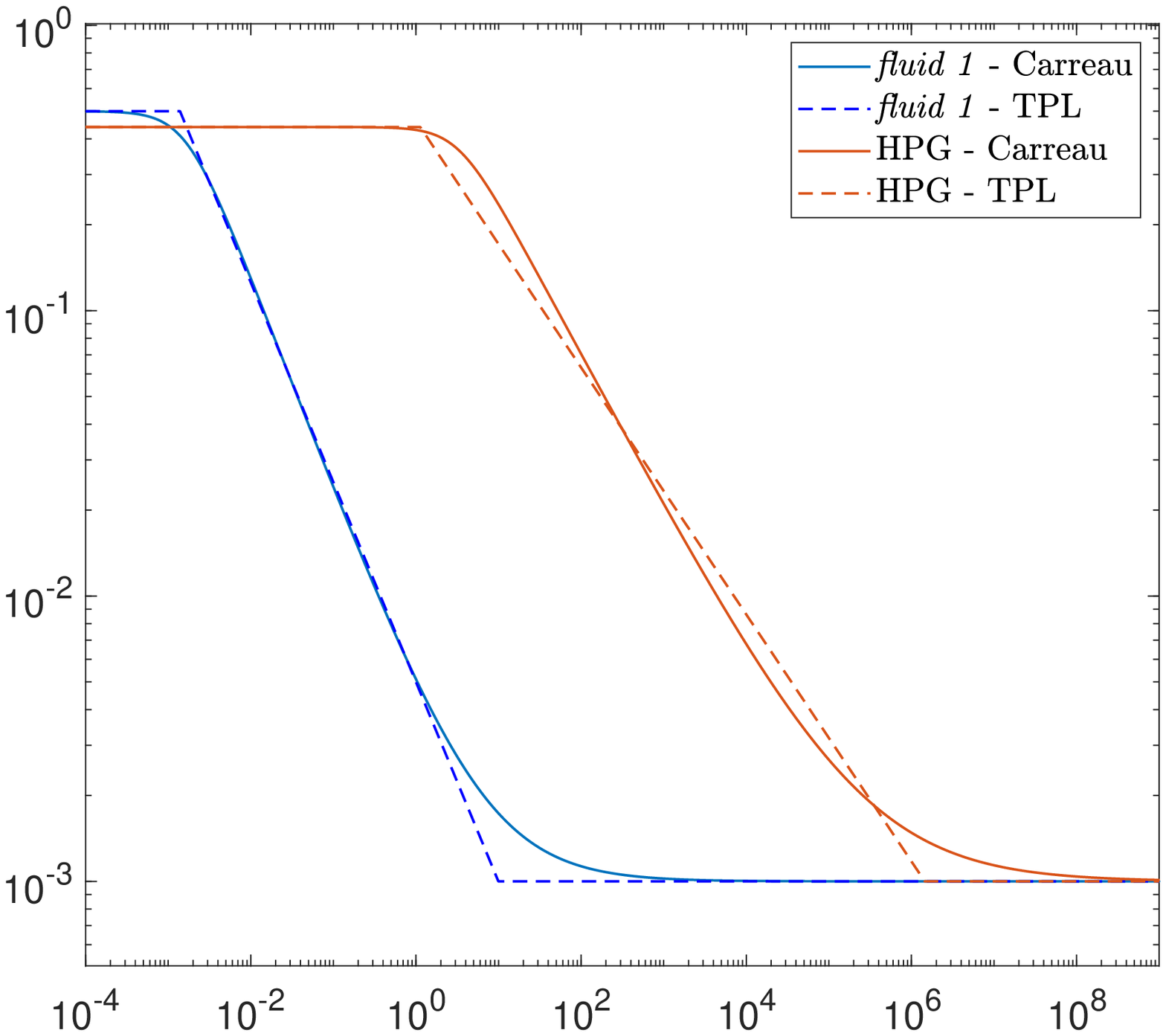}
%\hspace{0mm}
\includegraphics[scale=0.38]{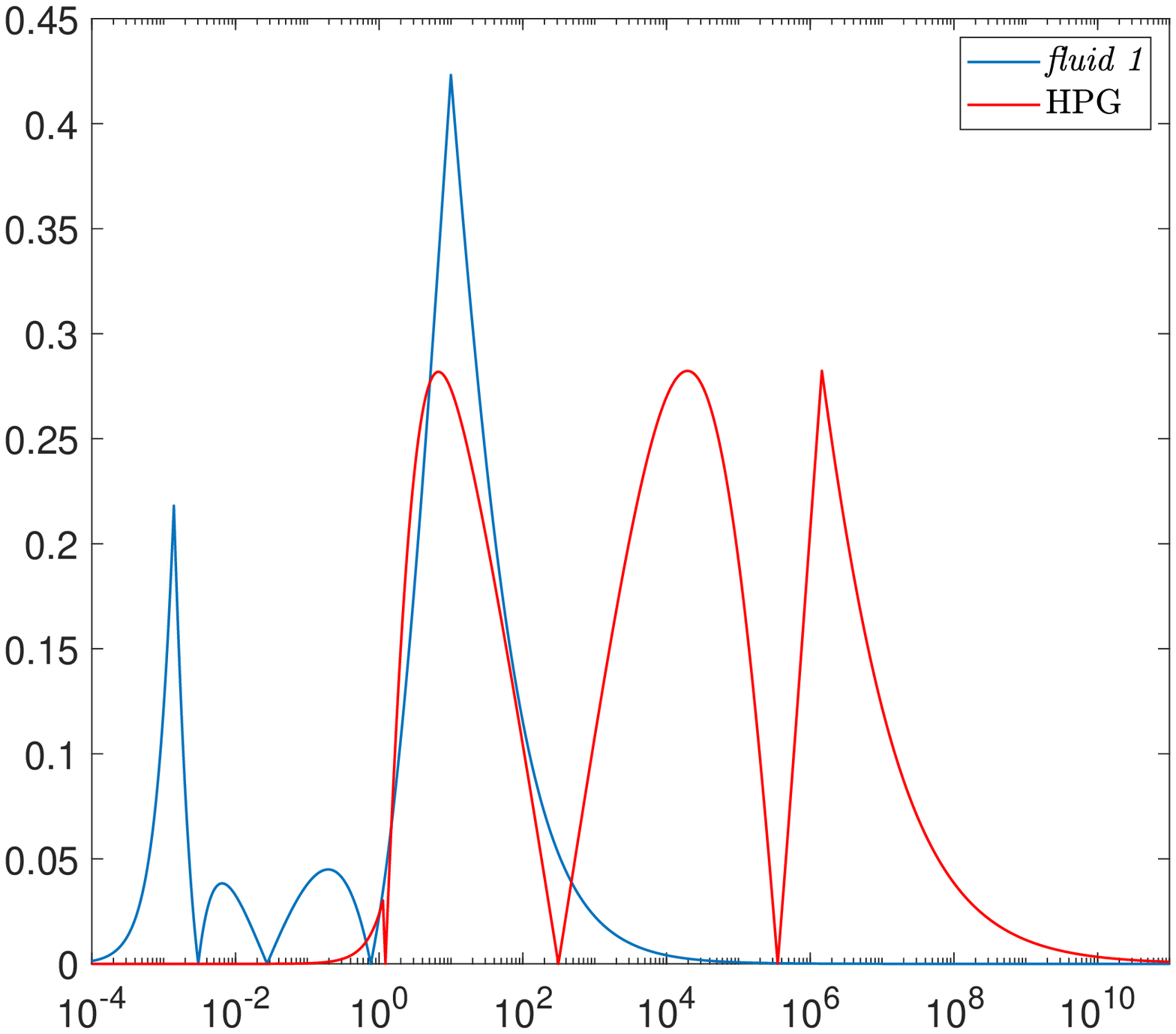}
\put(-338,-5){$|\dot \gamma|$}
\put(-110,-5){$|\dot \gamma|$}
\put(-446,165){$\textbf{a)}$}
\put(-220,165){$\textbf{b)}$}
\put(-440,90){$\eta_\text{a}$}
\put(-220,90){$\delta \eta_\text{a}$}
\caption{\textit{Fluid 1} and HPG fluid: a) apparent viscosities for Carreau and truncated power-law (TPL) rheologies, $\eta_\text{a}$ [Pa$\cdot$s], b)  relative deviations between the Carreau and the truncated power-law variants, $\delta \eta_\text{a}$.}
\label{Lav_HPG}
\end{center}
\end{figure}

We start our analysis with the first two fluids from Table \ref{tab_fluid}. The graphs of apparent viscosities, $\eta_\text{a}$, are shown in Fig. \ref{Lav_HPG}a) for the Carreau and the truncated power-law (TPL) rheologies. The quality of approximation of the respective Carreau characteristics by their truncated power-law imitations measured by the relative differences, $\delta \eta_\text{a}$, are depicted  Fig. \ref{Lav_HPG}b). We see that for \textit{fluid 1} the maximal error of approximation reaches over 40$\%$, while for the HPG the highest deviation is below 30$\%$.

\begin{figure}[htb!]
\begin{center}
\includegraphics[scale=0.38]{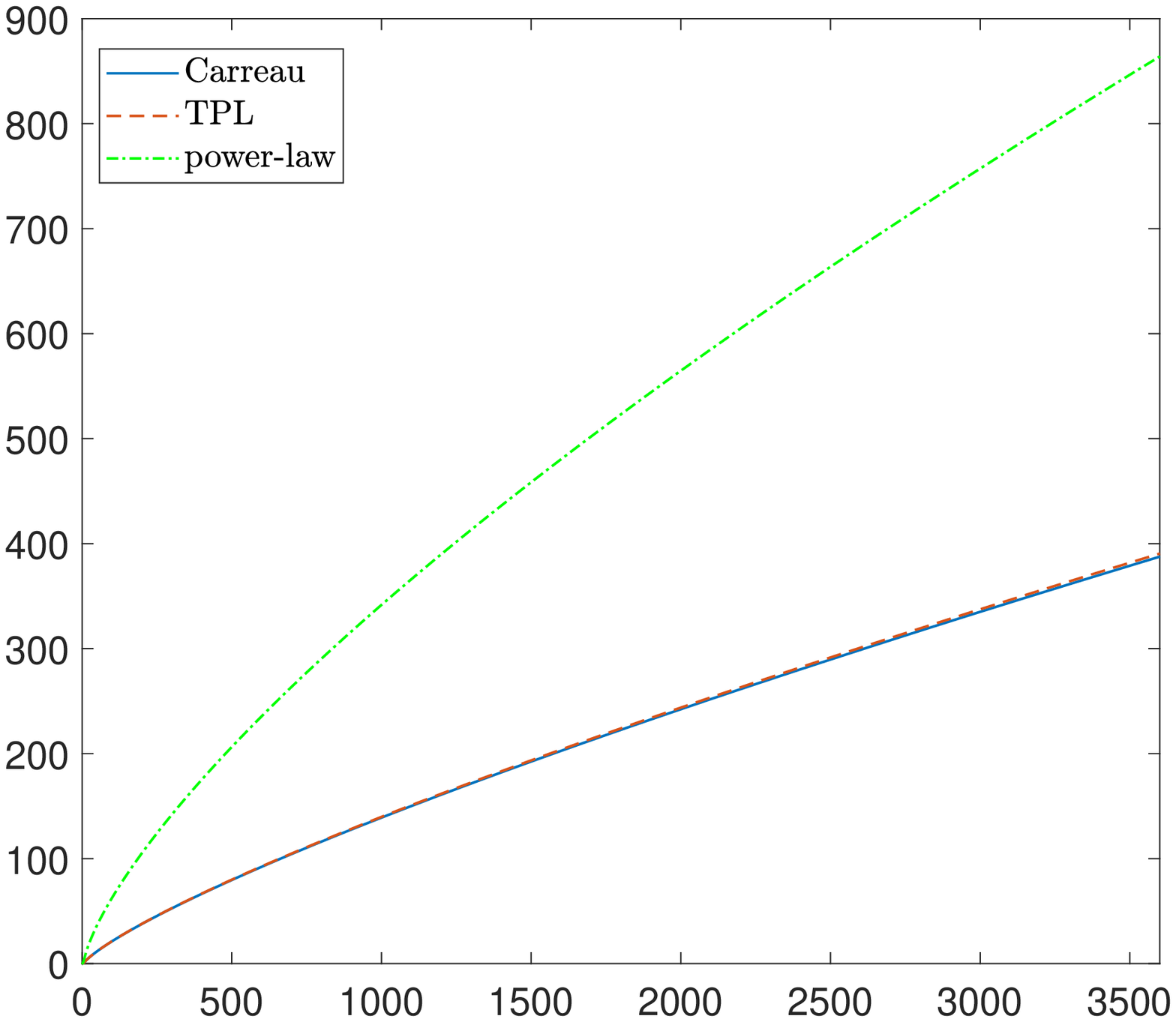}
%\hspace{0mm}
\includegraphics[scale=0.38]{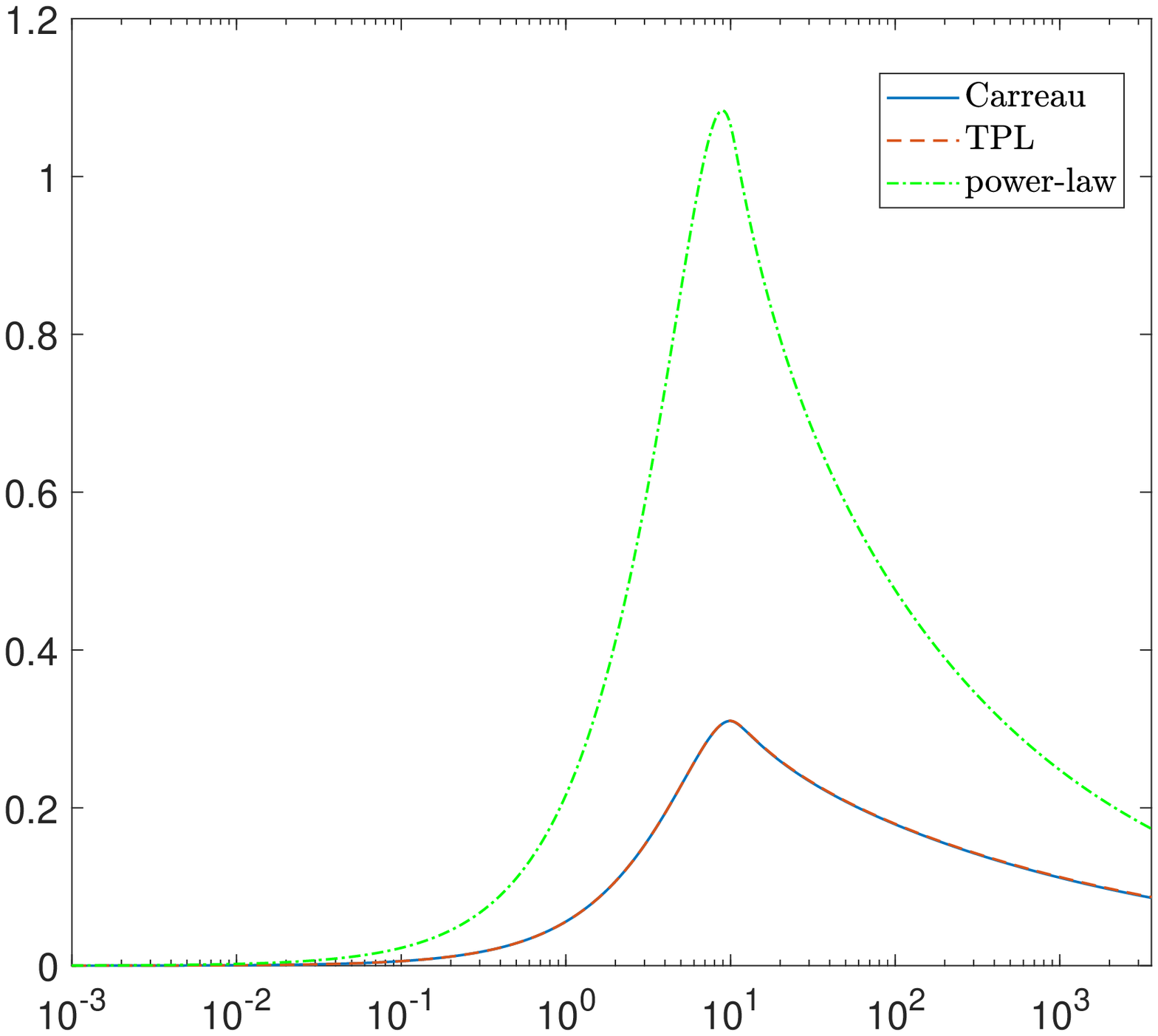}
\put(-338,-5){$t$}
\put(-110,-5){$t$}
\put(-446,165){$\textbf{a)}$}
\put(-220,165){$\textbf{b)}$}
\put(-440,90){$L$}
\put(-220,90){$v_0$}
\caption{Simulation results for \textit{fluid 1}: a) the crack length, $L$ [m], b) the crack propagation speed, $v_0$ $\left[\frac{\text{m}}{\text{s}}\right]$.}
\label{Lav_L}
\end{center}
\end{figure}

\begin{figure}[htb!]
\begin{center}
\includegraphics[scale=0.38]{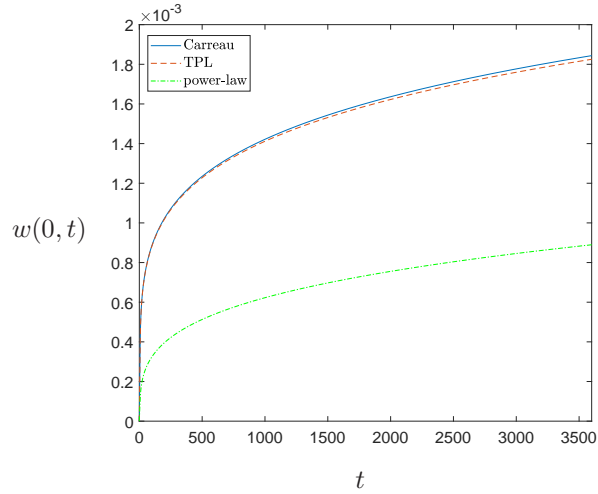}
\put(-110,-5){$t$}
\put(-240,90){$w(0,t)$}
\caption{Simulation results for \textit{fluid 1}: the fracture opening at the crack mouth, $w(0,t)$ [m].}
\label{W0_Lav}
\end{center}
\end{figure}

The simulation results for \textit{fluid 1} in terms of: i) the fracture length, $L$, ii) the crack propagation speed, $v_0$, and iii) the fracture opening at the crack mouth, $w(0,t)$, are shown in Figs. \ref{Lav_L}--\ref{W0_Lav}. As can be seen, the results obtained for the truncated power-law rheology are almost indistinguishable from those for the Carreau fluid (the relative deviations from the Carreau variant are well below 1$\%$ in almost the entire time interval). At the same time, for the power-law fluid one has huge overestimation of the crack length and the crack propagation speed with the substantial underestimation of the crack opening. This data suggests that the apparent viscosity of fluid obtained for the power-law model is much lower than that achieved with the remaining rheologies.

\begin{figure}[htb!]
\begin{center}
\includegraphics[scale=0.38]{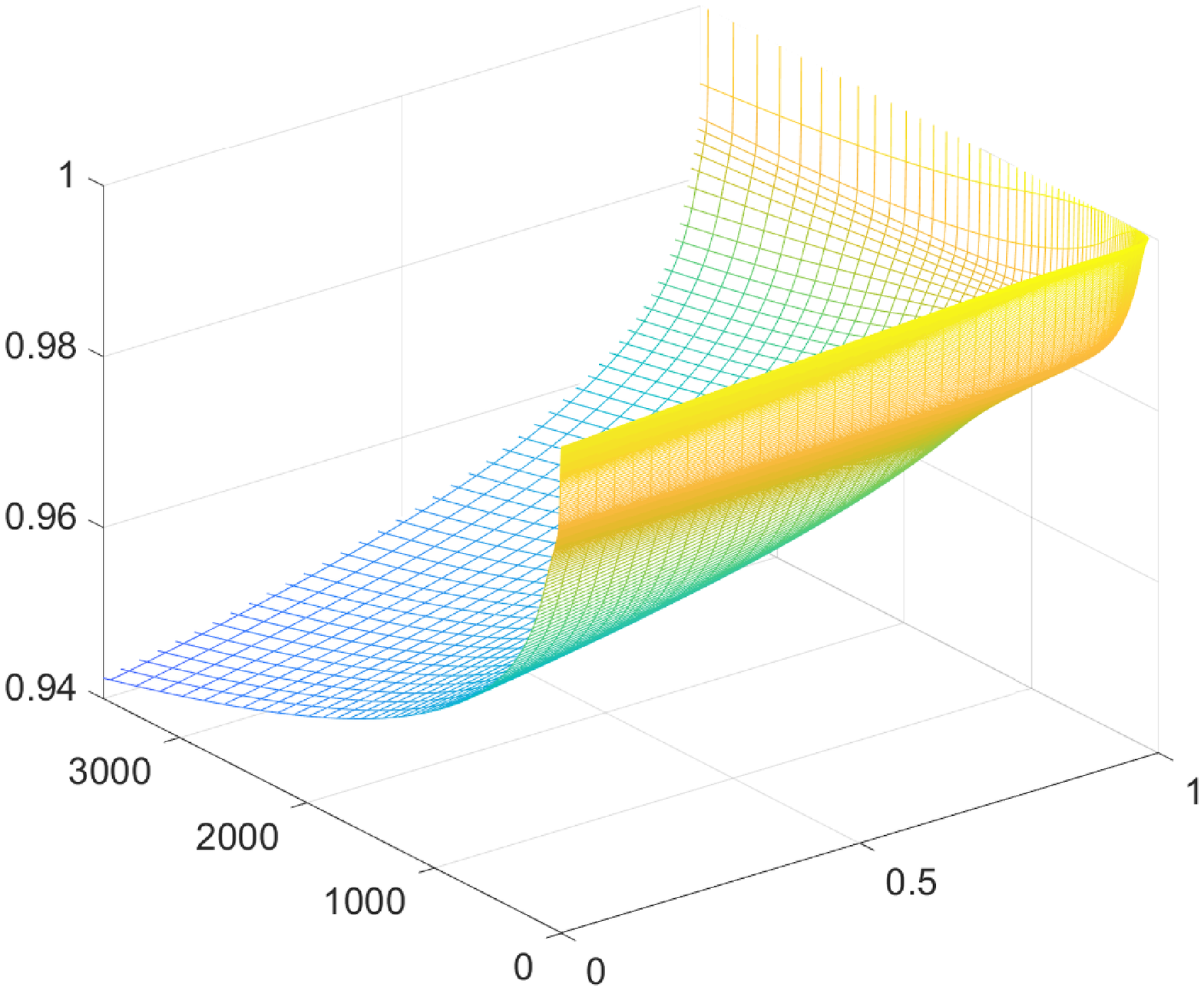}
%\hspace{0mm}
\includegraphics[scale=0.38]{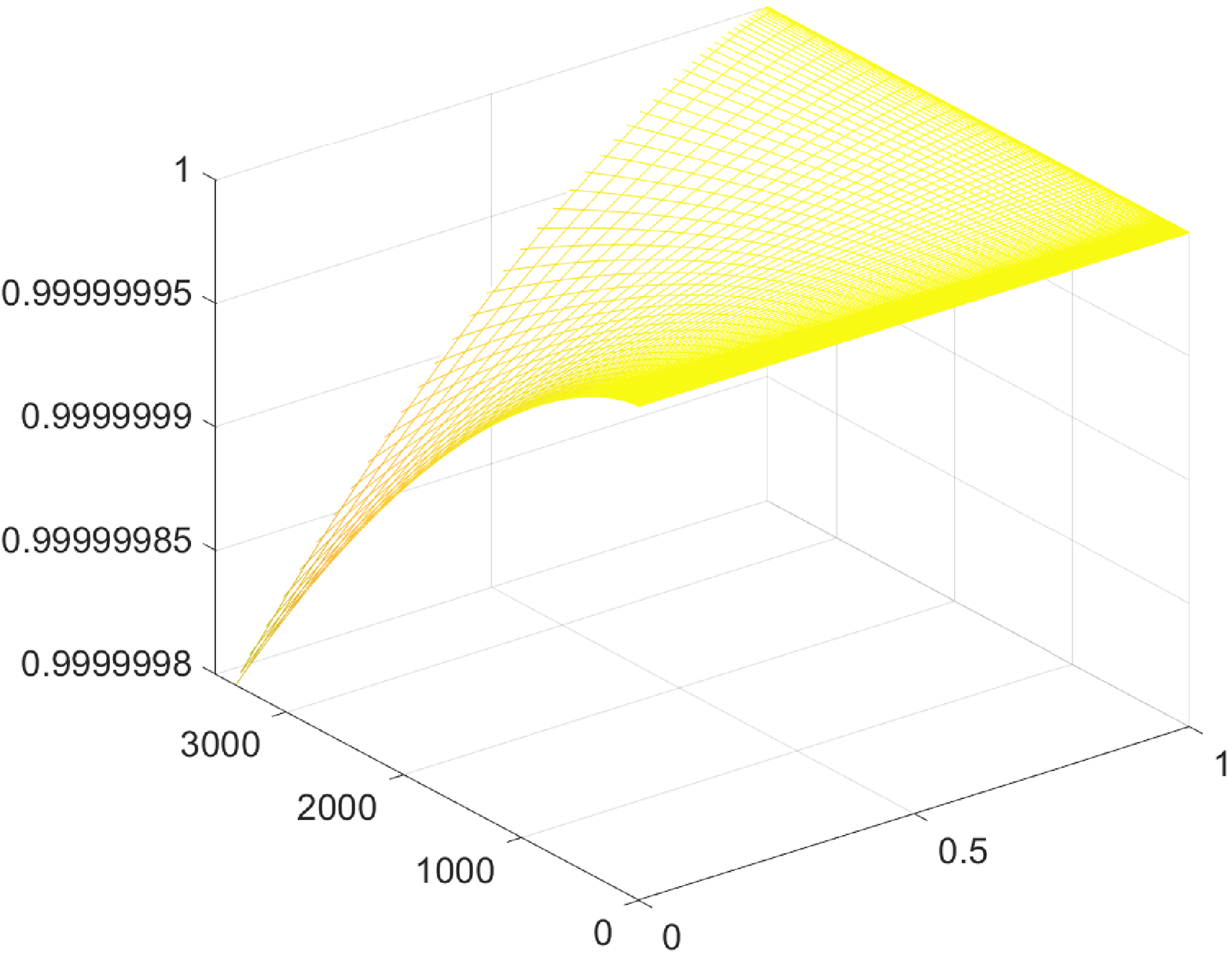}
\put(-405,20){$t$}
\put(-285,10){$x/L$}
\put(-170,20){$t$}
\put(-60,10){$x/L$}
\put(-446,165){$\textbf{a)}$}
\put(-220,165){$\textbf{b)}$}
\caption{Function $F(x,t)$ for \textit{fluid 1}: a) the Carreau rheology, b) the truncated power-law rheology.}
\label{F_Lav}
\end{center}
\end{figure}

In Fig. \ref{F_Lav} we show  distributions of the fluid flux component function $F$ (see equations \eqref{q_car_def}--\eqref{f_tip}) over space and time for the Carreau and the truncated power-law models. In both cases we see that with time growing the fracture deviates from the high shear rate Newtonian regime of flow with viscosity $\eta_\infty$. However, in the considered time span, this deviation is either very low (Carreau) or virtually negligible (truncated power-law). Thus, we can conclude that within the whole duration of the simulated process the fluid is subjected to the shear rates that are sufficient to yield the apparent viscosity very close to the limiting value  $\eta_\infty$. 

In order to quantify this trend let us introduce the following parameters:
\begin{itemize}
\item{The  fluid shear rate averaged over fracture cross-section (see Fig. \ref{PKN_geom} and Fig. \ref{channel} for schematic view of the integration area):
\begin{equation}
\label{Gam_def}
\Gamma(t)=\frac{2}{L(t)}\int_0^{L(t)}\frac{1}{w(x,t)}\int_0^{\frac{w(x,t)}{2}}\dot \gamma(x,y,t)\text{d}y\text{d}x.
\end{equation}
The values of shear rate, $\dot \gamma$, are obtained in post-processing computations. The basic computational relation here is equation \eqref{sol_2_cyl} treated as a non-linear algebraic equation with respect to $\dot \gamma$. The procedure to compute and integrate the shear rate function is described in \cite{Wrobel_Arxiv}.}
\item{The apparent viscosity averaged over fracture cross-section:
\begin{equation}
\label{E_def}
E(t)=\eta_\text{a}(\Gamma),
\end{equation}
where  formula for $\eta_\text{a}$ is taken according to the respective rheological model.}
\end{itemize}

\begin{figure}[htb!]
\begin{center}
\includegraphics[scale=0.38]{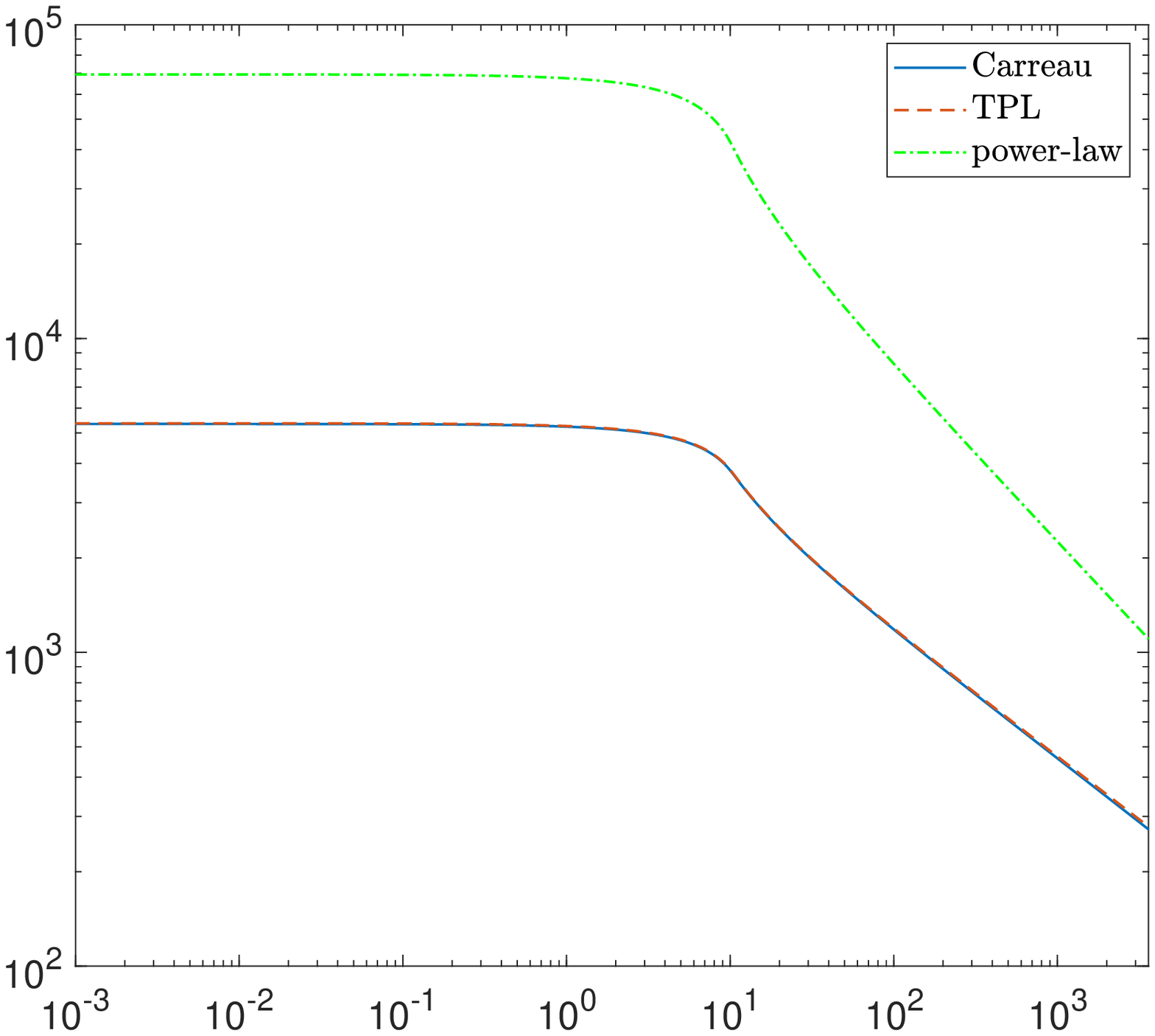}
%\hspace{0mm}
\includegraphics[scale=0.38]{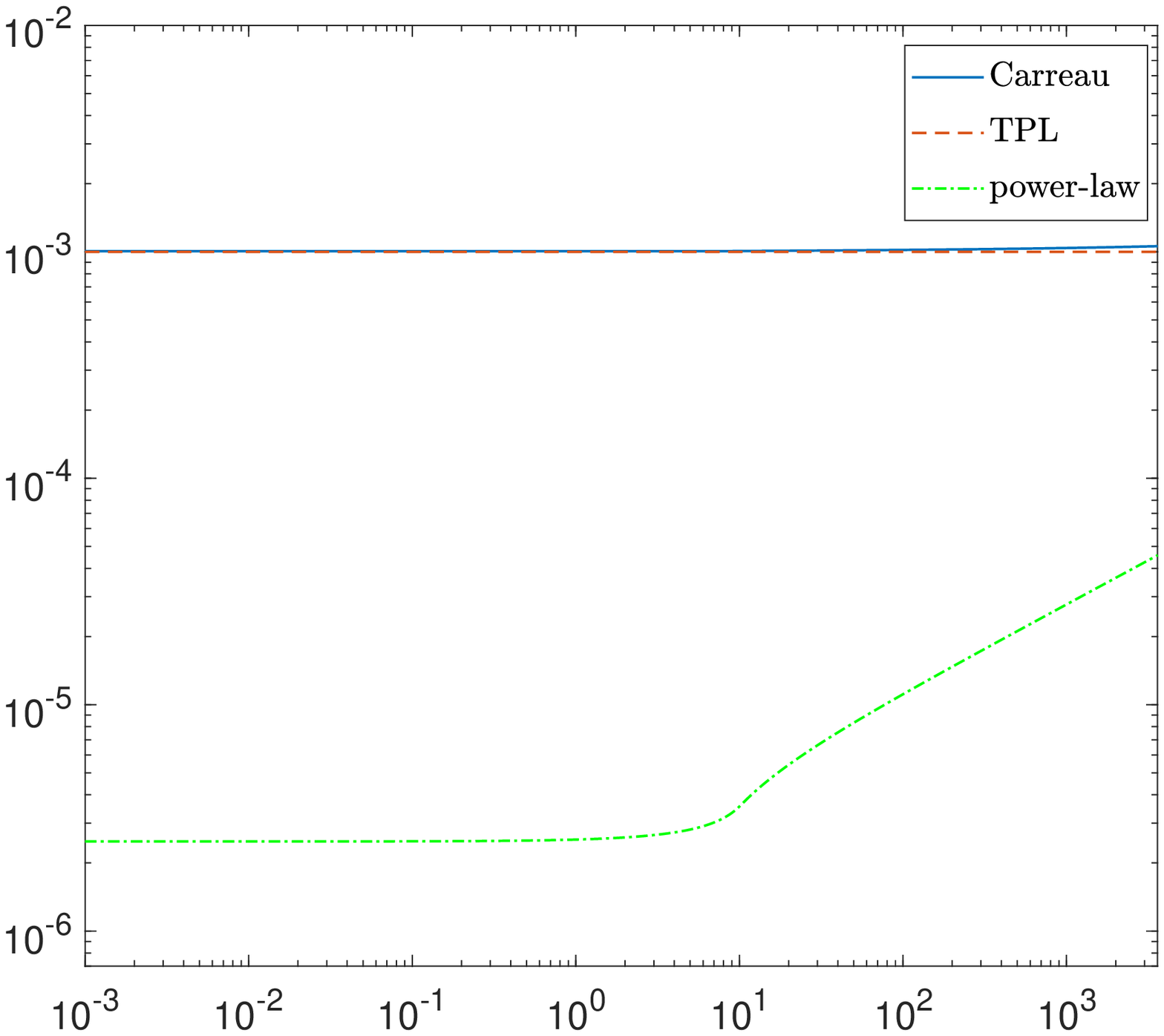}
\put(-338,-5){$t$}
\put(-110,-5){$t$}
\put(-446,165){$\textbf{a)}$}
\put(-220,165){$\textbf{b)}$}
\put(-440,90){$|\Gamma|$}
\put(-220,90){$E$}
\caption{The  average values of: a) shear rates, $\Gamma$ $\left[\frac{1}{\text{s}} \right]$, b) apparent viscosities, $E$ [Pa$\cdot$s], for \textit{fluid 1}.}
\label{Lav_mean}
\end{center}
\end{figure}

Temporal behaviours of $\Gamma$ and $E$ for the analysed rheologies are shown in Fig. \ref{Lav_mean}. As anticipated, the average shear rates for the Carreau and truncated power-law models are almost identical. The same holds for the respective viscosities. Notably, the minimum of $\Gamma$ for the Carreau variant was 197 $\frac{1}{\text{s}}$ (202 $\frac{1}{\text{s}}$ for the truncated power-law) while the maximal value amounted to $5.34 \cdot 10^{3}$ $\frac{1}{\text{s}}$ ($5.37 \cdot 10^{3}$ $\frac{1}{\text{s}}$ for the truncated power-law). It basically means that over the whole considered time interval the fluid flow is in the high shear rate regime. Even for $t_\text{end}$ the average shear rate is well above the higher cut-off value, $|\dot \gamma_2|$, assumed for the truncated power-law model (compare Table \ref{tab_fluid}). Thus, the resulting average viscosity for the truncated power-law is equal to $\eta_\infty$ for any $t$, and it barely deviates from $\eta_\infty$ when employing the Carreau rheology. On the other hand, for the pure power-law variant the averaged shear rates are highly overestimated, which results in very low values of the apparent viscosity. However, even if one used the shear rate values obtained for the Carreau or truncated power-law models in the power-law relation \eqref{PL_eta}, the resulting viscosities would be much underestimated.

\begin{figure}[htb!]
\begin{center}
\includegraphics[scale=0.38]{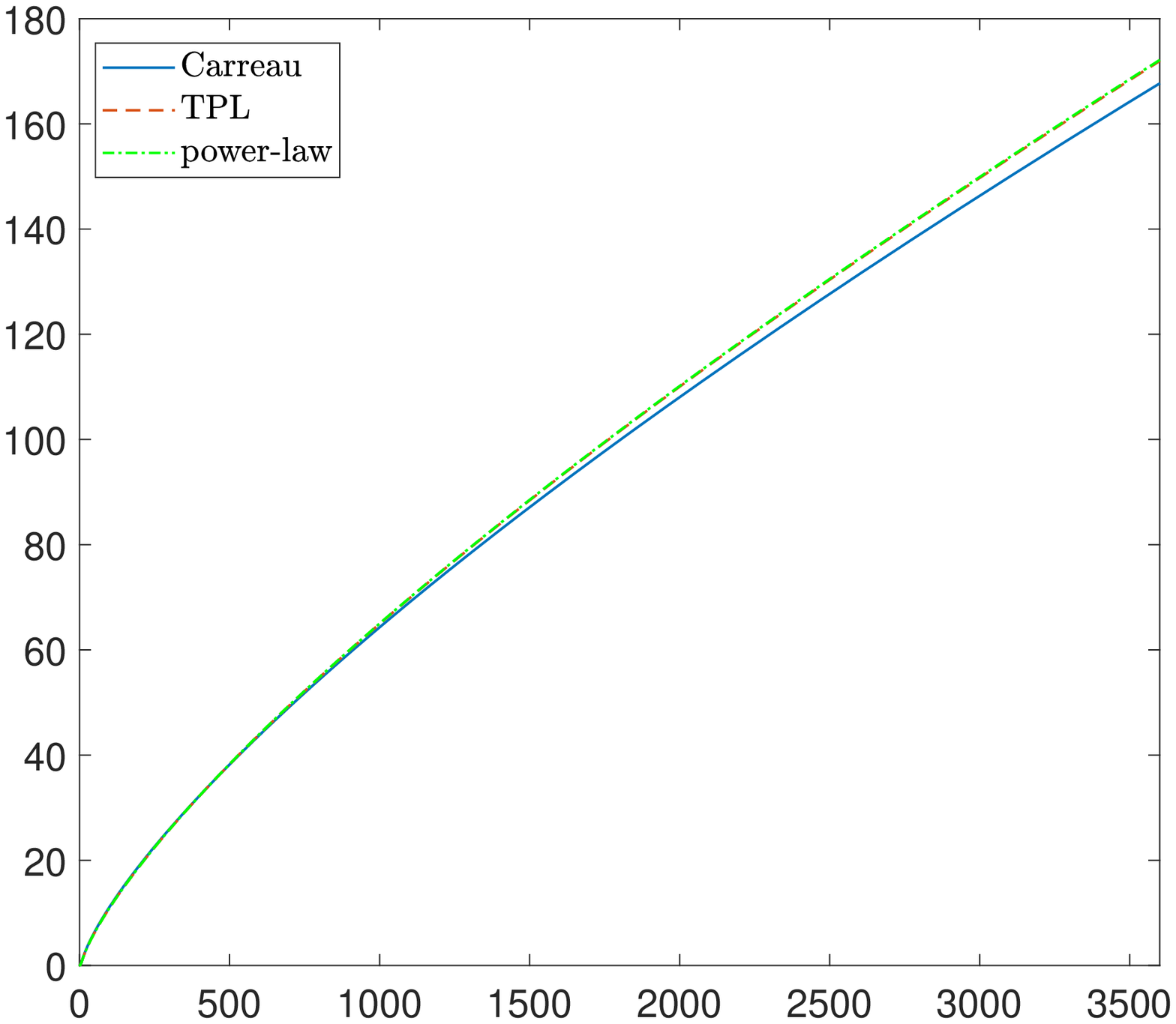}
%\hspace{0mm}
\includegraphics[scale=0.38]{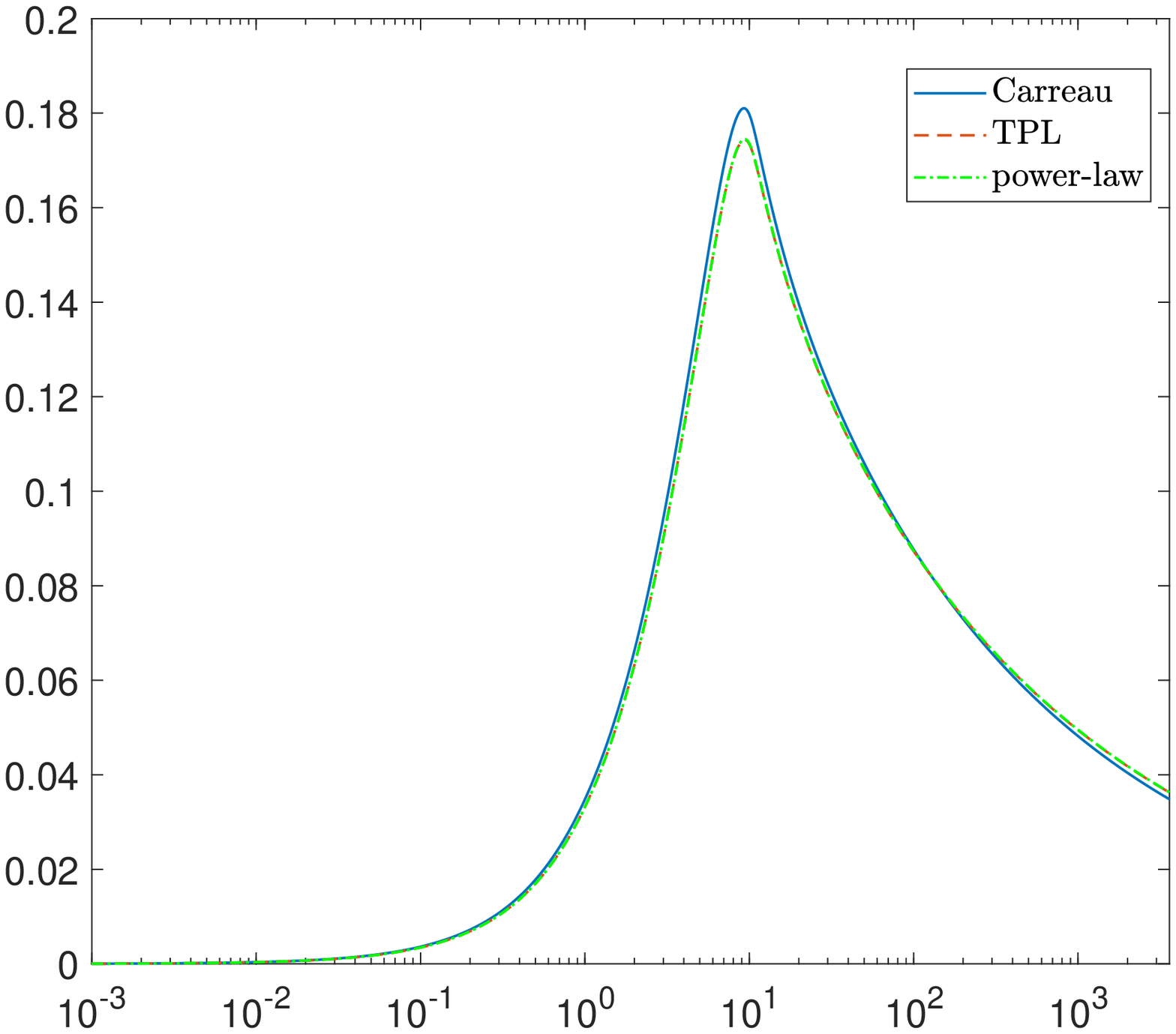}
\put(-338,-5){$t$}
\put(-110,-5){$t$}
\put(-446,165){$\textbf{a)}$}
\put(-220,165){$\textbf{b)}$}
\put(-440,90){$L$}
\put(-220,90){$v_0$}
\caption{Simulation results for the HPG fluid: a) the crack length, $L$ [m], b) the crack propagation speed, $v_0$ $\left[\frac{\text{m}}{\text{s}}\right]$.}
\label{HPG_L}
\end{center}
\end{figure}

\begin{figure}[htb!]
\begin{center}
\includegraphics[scale=0.38]{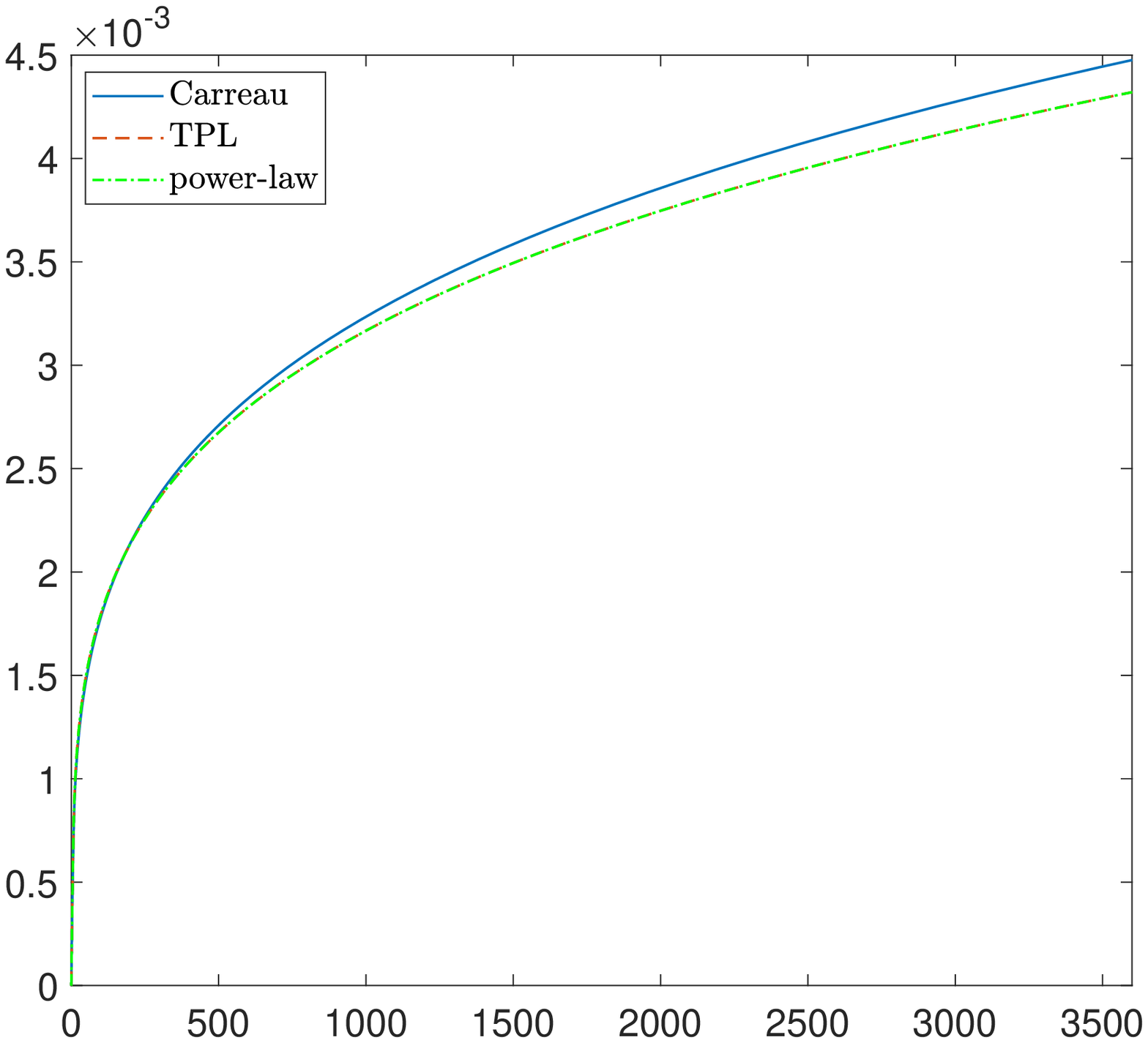}
\put(-110,-5){$t$}
\put(-240,90){$w(0,t)$}
\caption{Simulation results for the HPG fluid: the fracture opening at the crack mouth, $w(0,t)$ [m].}
\label{W0_HPG}
\end{center}
\end{figure}

Now let us analyse  results obtained for the HPG fluid. The graphs for $L(t)$, $v_0(t)$ and $w(0,t)$ are shown in Figs. \ref{HPG_L}--\ref{W0_HPG}. As can be seen, this time the results for respective rheologies are very close to each other, with the truncated power-law characteristics being indistinguishable from those produced with the pure power law. 

\begin{figure}[htb!]
\begin{center}
\includegraphics[scale=0.38]{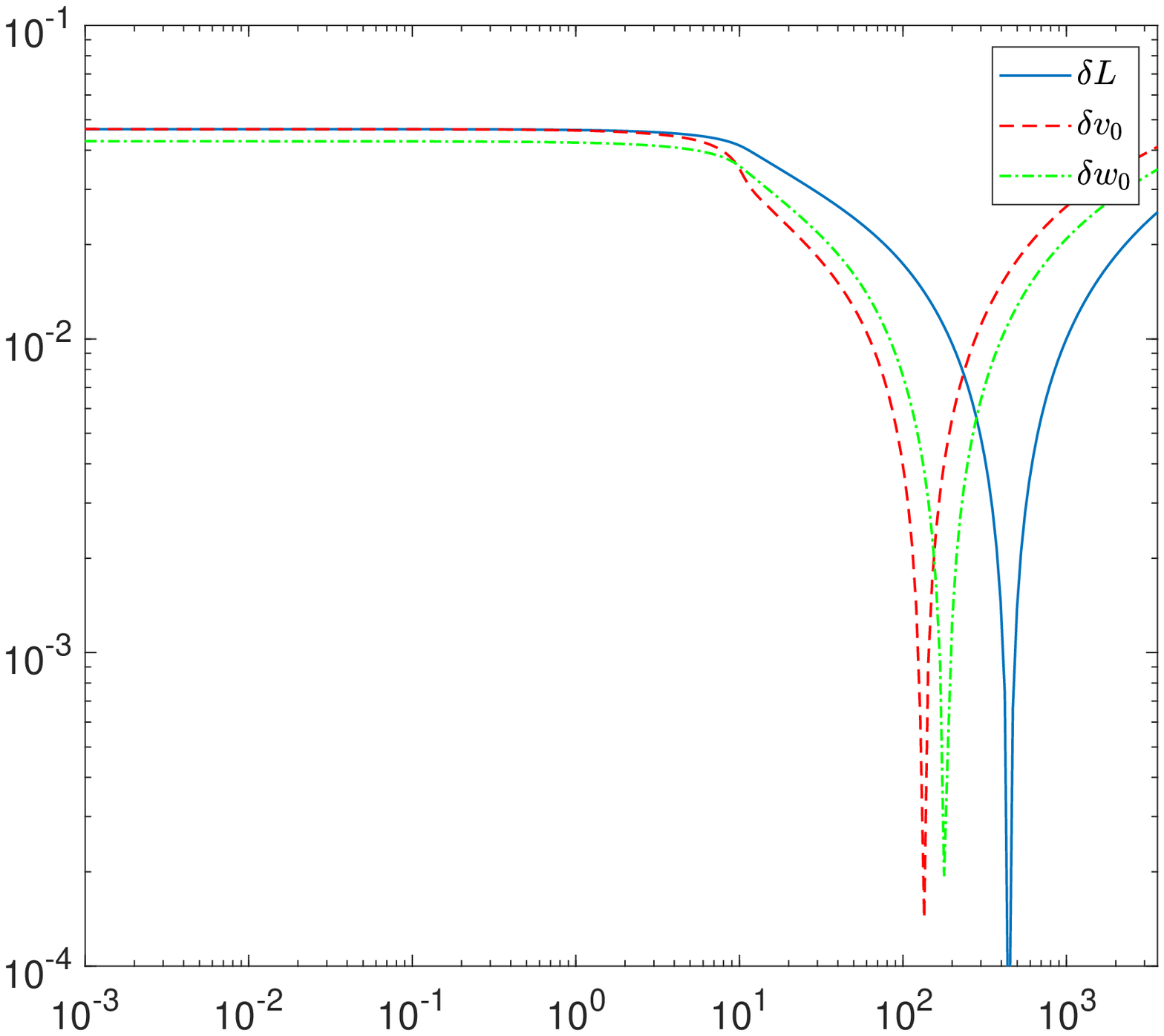}
%\hspace{0mm}
\includegraphics[scale=0.38]{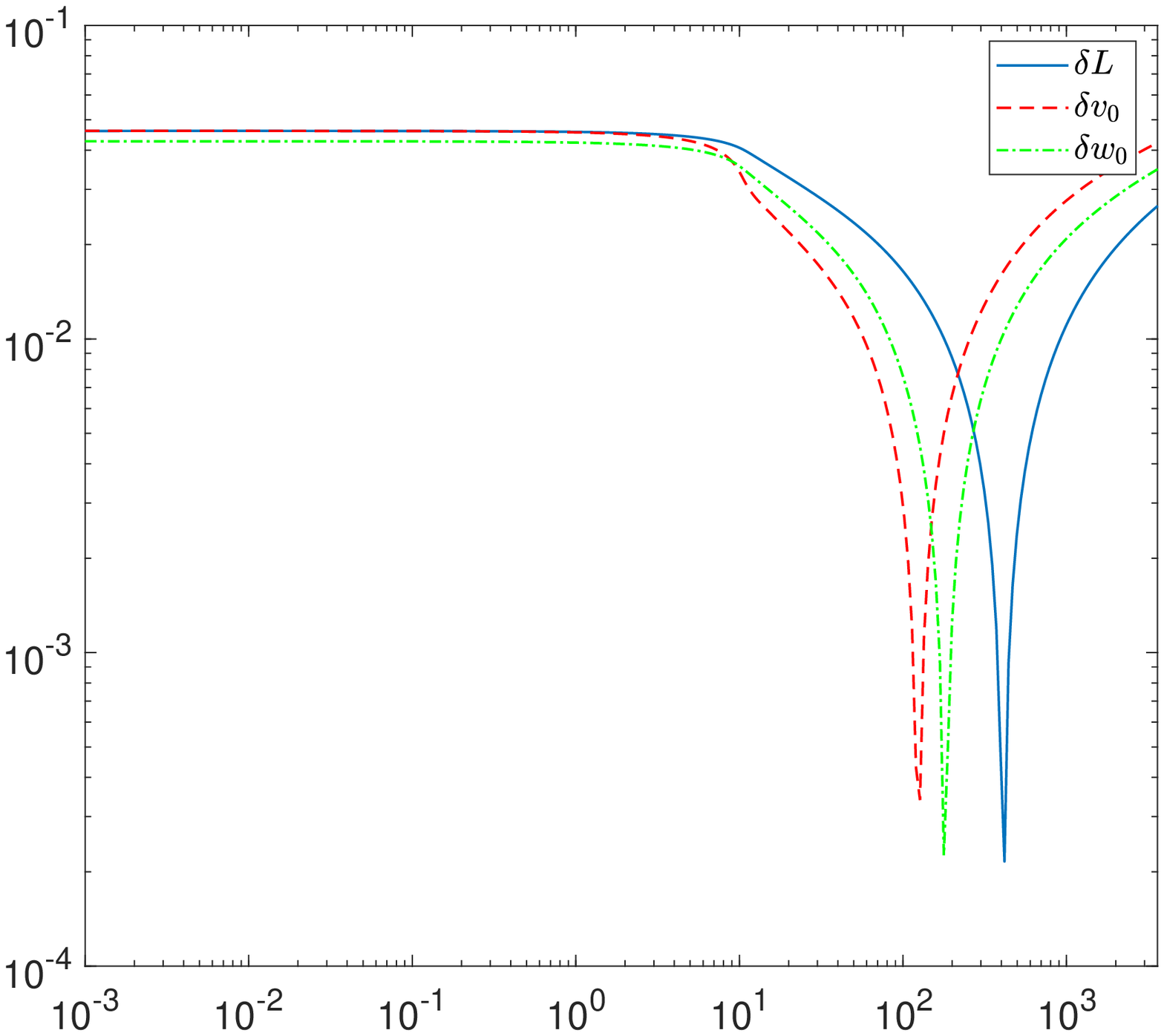}
\put(-338,-5){$t$}
\put(-110,-5){$t$}
\put(-446,165){$\textbf{a)}$}
\put(-220,165){$\textbf{b)}$}
\caption{The relative deviation of solution from the Carreau variant for the HPG fluid in the case of: a) the truncated power-law rheology b) the power-law rheology.}
\label{HPG_bledy}
\end{center}
\end{figure}

The relative deviations of: i) the crack length, $\delta L$, ii) the crack propagation speed, $\delta v_0$, iii) the fracture opening at the crack mouth, $\delta w_0$, from respective values obtained for the Carreau variant  are shown in Fig. \ref{HPG_bledy} for the truncated power-law and the power-law rheologies. In the analysed time span neither of the shown parameters exceeds 5$\%$, which constitutes a very good approximation of the Carreau solution for practical purposes.

\begin{figure}[htb!]
\begin{center}
\includegraphics[scale=0.38]{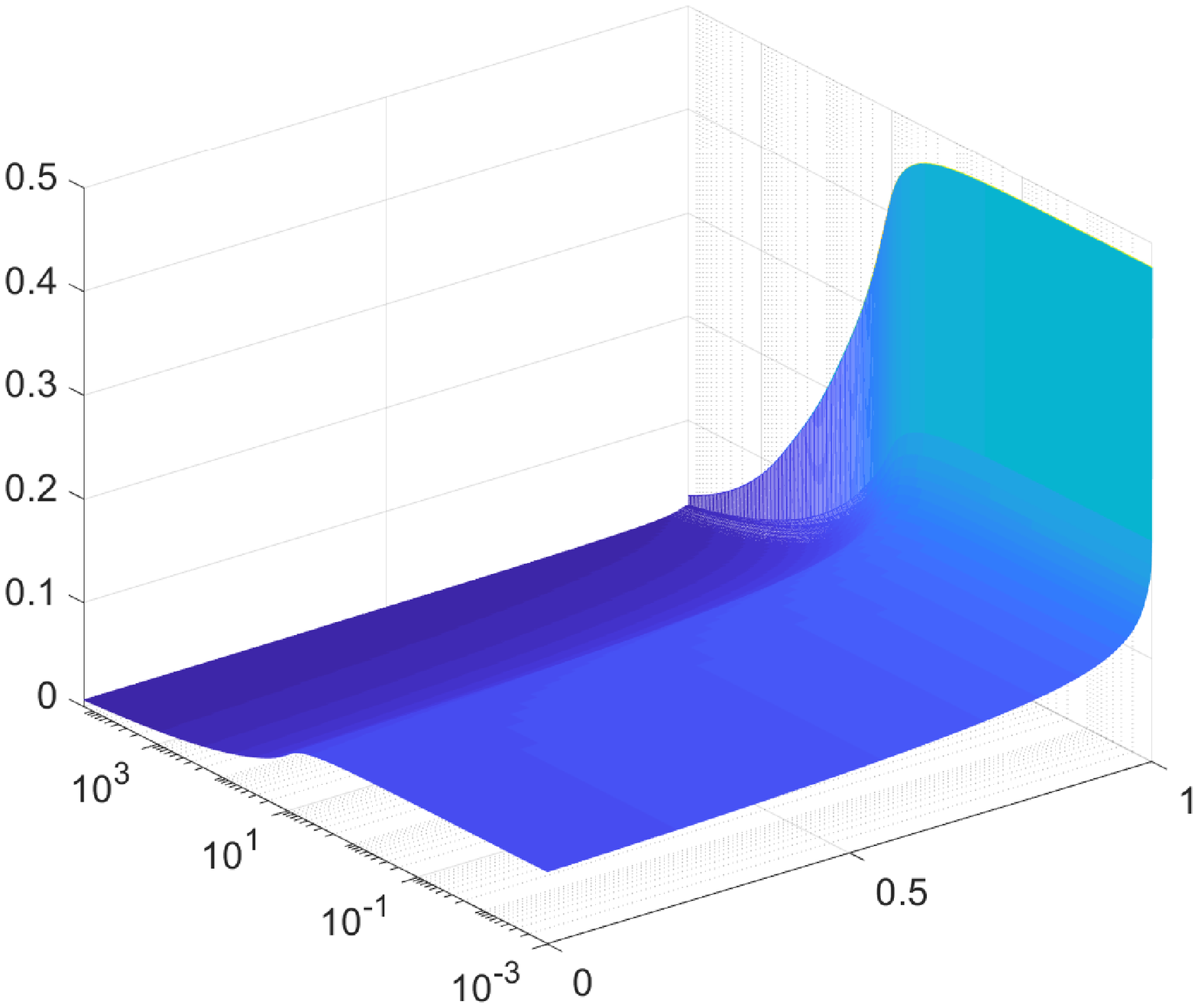}
%\hspace{0mm}
\includegraphics[scale=0.38]{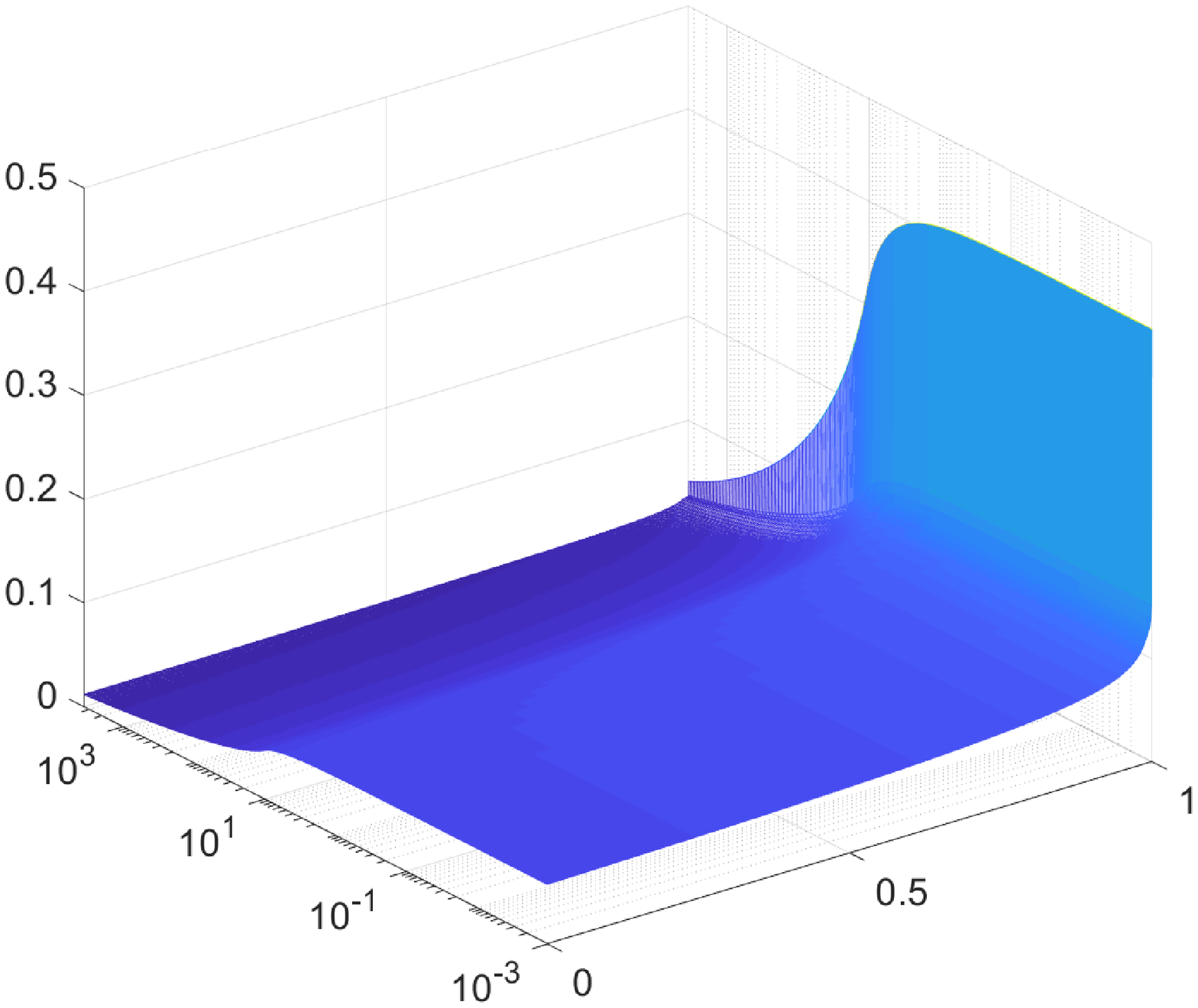}
\put(-405,20){$t$}
\put(-285,10){$x/L$}
\put(-170,20){$t$}
\put(-60,10){$x/L$}
\put(-446,165){$\textbf{a)}$}
\put(-220,165){$\textbf{b)}$}
\caption{Function $F(x,t)$ for the HPG fluid: a) the Carreau rheology, b) the truncated power-law rheology.}
\label{F_HPG}
\end{center}
\end{figure}

The fluid flux component function, $F$, is depicted in Fig. \ref{F_HPG} for the Carreau and the truncated power-law models. We see that with  growing time the fluid flow evolves towards the low shear rate regime. However, unlike the \textit{fluid 1} example, even for small time we  are relatively far away from the high shear rate regime. It is only the near-tip zone where $F$ grows appreciably.\footnote{The employed computational algorithm assumes that  the spatial domain is truncated to the dimension $x \in [0, L(1-\varepsilon)]$ where $\varepsilon$ is a regularisation parameter. For this reason the results displayed in Fig. \ref{F_HPG} do not cover the small near-tip region $x \in [L(1-\varepsilon),L]$ and thus the limiting value $F=1$ at the crack tip is not visible here. In the computations we assumed $\varepsilon=10^{-6}$. For more details on the regularisation technique see \cite{Wrobel_2015,Perkowska_2016}.} This, together with very good coincidence between the results obtained for various fluid rheologies, suggests that under the analysed values of the HF process parameters it is the intermediate (power law) part of the viscosity characteristics that affects final solution the most.

\begin{figure}[htb!]
\begin{center}
\includegraphics[scale=0.38]{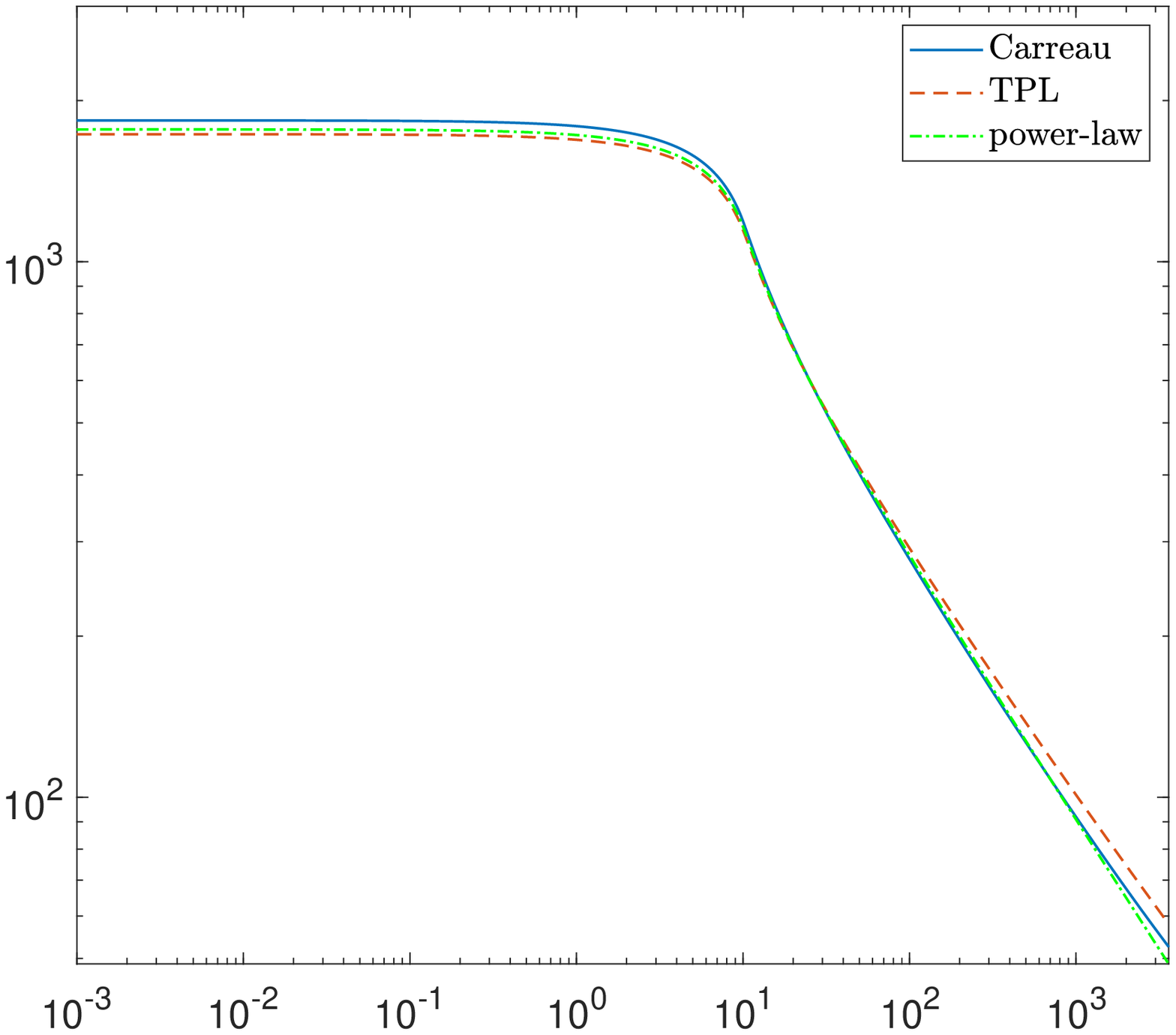}
%\hspace{0mm}
\includegraphics[scale=0.38]{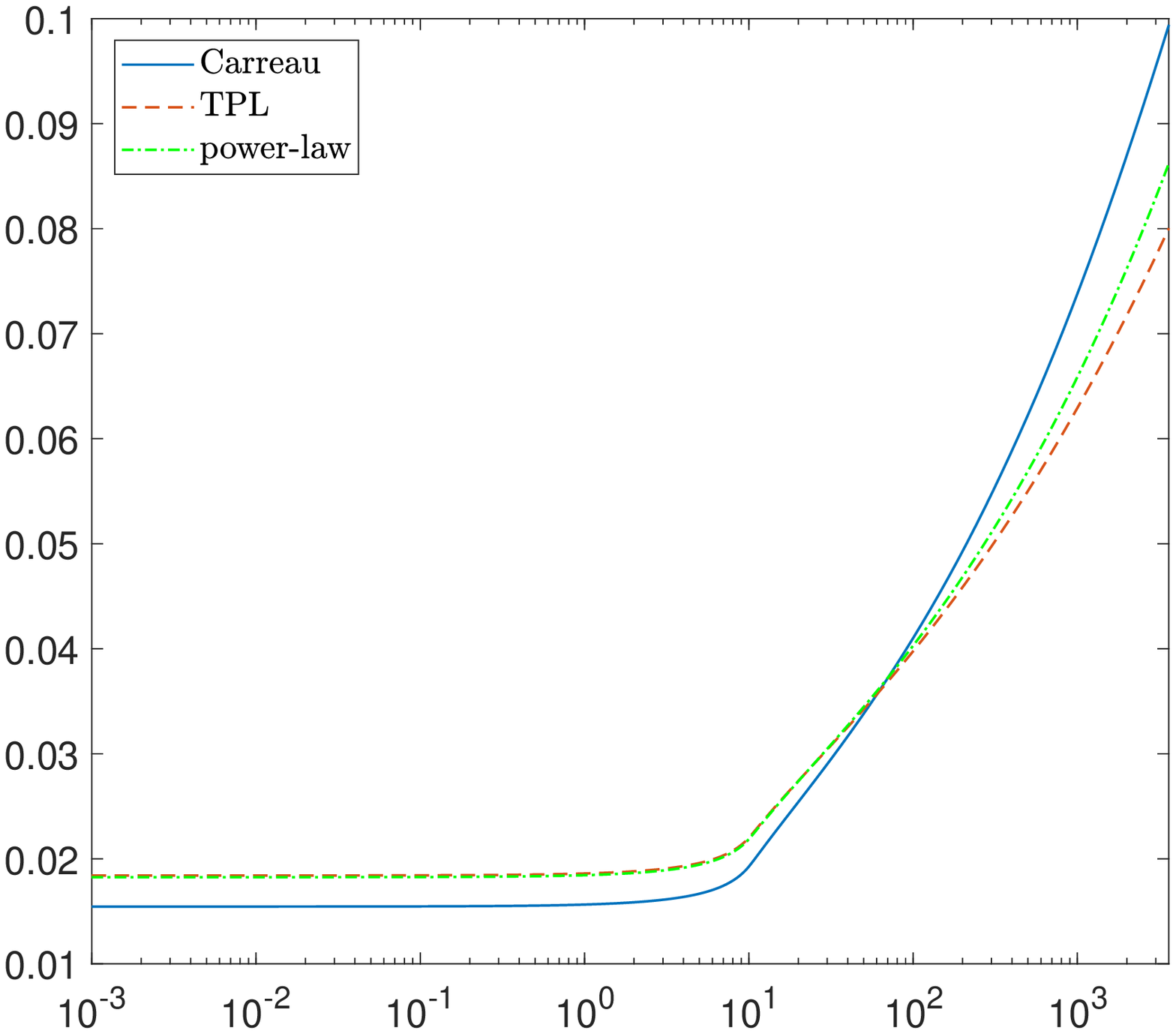}
\put(-338,-5){$t$}
\put(-110,-5){$t$}
\put(-446,165){$\textbf{a)}$}
\put(-220,165){$\textbf{b)}$}
\put(-440,90){$|\Gamma|$}
\put(-220,90){$E$}
\caption{The average values of: a) shear rates, $\Gamma$ $\left[\frac{1}{\text{s}} \right]$, b) apparent viscosities, $E$ [Pa$\cdot$s], for the HPG fluid.}
\label{HPG_mean}
\end{center}
\end{figure}

In order to verify this claim we present in Fig. \ref{HPG_mean} the average values of the fluid shear rate, $\Gamma$ \eqref{Gam_def}, and the averaged viscosities, $E$ \eqref{E_def}. It shows that the respective shear rate values are very close to each other. The resulting viscosities are also very similar, especially those obtained for the truncated power-law and the  pure power-law. The minima of $|\Gamma|$ are: 31  $\frac{1}{\text{s}}$ for the Carreau rheology, 32  $\frac{1}{\text{s}}$ for the truncated power-law and 24  $\frac{1}{\text{s}}$ for the pure power-law. The maximal values of $|\Gamma|$ yield: $1.83\cdot 10^{3}$  $\frac{1}{\text{s}}$ for the Carreau model, $1.73\cdot 10^{3}$  $\frac{1}{\text{s}}$ for the truncated power-law and $1.76\cdot 10^{3}$  $\frac{1}{\text{s}}$ for the pure power-law. When comparing the above figures with data from Table \ref{tab_fluid} it is evident that the range of the averaged shear rates fits well inside the interval defined by the limiting values of $\dot \gamma$ assumed for the truncated power-law ($|\dot \gamma_1|=1.128$ $\frac{1}{\text{s}}$, $|\dot \gamma_2|=1.45\cdot 10^6$ $\frac{1}{\text{s}}$). In other words, for no value of $\Gamma$ the limiting viscosities $\eta_0$ and $\eta_\infty$ are obtained  (although they are naturally achieved locally for the Carreau and the truncated power-law models). This explains a very good coincidence between the results produced for different rheological models. 

\subsection{150 wppm HPAM and 600 wppm XG }

\begin{figure}[htb!]
\begin{center}
\includegraphics[scale=0.38]{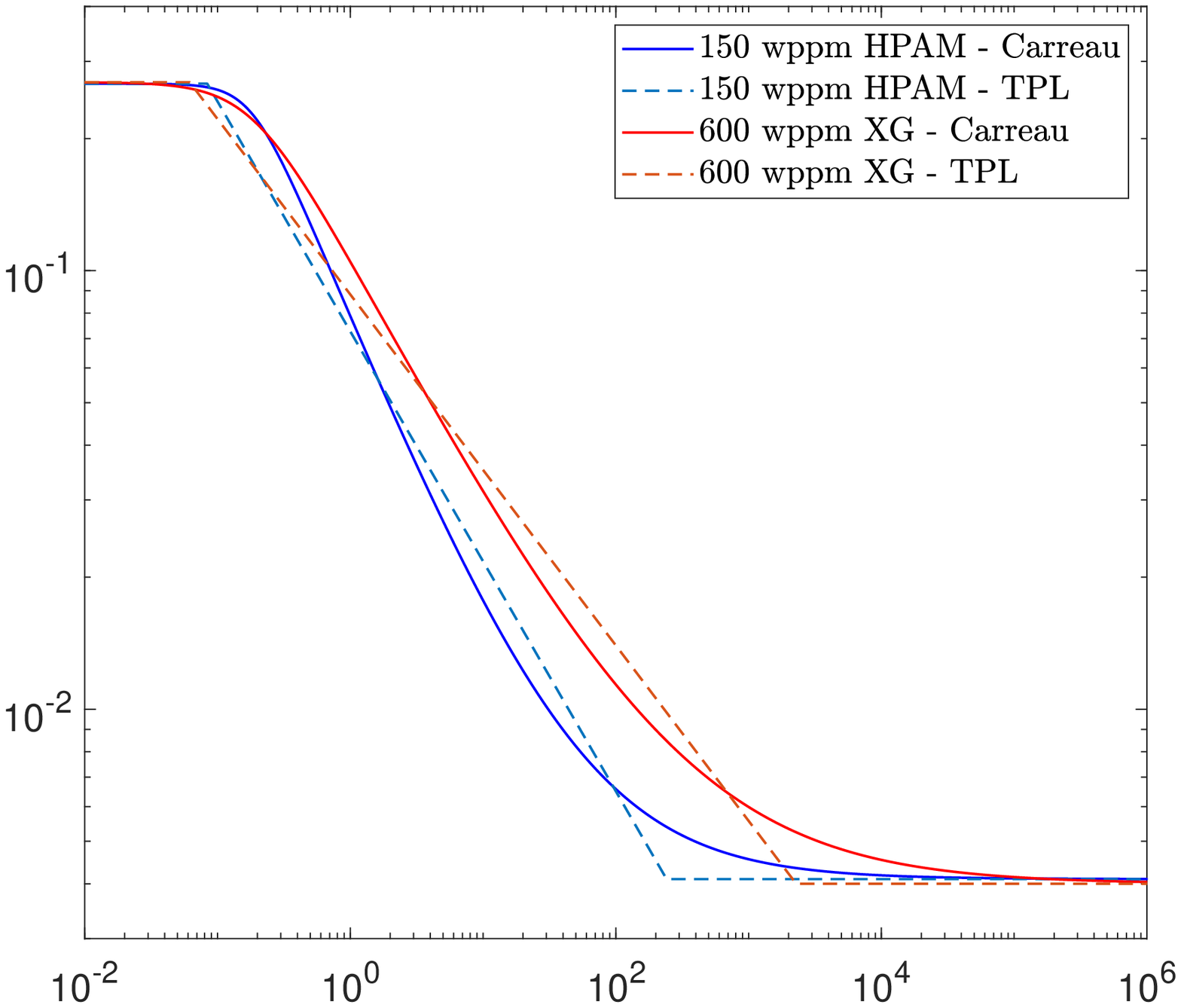}
%\hspace{0mm}
\includegraphics[scale=0.38]{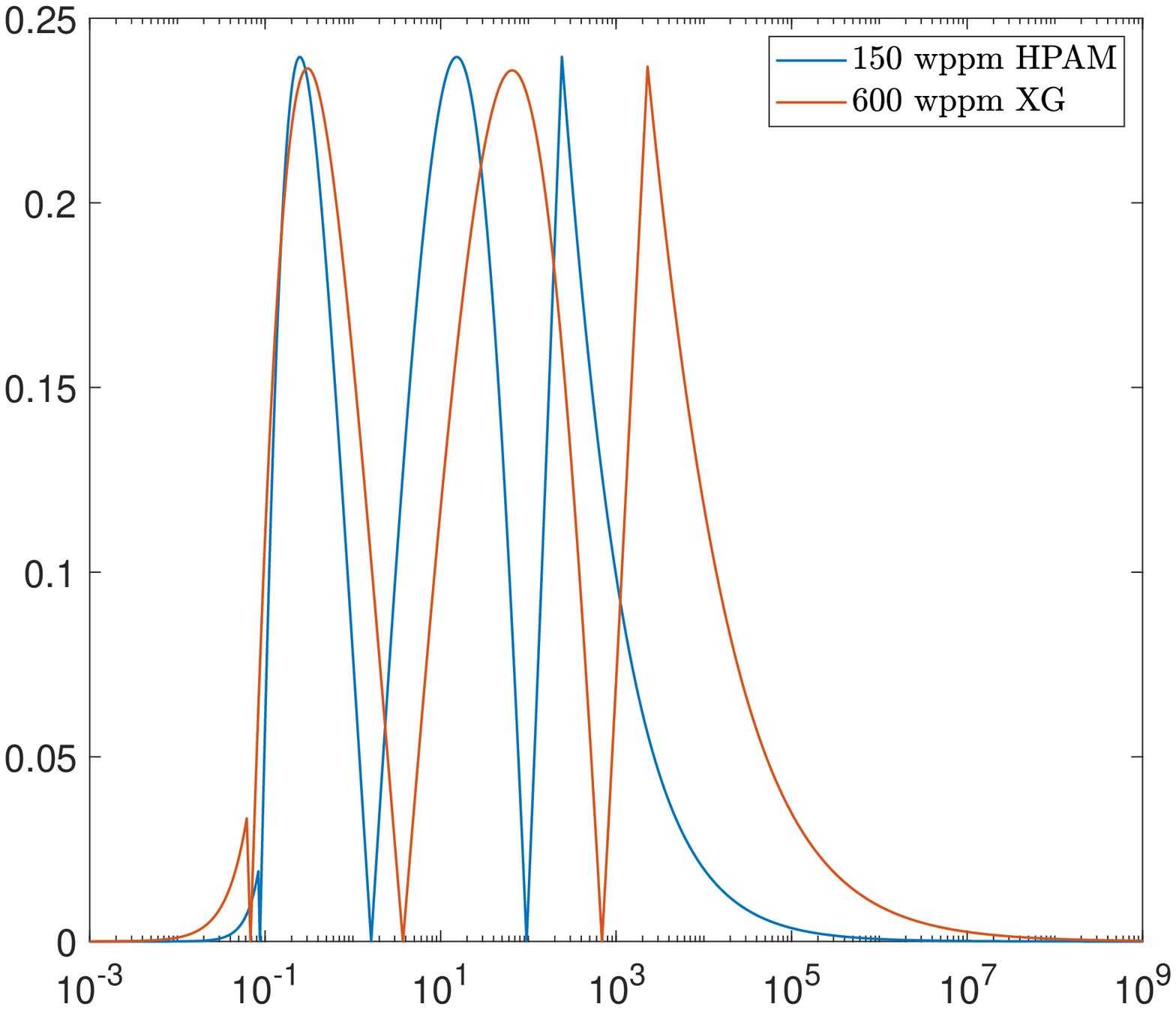}
\put(-338,-5){$|\dot \gamma|$}
\put(-110,-5){$|\dot \gamma|$}
\put(-446,165){$\textbf{a)}$}
\put(-220,165){$\textbf{b)}$}
\put(-440,90){$\eta_\text{a}$}
\put(-220,90){$\delta \eta_\text{a}$}
\caption{150 wppm HPAM fluid and 600 wppm XG fluid: a) apparent viscosities for Carreau and truncated power-law (TPL) rheologies, $\eta_\text{a}$ [Pa$\cdot$s], b) the relative deviations between the Carreau and the truncated power-law variants, $\delta \eta_\text{a}$.}
\label{150_600}
\end{center}
\end{figure}

Let us perform an analysis similar to that from the previous subsection for the two remaining fluids: 150 wppm HPAM and 600 wppm XG. The characteristics of respective apparent viscosities, $\eta_\text{a}$, described by the Carreau-Yasuda model and their truncated power-law approximations are depicted in Fig. \ref{150_600}a). The relative deviations between the Carreau and truncated power-law variants of $\eta_\text{a}$ are shown in Fig. \ref{150_600}b). As can be seen, the maximal discrepancies between corresponding viscosities are below $25\%$.

\begin{figure}[htb!]
\begin{center}
\includegraphics[scale=0.38]{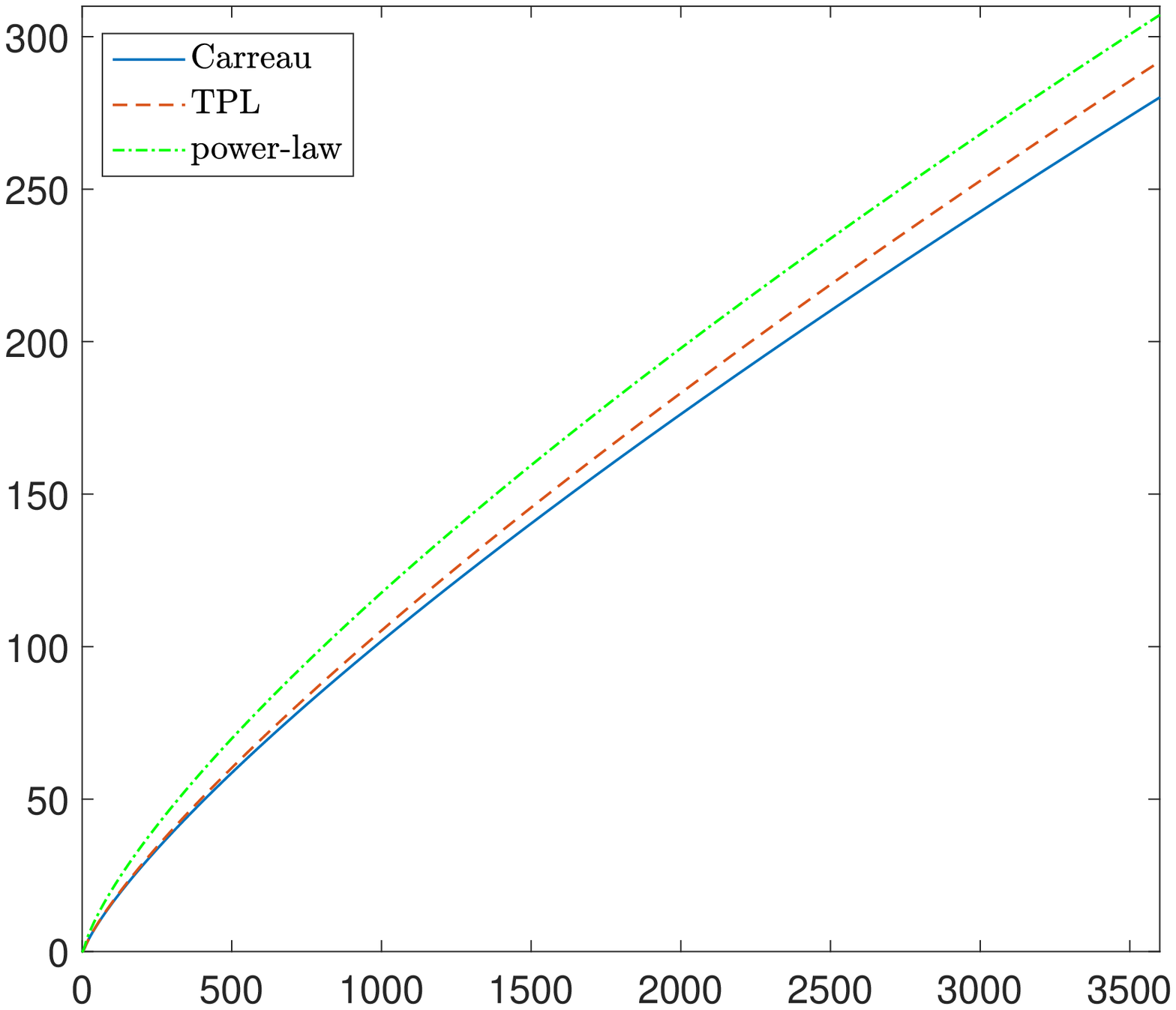}
%\hspace{0mm}
\includegraphics[scale=0.38]{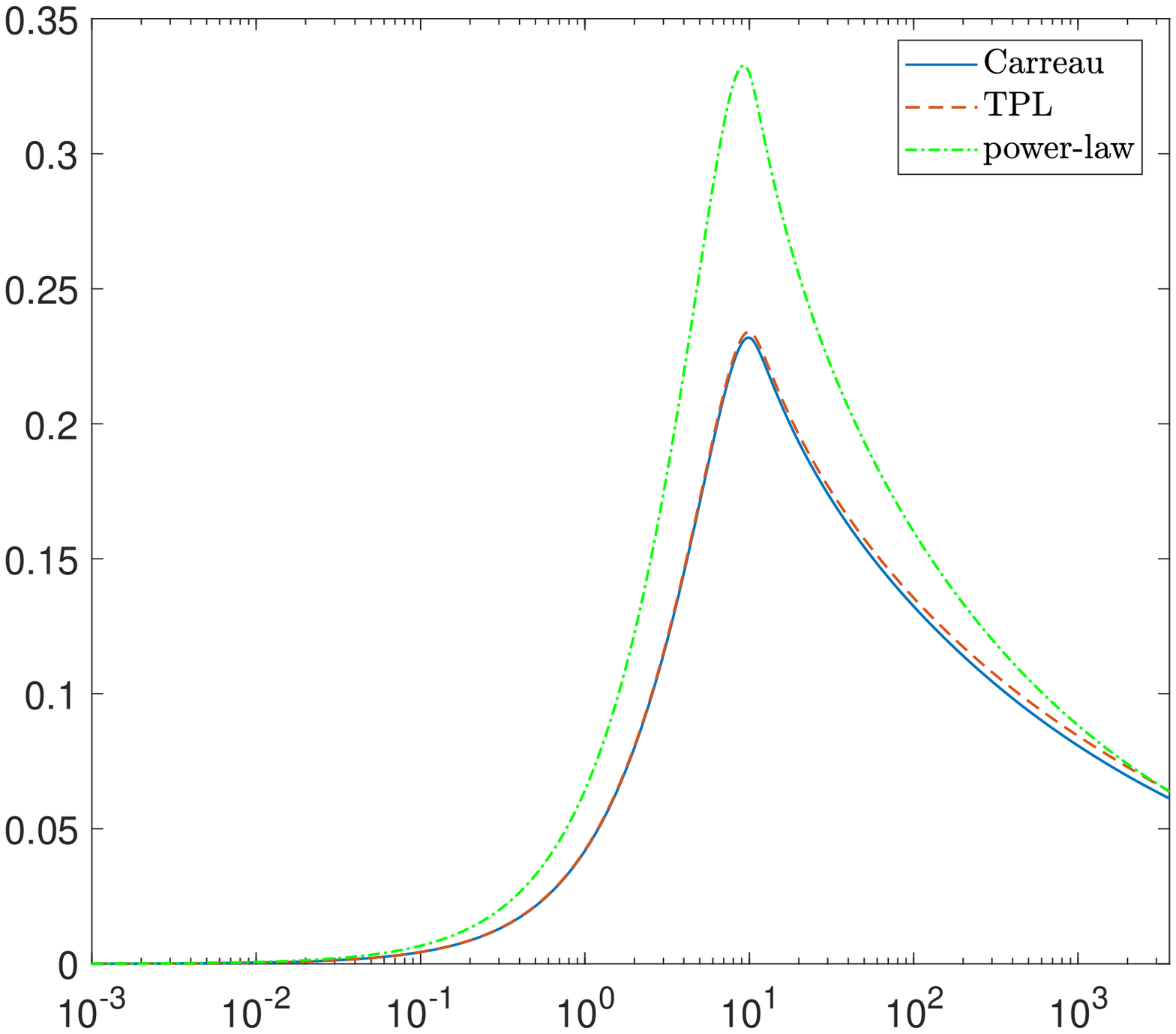}
\put(-338,-5){$t$}
\put(-110,-5){$t$}
\put(-446,165){$\textbf{a)}$}
\put(-220,165){$\textbf{b)}$}
\put(-440,90){$L$}
\put(-220,90){$v_0$}
\caption{Simulation results for the 150 wppm HPAM fluid: a) the crack length, $L$ [m], b) the crack propagation speed, $v_0$ $\left[\frac{\text{m}}{\text{s}}\right]$.}
\label{HPAM_L}
\end{center}
\end{figure}

\begin{figure}[htb!]
\begin{center}
\includegraphics[scale=0.38]{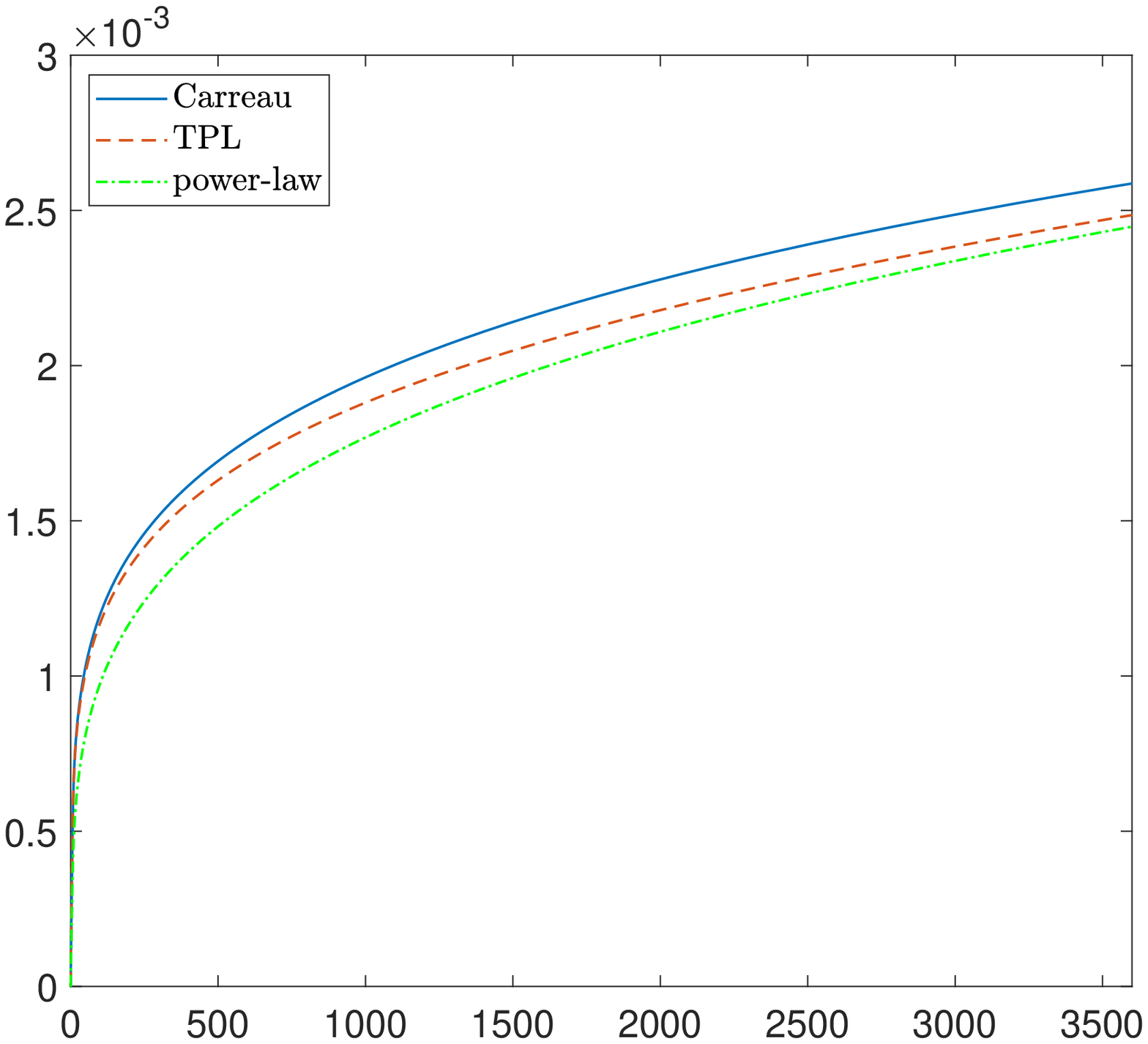}
\put(-110,-5){$t$}
\put(-240,90){$w(0,t)$}
\caption{Simulation results for the 150 wppm HPAM fluid: the fracture opening at the crack mouth, $w(0,t)$ [m].}
\label{W0_HPAM}
\end{center}
\end{figure}

For 150 wppm HPAM fluid the simulation results in terms of the crack length, $L$, the crack propagation speed, $v_0$, and the fracture opening at the crack mouth, $w(0,t)$, are displayed in Figs. \ref{HPAM_L}--\ref{W0_HPAM}. The solution for the power-law fluid rheology deviates notably from the one obtained for the Carreau fluid, overestimating the fracture length and the crack propagation speed and underestimating the fracture opening. The truncated power-law variant yields very good resemblance of the Carreau results for early times, but  departs later  towards the power-law solution. 

\begin{figure}[htb!]
\begin{center}
\includegraphics[scale=0.38]{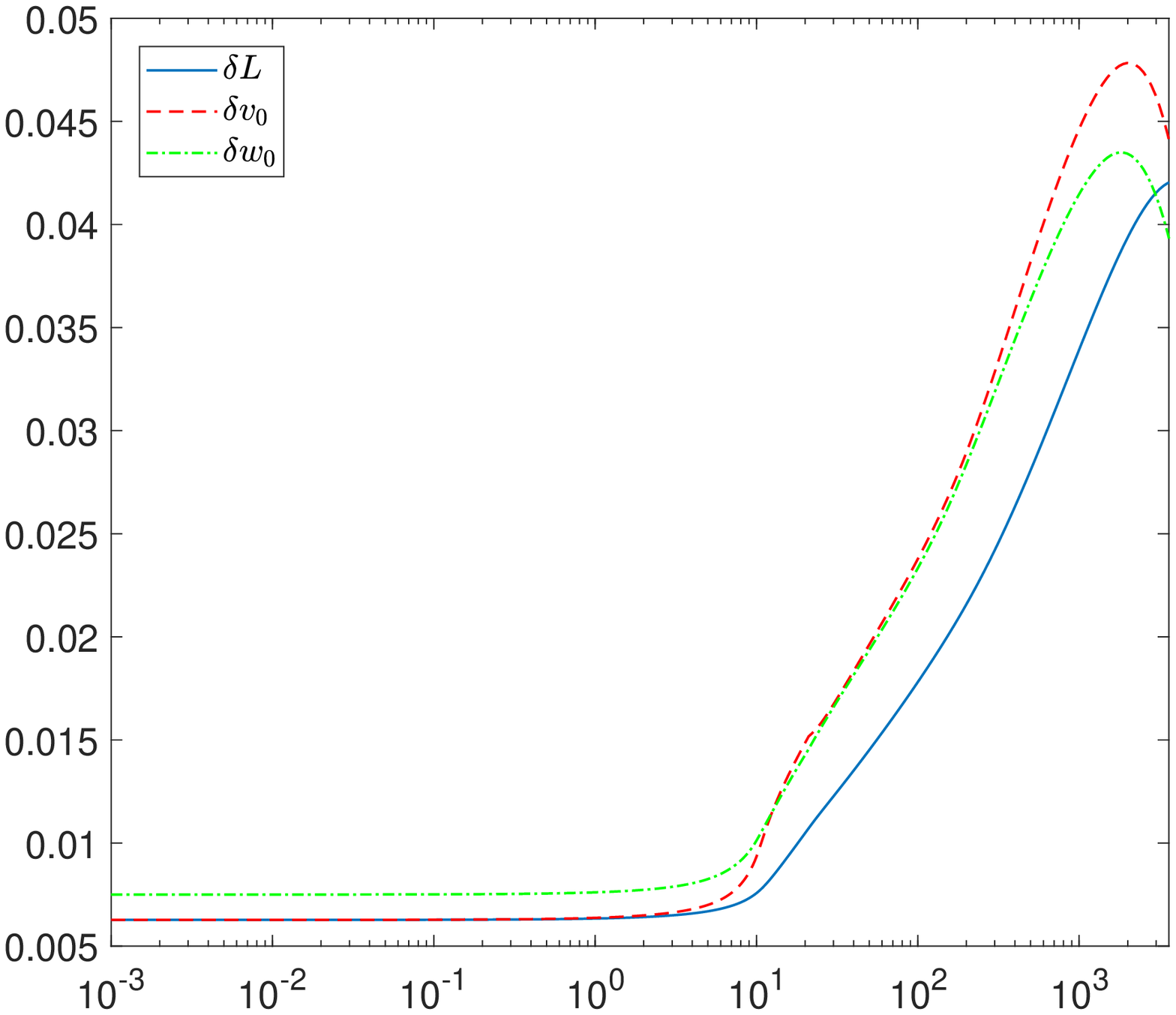}
%\hspace{0mm}
\includegraphics[scale=0.38]{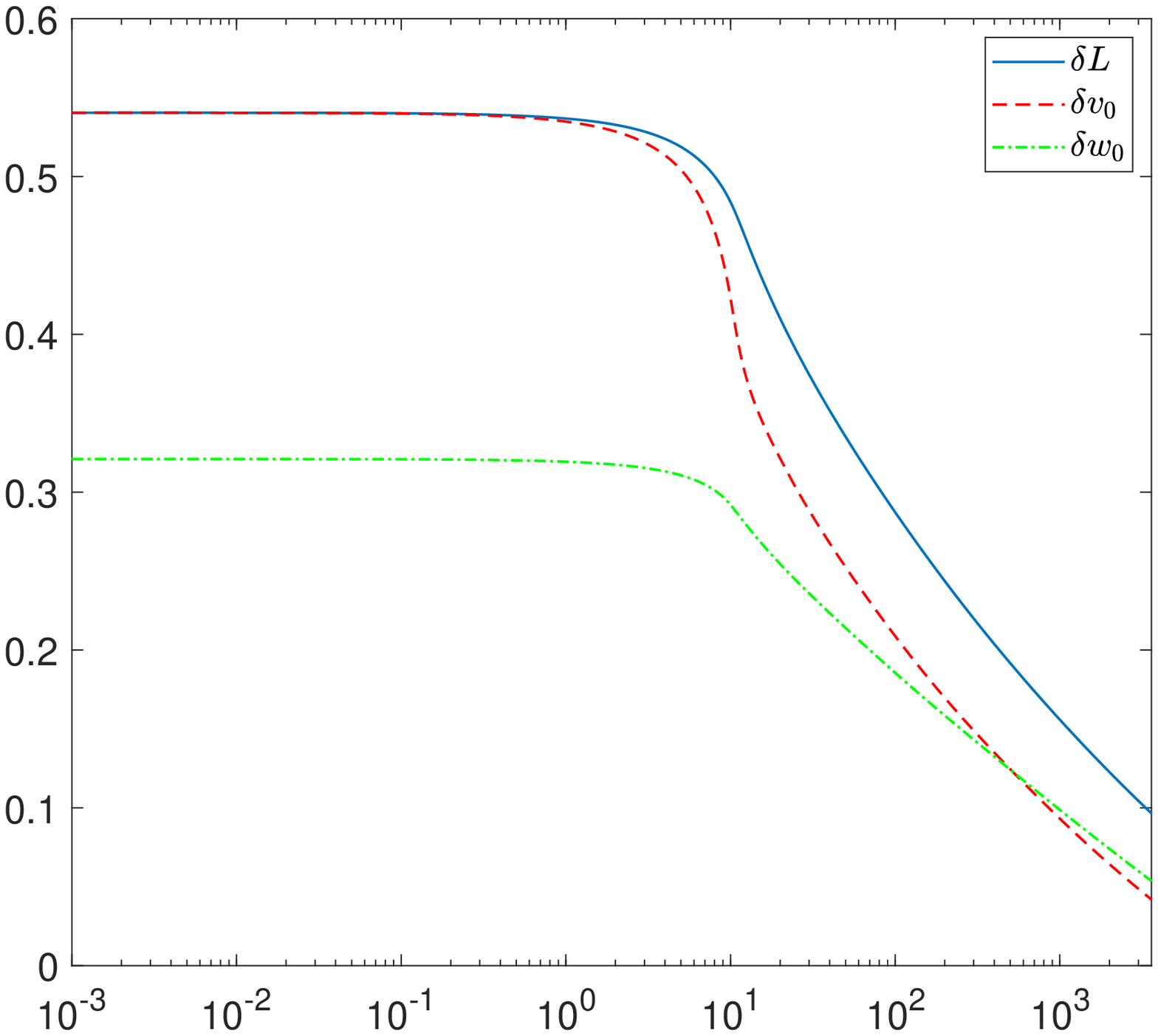}
\put(-338,-5){$t$}
\put(-110,-5){$t$}
\put(-446,165){$\textbf{a)}$}
\put(-220,165){$\textbf{b)}$}
\caption{The relative deviation of solution from the Carreau variant for the 150 wppm HPAM fluid in the case of: a) the truncated power-law rheology b) the power-law rheology.}
\label{HPAM_bledy}
\end{center}
\end{figure}

The relative deviations of the truncated power-law and the power-law solutions from the Carreau variant are depicted in Fig. \ref{HPAM_bledy}. It shows that the truncated power-law rheology yields the relative errors that grow with time to almost 5$\%$ at $t_\text{end}$. Conversely, the pure power-law case produces the highest errors of over 50$\%$ at the initial times. Then the errors are reduced to achieve the level 5-10$\%$ for the final time instant.

\begin{figure}[htb!]
\begin{center}
\includegraphics[scale=0.38]{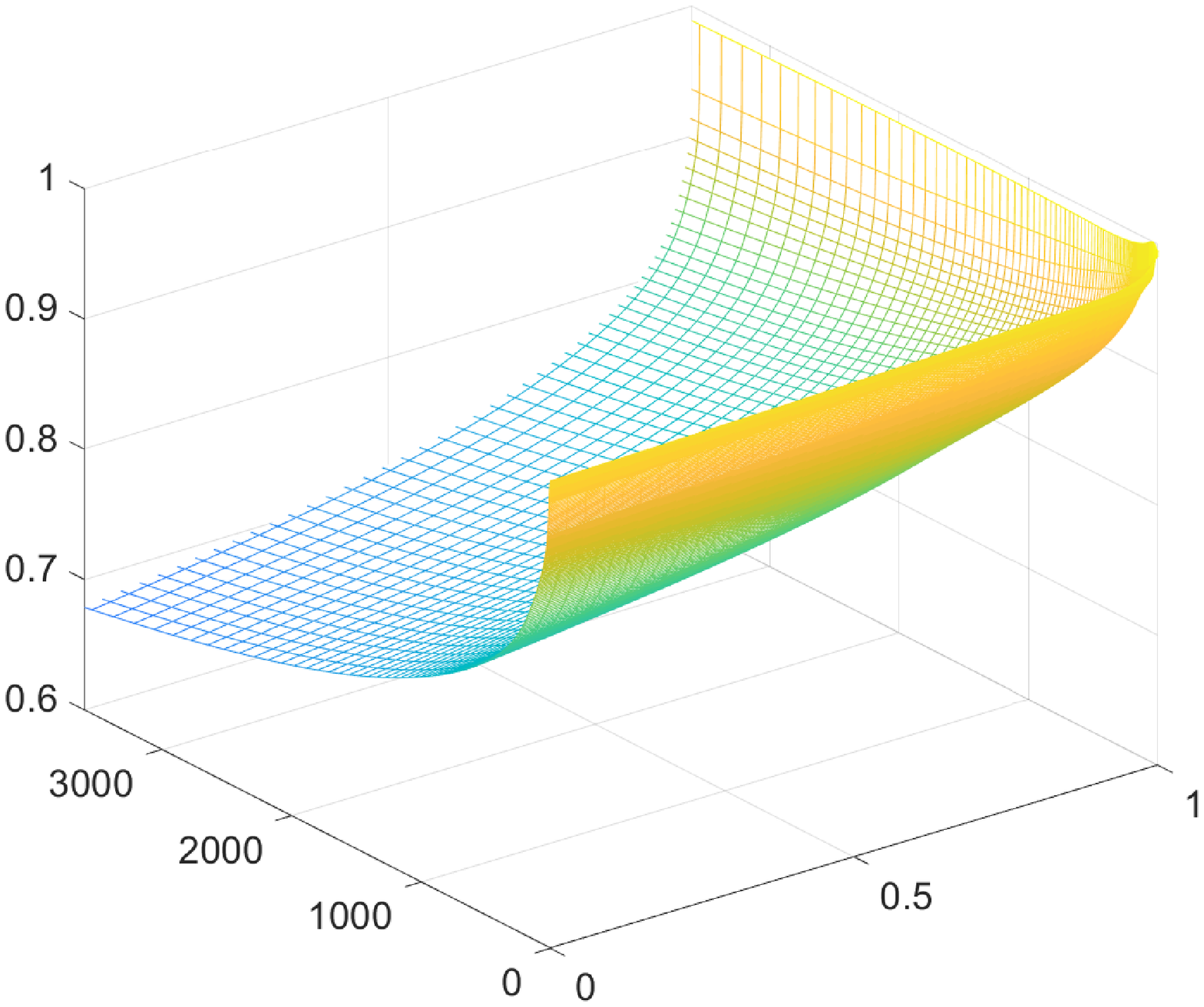}
%\hspace{0mm}
\includegraphics[scale=0.38]{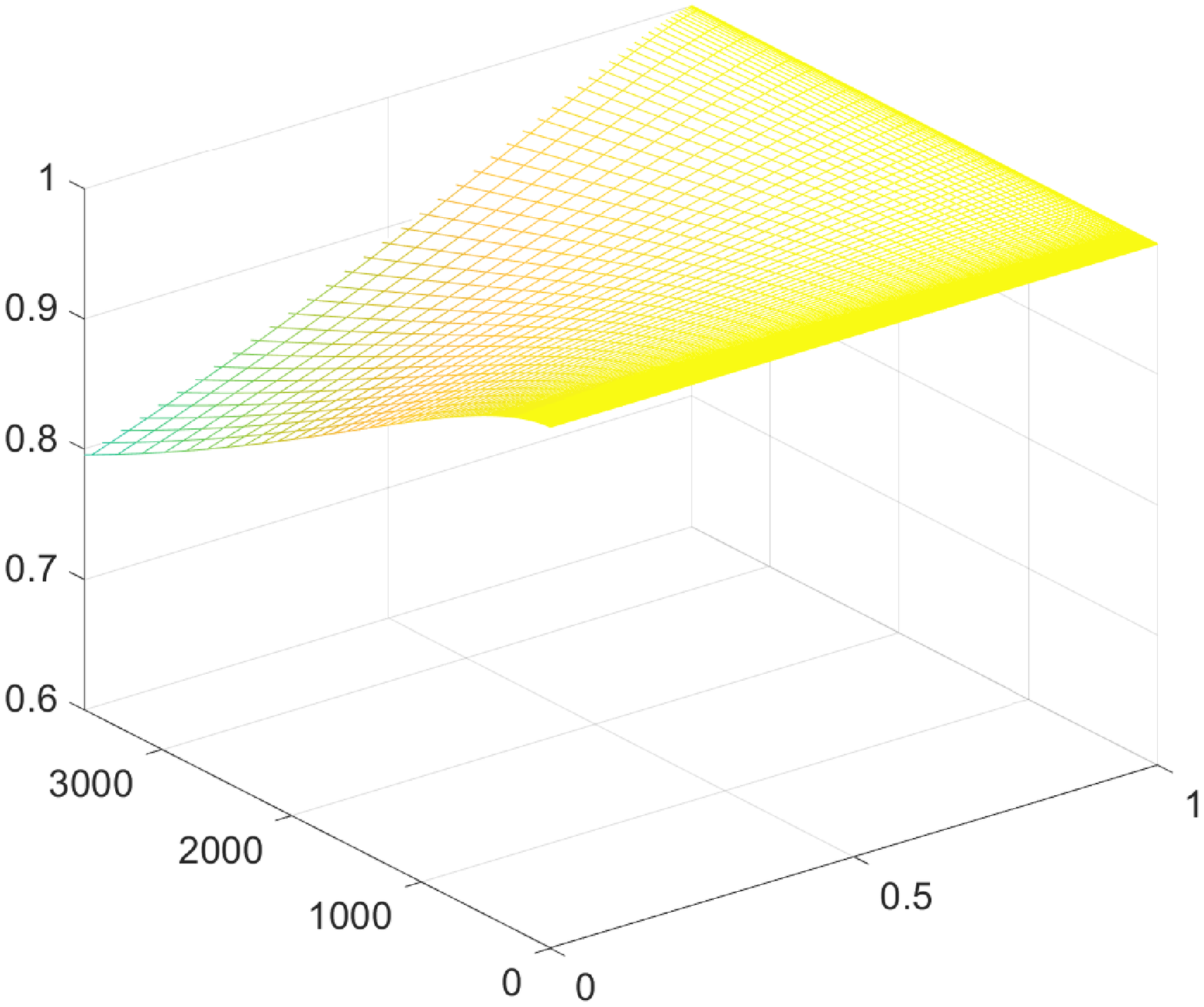}
\put(-405,20){$t$}
\put(-285,10){$x/L$}
\put(-170,20){$t$}
\put(-60,10){$x/L$}
\put(-446,165){$\textbf{a)}$}
\put(-220,165){$\textbf{b)}$}
\caption{Function $F(x,t)$ for the 150 wppm HPAM fluid: a) the Carreau rheology, b) the truncated power-law rheology.}
\label{F_HPAM}
\end{center}
\end{figure}

The component function, $F$, of the fluid flux is depicted in Fig. \ref{F_HPAM} for the Carreau and the truncated power-law models. In both cases the solutions for $q$ are very close to the Newtonian high shear rate variant ($\eta_\text{a}=\eta_\infty$) in the early time range. Then, with time growth, both fluxes depart from this regime, however for the Carreau solution this deviation is much more pronounced. 

\begin{figure}[htb!]
\begin{center}
\includegraphics[scale=0.38]{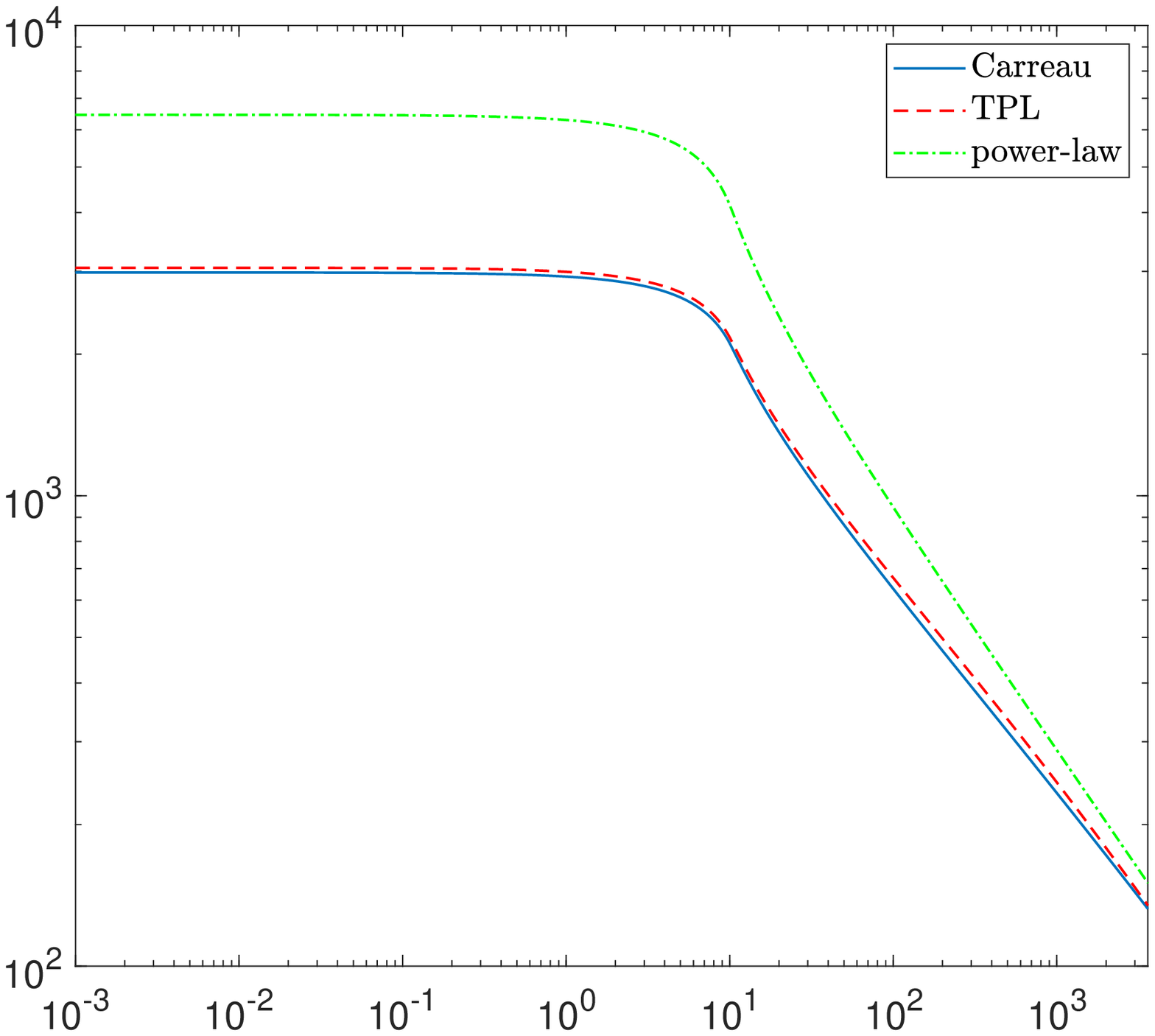}
%\hspace{0mm}
\includegraphics[scale=0.38]{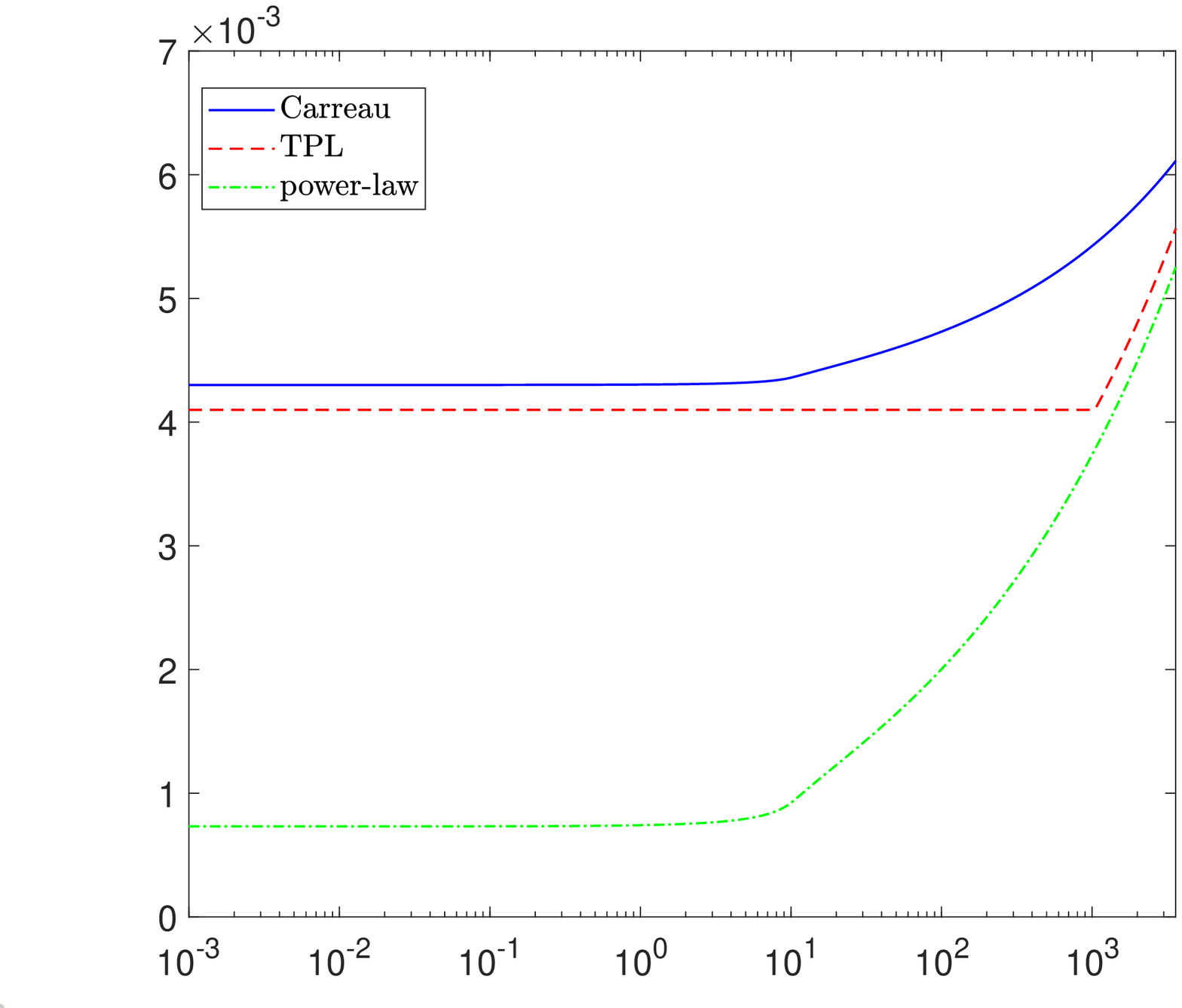}
\put(-338,-5){$t$}
\put(-110,-5){$t$}
\put(-446,165){$\textbf{a)}$}
\put(-220,165){$\textbf{b)}$}
\put(-440,90){$|\Gamma|$}
\put(-220,90){$E$}
\caption{The average values of: a) shear rates, $\Gamma$ $\left[\frac{1}{\text{s}} \right]$, b) apparent viscosities, $E$ [Pa$\cdot$s], for the 150 wppm HPAM fluid.}
\label{HPAM_mean}
\end{center}
\end{figure}

In order to better understand  these trends let us analyse the temporal evolution of the average shear rates, $\Gamma$, and the average viscosities, $E$. Respective graphs are shown in Fig. \ref{HPAM_mean}. This time the minimal values of $|\Gamma|$ are: 69  $\frac{1}{\text{s}}$ for the Carreau rheology, 67  $\frac{1}{\text{s}}$ for the truncated power-law and 73 $\frac{1}{\text{s}}$ for the pure power-law. The maxima of $|\Gamma|$ yield: $2.98\cdot 10^3$  $\frac{1}{\text{s}}$ for the Carreau variant, $3.05\cdot 10^3$  $\frac{1}{\text{s}}$ for the truncated power-law and $6.46\cdot 10^3$ $\frac{1}{\text{s}}$ for the pure power-law. When comparing these figures with the data from Table \ref{tab_fluid} one concludes that the obtained average shear rates are well above the lower limiting value for the truncated power-law model ($|\dot \gamma_1|=8.37\cdot 10^{-2}$ $\frac{1}{\text{s}}$). Moreover, they are also above the higher limiting shear rate ($|\dot \gamma_2|=241$ $\frac{1}{\text{s}}$) up to approximately $t=10^3$ s for any of the analysed models. This is clearly reflected in Fig. \ref{HPAM_mean}b), where the truncated power-law rheology yields $\eta_\text{a}=\eta_\infty$ up to the instant $t=991$ s. Such a behaviour of $\Gamma$ and $E$ explains our previous observations  on the relations between respective results. Firstly, the power-law rheology produces credible viscosity values only  above the aforementioned time limit (see Fig. \ref{HPAM_mean}b)). When analyzing the relative errors from Fig. \ref{HPAM_bledy}b) we see that for $t>1000$ s the quality of power law approximation becomes indeed sufficiently good for practical purposes. Secondly, for small times the truncated power-law rheology yields the viscosity values very close to those obtained with the Carreau-Yasuda model and thus the coincidence of respective results is very good. Then, with time growing, the latter model produces increasingly larger $\eta_\text{a}$, while the truncated power-law retains the value $\eta_\text{a}=\eta_\infty$ up to the moment $t=991$ s. For this reason one can observe a gradual divergence of respective results. Next, for $t>991$ the truncated power-law viscosity increases $\eta_\text{a}$ trying to match original Carreau characteristics. This contributes again to the error reduction, which can be noted in  Fig. \ref{HPAM_bledy}a).

\begin{figure}[htb!]
\begin{center}
\includegraphics[scale=0.38]{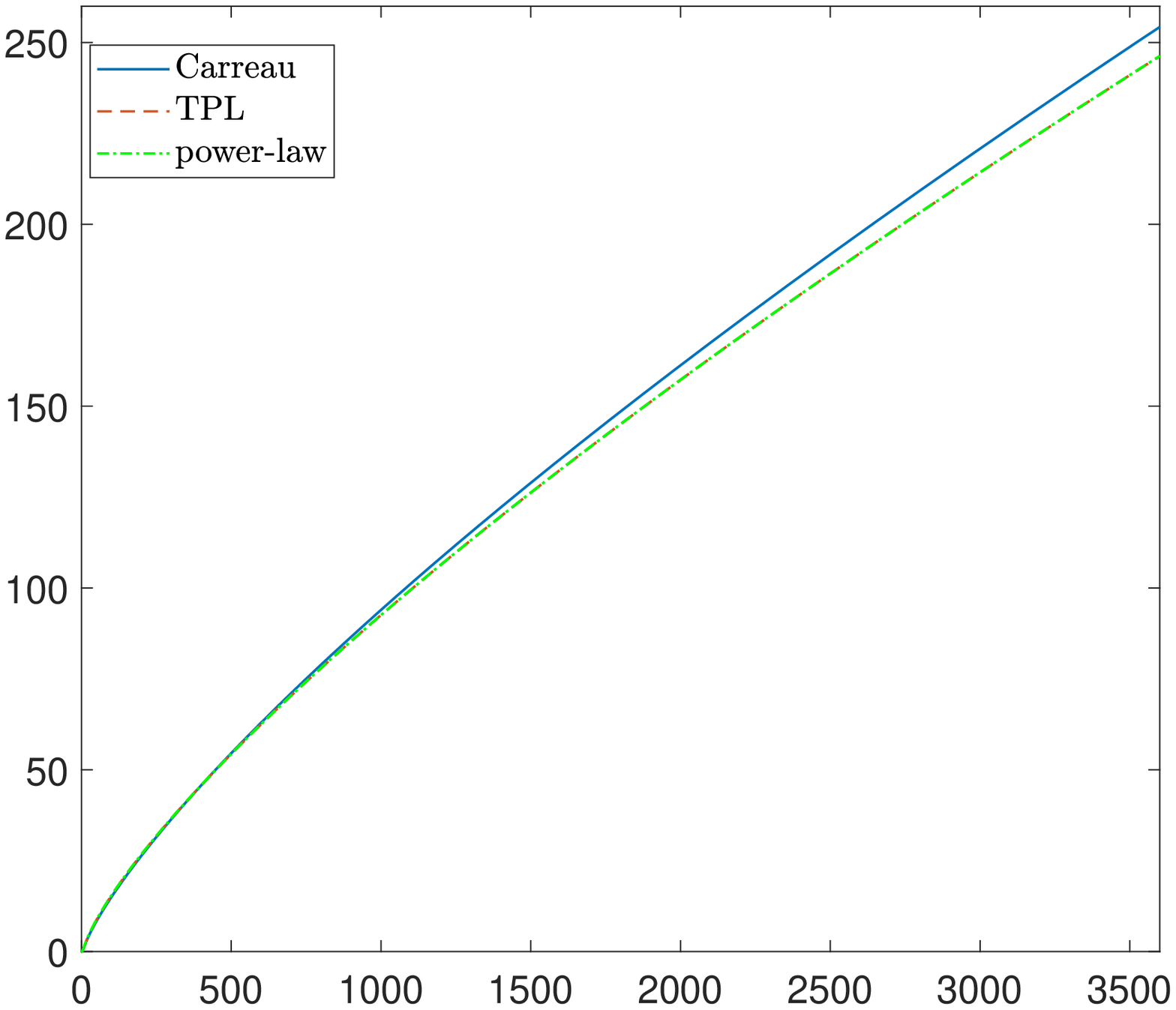}
%\hspace{0mm}
\includegraphics[scale=0.38]{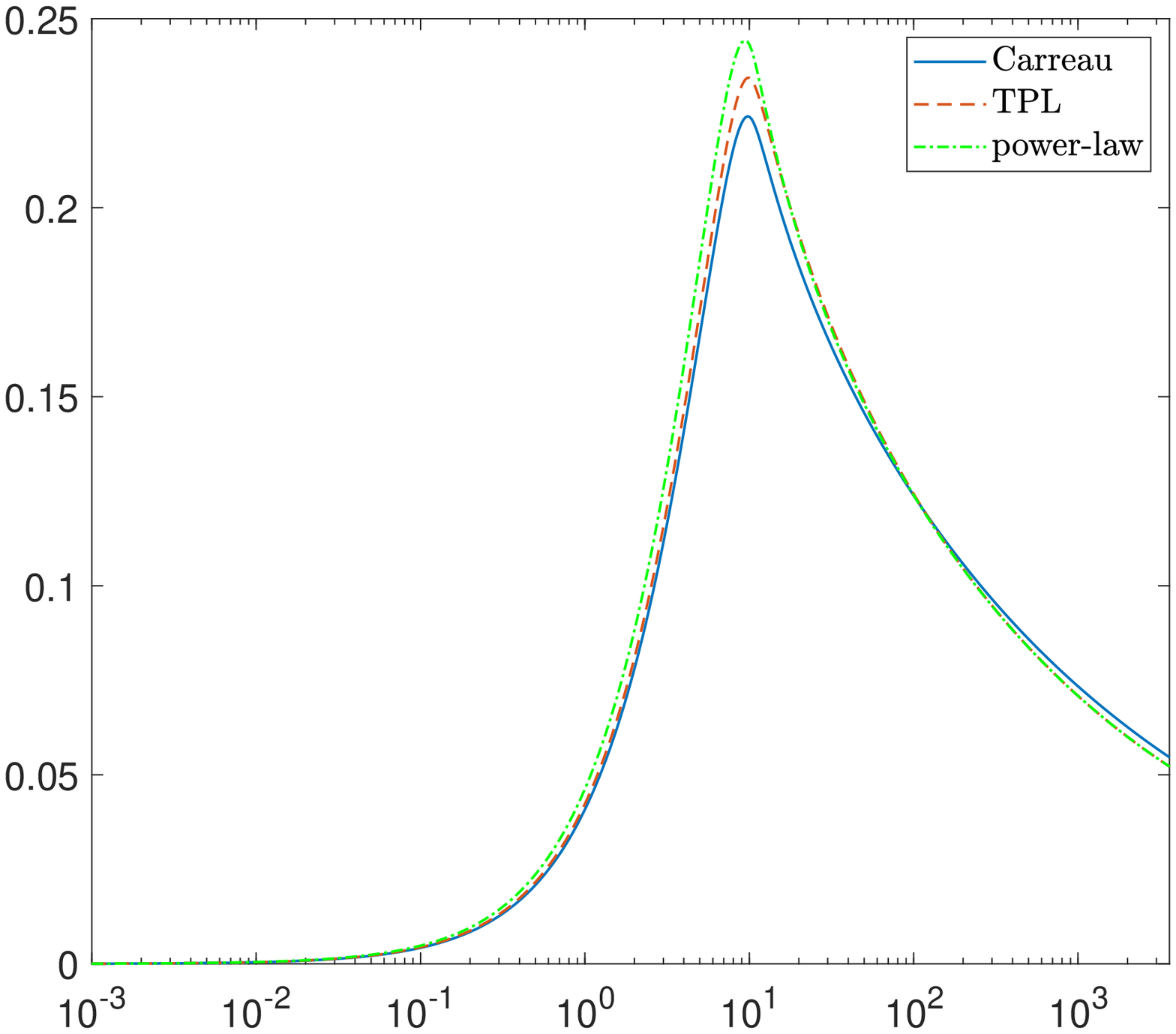}
\put(-338,-5){$t$}
\put(-110,-5){$t$}
\put(-446,165){$\textbf{a)}$}
\put(-220,165){$\textbf{b)}$}
\put(-440,90){$L$}
\put(-220,90){$v_0$}
\caption{Simulation results for the 600 wppm XG fluid: a) the crack length, $L$ [m], b) the crack propagation speed, $v_0$ $\left[\frac{\text{m}}{\text{s}}\right]$.}
\label{XG_L}
\end{center}
\end{figure}

\begin{figure}[htb!]
\begin{center}
\includegraphics[scale=0.38]{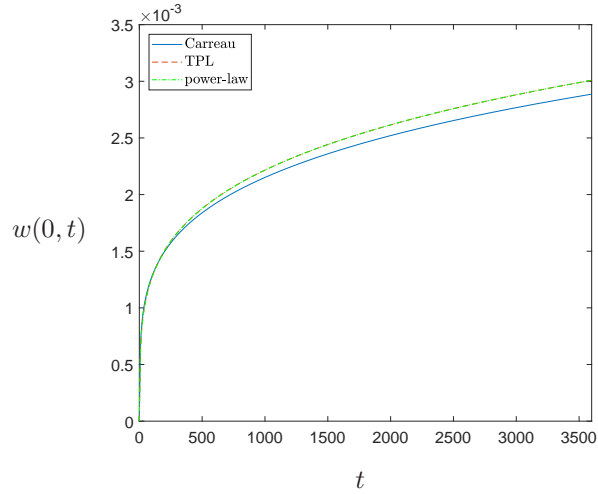}
\put(-110,-5){$t$}
\put(-240,90){$w(0,t)$}
\caption{Simulation results for the 600 wppm XG fluid: the fracture opening at the crack mouth, $w(0,t)$ [m].}
\label{W0_XG}
\end{center}
\end{figure}

The simulation results for the 600 wppm XG fluid in terms of the crack lenght, $L$, the crack propagation speed, $v_0$, and the crack opening at $x=0$, $w(0,t)$ are shown in Figs. \ref{XG_L}--\ref{W0_XG}. Again, as it was in the case of the HPG fluid, we see a very good coincidence of respective results, with the curves obtained for the truncated power-law and power-law models virtually indistinguishable from each other.

\begin{figure}[htb!]
\begin{center}
\includegraphics[scale=0.38]{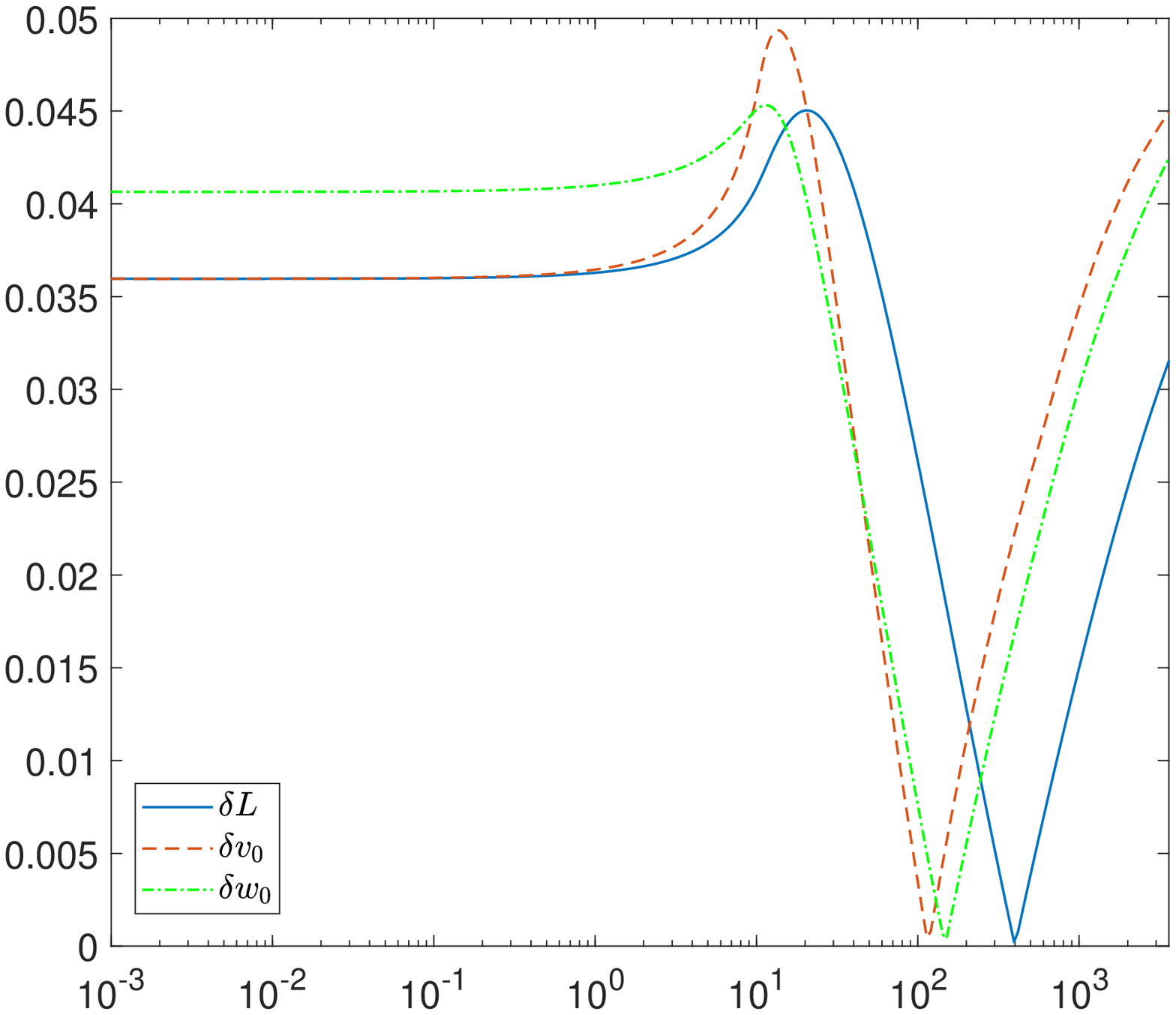}
%\hspace{0mm}
\includegraphics[scale=0.38]{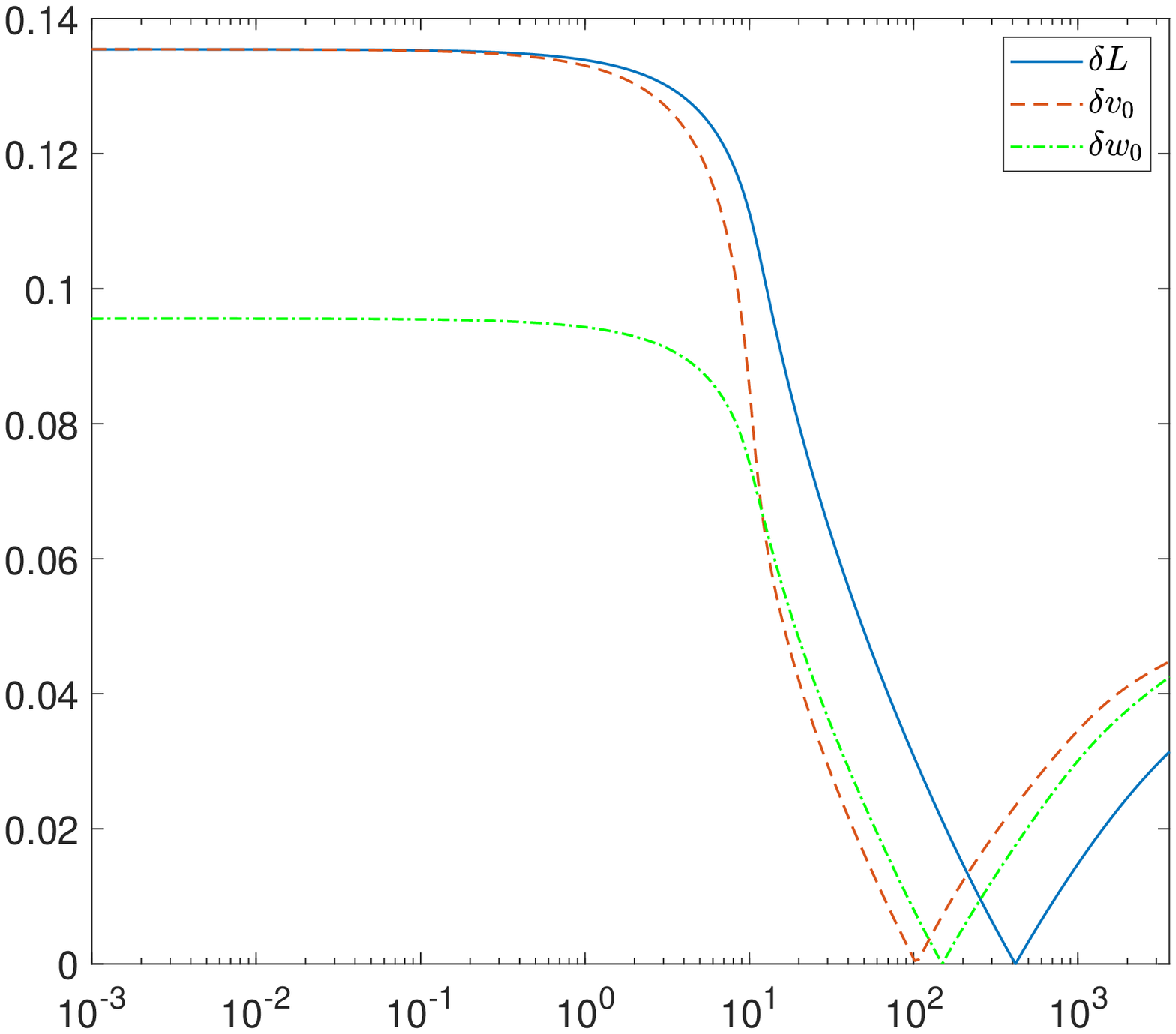}
\put(-338,-5){$t$}
\put(-110,-5){$t$}
\put(-446,165){$\textbf{a)}$}
\put(-220,165){$\textbf{b)}$}
\caption{The relative deviation of solution from the Carreau variant for the 600 wppm XG fluid in the case of: a) the truncated power-law rheology b) the power-law rheology.}
\label{XG_bledy}
\end{center}
\end{figure}

The relative deviations of the truncated power-law and power-law solutions from the Carreau variant are depicted in Fig. \ref{XG_bledy}. It shows that only in the power-law case for initial times the relative deviations exceed 10$\%$. For the truncated power-law rheology the solution diverges from the Carreau results by no more than $5\%$ over the whole temporal interval.

\begin{figure}[htb!]
\begin{center}
\includegraphics[scale=0.38]{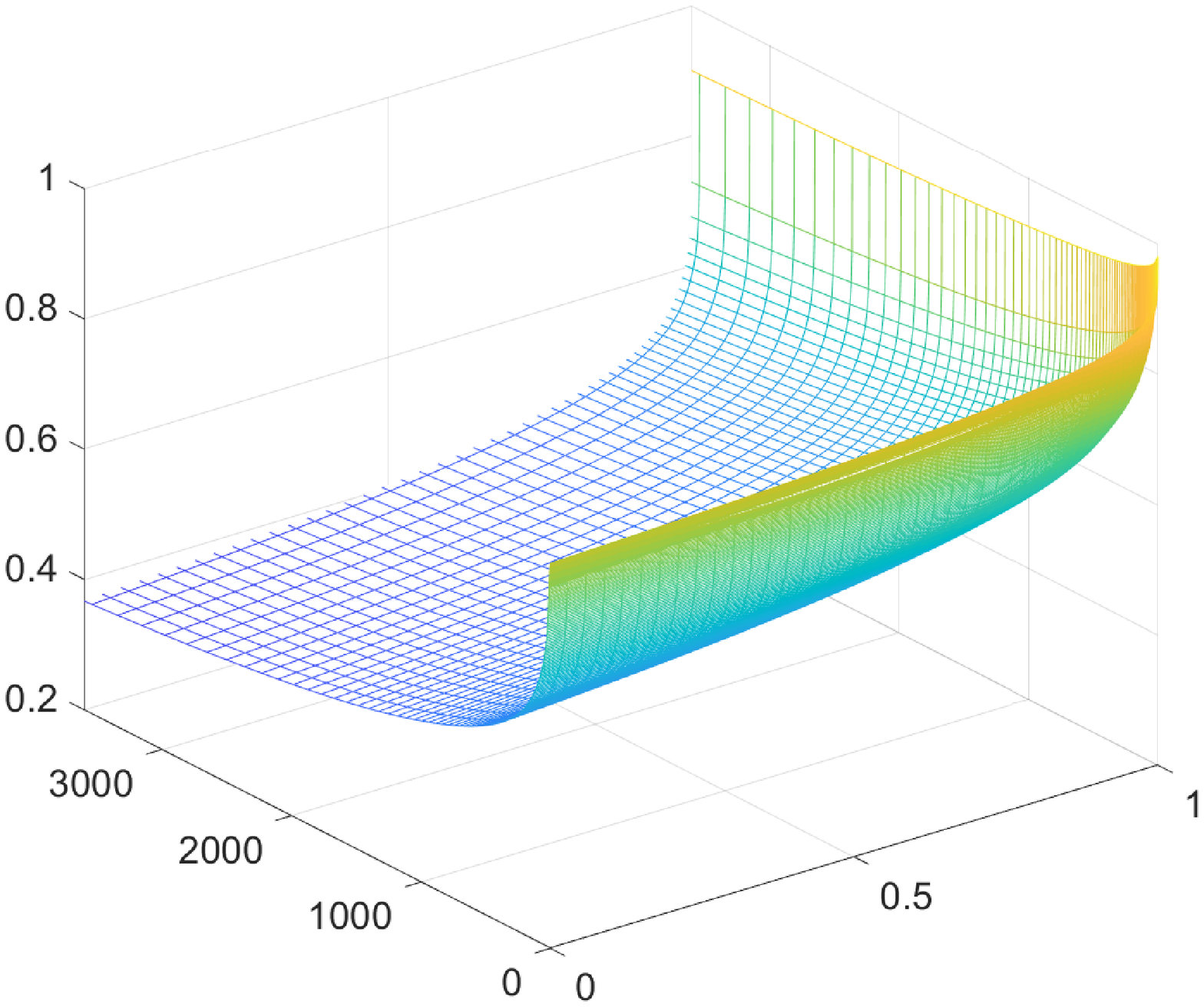}
%\hspace{0mm}
\includegraphics[scale=0.38]{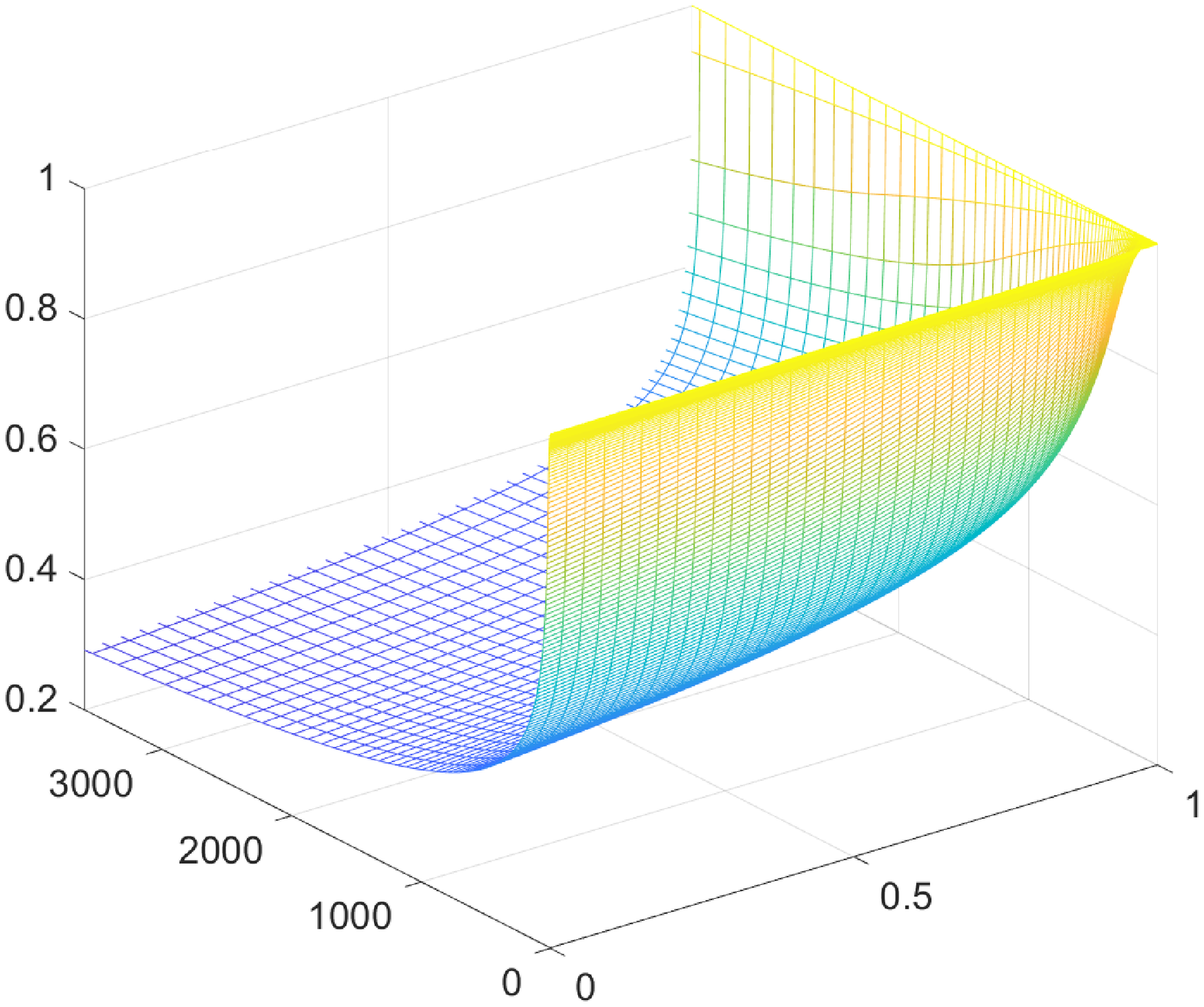}
\put(-405,20){$t$}
\put(-285,10){$x/L$}
\put(-170,20){$t$}
\put(-60,10){$x/L$}
\put(-446,165){$\textbf{a)}$}
\put(-220,165){$\textbf{b)}$}
\caption{Function $F(x,t)$ for the 600 wppm XG fluid: a) the Carreau rheology, b) the truncated power-law rheology.}
\label{F_XG}
\end{center}
\end{figure}

The distributions of the component function $F$ of the fluid flux are displayed in Fig. \ref{F_XG} for the Carreau and the truncated power-law rheologies. In the early time range the fluid flow regime is close to the Newtonian high shear variant ($F\to 1$), however with time growing the results depart swiftly from this mode. Thus, one can expect that in the case of 600 wppm XG fluid the high shear rate part of the viscosity characteristics very quickly ceases to play an important role in the HF process.

\begin{figure}[htb!]
\begin{center}
\includegraphics[scale=0.38]{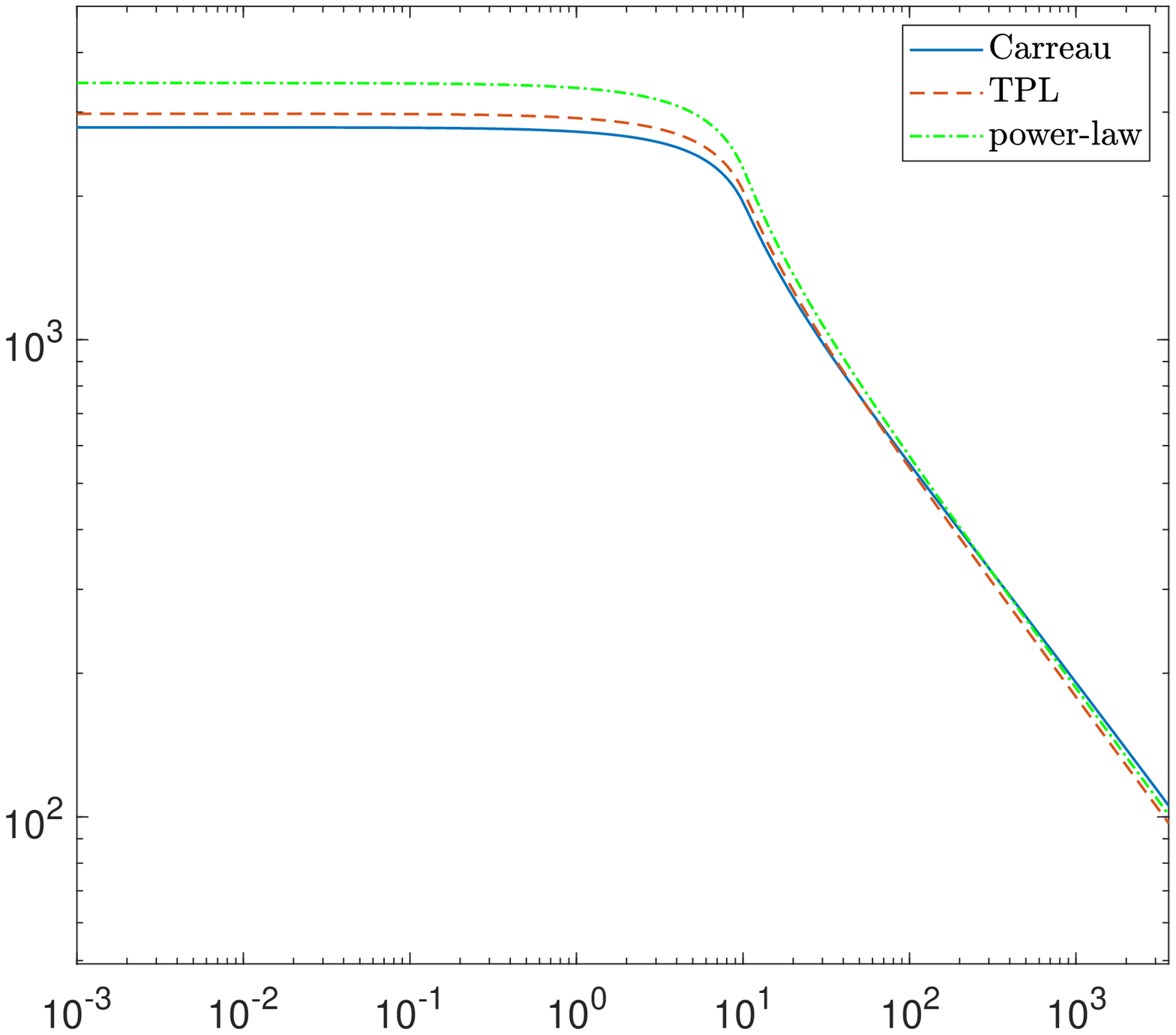}
%\hspace{0mm}
\includegraphics[scale=0.38]{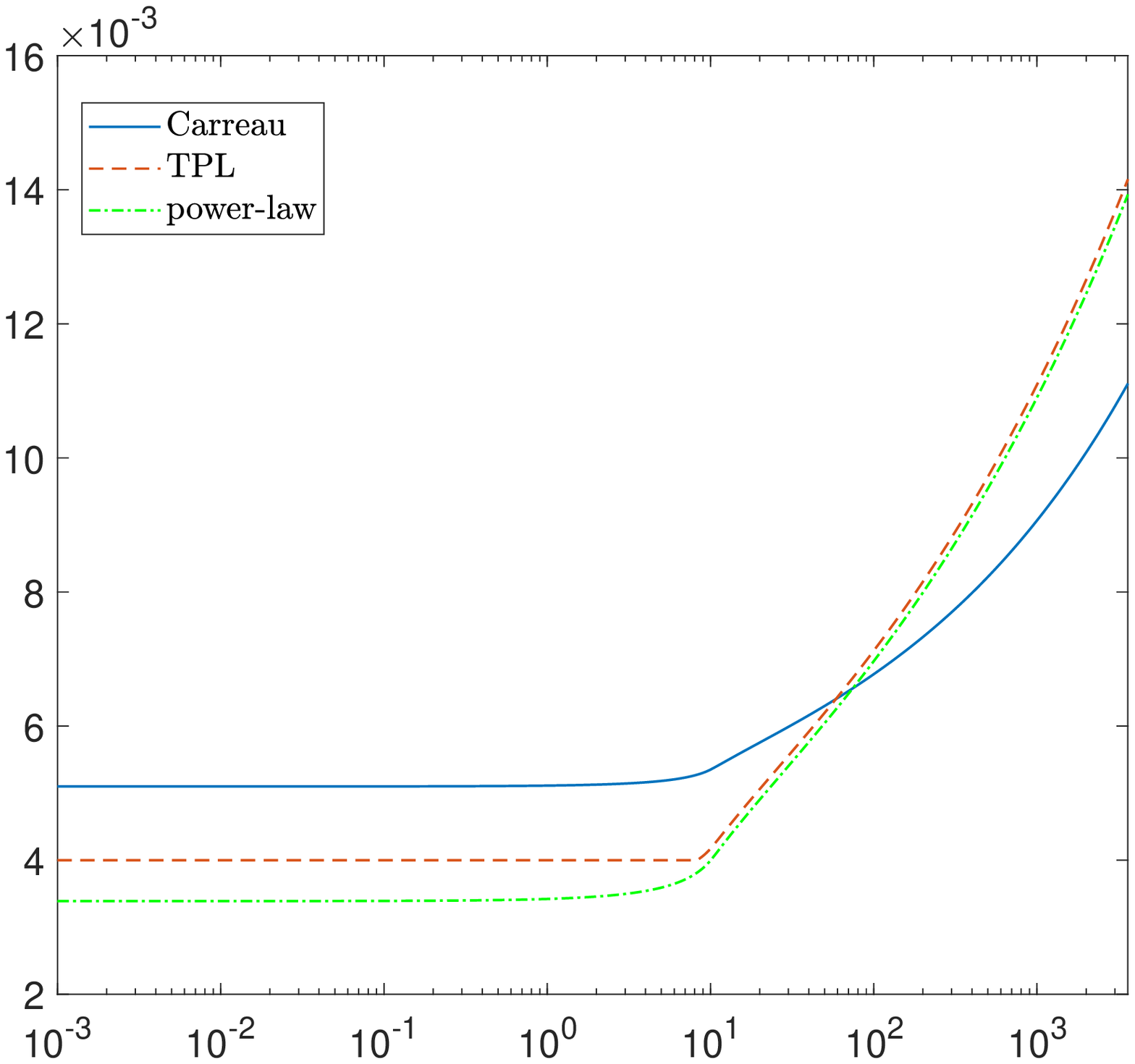}
\put(-338,-5){$t$}
\put(-110,-5){$t$}
\put(-446,165){$\textbf{a)}$}
\put(-220,165){$\textbf{b)}$}
\put(-440,90){$|\Gamma|$}
\put(-220,90){$E$}
\caption{The average values of: a) shear rates, $\Gamma$ $\left[\frac{1}{\text{s}}\right]$, b) apparent viscosities, $E$ [Pa$\cdot$s], for the 600 wppm XG fluid.}
\label{XG_mean}
\end{center}
\end{figure}

In order to substantiate this claim let us analyse the temporal behaviours of the average fluid shear rate, $\Gamma$, and the average fluid viscosity, $E$. Their graphs for respective rheologies are depicted in Fig. \ref{XG_mean}.
The minimal values of $|\Gamma|$ are: 54  $\frac{1}{\text{s}}$ for the Carreau rheology, 49  $\frac{1}{\text{s}}$ for the truncated power-law and 51 $\frac{1}{\text{s}}$ for the pure power-law. The maxima of $|\Gamma|$ yield: $2.79\cdot 10^3$  $\frac{1}{\text{s}}$ for the Carreau rheology, $2.97\cdot 10^3$  $\frac{1}{\text{s}}$ for the truncated power-law and $3.45\cdot 10^3$  $\frac{1}{\text{s}}$ for the pure power-law. The maximal values of $|\Gamma|$ are above the limiting high shear viscosity $|\dot \gamma_2|=2283$ $\frac{1}{\text{s}}$ assumed for the truncated power-law model (see Table \ref{tab_fluid}). The relation $|\Gamma| \geq |\dot \gamma_2|$ holds for all considered rheological models  for at least $t<7$ s. As can be seen in Fig. \ref{XG_mean}b), the limiting viscosity $\eta_\infty$ is retained over this time interval when using the truncated power-law rheology. On the other hand, even in this initial period the average shear rates obtained for different models are relatively close to each other. So are the corresponding average viscosities. Furthermore, for all considered variants the minimal values of the average shear rates are still far away from the limiting low shear rate  $|\dot \gamma_1|=6.17 \cdot 10^{-2}$ $\frac{1}{\text{s}}$. All these facts explain  why the respective solutions are in  good agreement with each other even in the early time range.

\section{Discussion of results}
\label{disc}

In the preceding section we conducted a number of simulations for three different rheological models of fluids: the Carreau model, the truncated power-law model and the power-law model. Using the examples of four fracturing fluids we investigated to what degree the simpler rheologies (truncated power-law and pure power-law) can be considered a reasonable substitute for the original Carreau variant. In every particular case we analysed why the respective solutions are or are not in a good agreement.  We showed that such an analysis can be performed based on the values (computed a posteriori) of the  shear rates averaged over the fracture cross section, $\Gamma$ \eqref{Gam_def}, combined with the  viscosity characteristics $\eta_\text{a}(\dot \gamma)$. 

When comparing \textit{fluid 1} and the HPG fluid we see that the viscosity characteristics of the former is moved towards the low fluid shear rate values with respect to the characteristics of the latter (the limiting viscosities $\eta_0$ and $\eta_\infty$ are virtually the same). This translation amounts to a few orders of magnitude in $\dot \gamma$. As a result, in the case of \textit{fluid 1} the obtained numerical solution for the fluid flow inside the fracture (quantified by the average shear rates $\Gamma$) is close to the Newtonian high shear rate variant with viscosity $\eta_\infty$. Naturally, in this range the Carreau and truncated power-law rheologies yield very similar results. Note that even though the maximal deviations between $\eta_\text{a}$  for these models are  over 40$\%$ (see Fig. \ref{Lav_HPG}b)), the relative differences between the respective solutions do not exceed 1$\%$. On the other hand, in the resulting shear rates range the power-law rheology greatly underestimates the apparent viscosity and thus the respective solution can not be considered a substitute for the Carreau variant at all.

A completely different situation is reported for the HPG fluid. Here, regardless of the rheological model, the obtained average shear rate values fit very well inside the interval defined by the limiting shear rates for the truncated power-law ($\dot \gamma_1$ and $\dot \gamma_2$). Thus, almost during the entire time of fracture evolution the average shear rates produce the viscosities from the interim between the plateaus of $\eta_0$ and $\eta_\infty$. For this reason the results obtained for the truncated power-law and the pure power-law models are virtually the same and simultaneously very close to the solution of the Carreau variant of the problem. In this case even the power-law rheology can be considered a credible substitute for the Carreau law.

The above two trends could be easily identified due to the respective viscosity characteristics being essentially different from each other in terms of the intermediate behaviour between $\eta_0$ and $\eta_\infty$. For the second analysed pair of fluids, the 150 wppm HPAM fluid and the 600 wppm XG fluid, the viscosity curves are much closer to each other (again the limiting viscosities  $\eta_0$ and $\eta_\infty$ are practically the same). However, the relations between the solutions obtained for various rheological models for each of these fluids are quite different. 

For the 150 wppm HPAM fluid the truncated power-law solution is close to the Carreau variant throughout the whole duration of fracture evolution. On the other hand, for the power-law rheology one has substantial deviations from the Carreu results in the initial stage of the crack propagation, with the relative difference minimised with time growth (see Fig. \ref{HPAM_bledy}b)). The explanation of this issue can be deduced from the graphs in Fig. \ref{HPAM_mean}. We see that in the aforementioned initial stage the average shear rates, $\Gamma$, are greater by one order of magnitude than $\dot \gamma_2$.  Therefore, a high shear rate regime of flow is achieved, which yields a good coincidence of results between the Carreau and truncated power-law models and simultaneously a large underestimation of the apparent viscosity by the pure power-law rheology. As a result, the power-law solution does not mimic well its Carreau counterpart, especially in the initial stage of crack propagation.

When considering the results obtained for the 600 wppm XG fluid we have a situation very similar to that reported for the HPG fluid. Again, a good coincidence of results obtained for different rheological models is observed over the entire time interval. The power-law solution is barely distinguishable from the truncated power-law variant. This can be a bit surprising if one recalls that in the initial stage of the fracture extension the average shear rates for all rheological models exceed the value of $\dot \gamma_2$ (compare Fig. \ref{XG_mean}a)), just as was the case of the 150 wppm HPAM fluid. However, this time the differences between the values of $\Gamma$ and $\dot \gamma_2$ are much smaller than previously, for both considered quantities being of the same order of magnitude. For this reason, the resulting viscosities are relatively close to each other so are the respective solutions. Therefore, even the power-law model can be confidently adopted in this case to approximate the Carreau rheology and simulate the HF process.

\section{Conclusions}
\label{conc}

In the paper a problem of a hydraulic fracture driven by a non-Newtonian shear-thinning fluid was analysed. For the PKN fracture geometry three different rheological models of fluid were used: the Carreau model, the truncated power-law model, the power-law model. Each of these models was employed to describe the apparent viscosity of four fracturing fluids where the truncated power-law and power-law rheologies were considered approximations of respective Carreau characteristics. For some typical values of the HF process a comparative analysis was performed in order to verify whether the simplified rheologies (trunctaed power-law and power-law) can be considered credible substitutes for the Carreau model. 

The following conclusions can be drawn from the conducted analysis:
\begin{itemize}
\item{The shear rate dependent rheological properties of fracturing fluids affect the HF process in various ways and with varying intensities at different stages of  crack propagation. In all  analysed cases the fluid flow inside the fracture evolved from the high shear rate  regime at the initial time towards low shear rate modes at later stages. Nevertheless, for fixed HF process parameters such a transition between the respective modes depends on the particular viscosity characteristics $\eta_\text{a}(\dot \gamma)$. For some fluids the low shear rate regime of flow  could be achieved only for times beyond the values typical to the HF treatment.  }
\item{The truncated power-law rheology is a good substitute for the Carreau model in the HF problems. It provides a good coincidence (sufficient for any practical application) of the computational results with those obtained for the equivalent Carreau fluid. Simultaneously, this model offers a relative simplicity in numerical implementation.}
\item{The power-law model can be used in some cases as a substitute for the Carreau rheology. However, the credible results are produced only if the average values of the fluid shear rates are within the interval defined by the limiting  viscosities $\eta_0$ and $\eta_\infty$. This interval can be approximated by the limiting shear rates of the truncated power-law model ($\dot \gamma_1$ and $\dot \gamma_2$). Unfortunately, no a priori estimation of the applicability of the power-law model can be done. On the other hand, a posteriori evaluation of the average fluid shear rates can verify the credibility of the obtained results.}
\item{The values of the fluid shear rates averaged over fracture cross section, $\Gamma$, combined  with the respective viscosity characteristics, $\eta_\text{a}(\dot \gamma)$, constitute a good tool to verify credibility of the results obtained for simplified rheological models (such as power-law model).}
\item{The employed methodology and numerical scheme can be used to investigate the HF problem for any generalised Newtonian fluid.}
\end{itemize}

\section*{Acknowledgments}
\noindent
The author is thankful to Prof. Panos Papanastasiou, Prof. Gennady Mishuris and Dr. Monika Perkowska for their useful comments and discussions.

\vspace{5mm}
\noindent
{\bf Funding:} 
This work was funded by European Regional Development Fund and the Republic of Cyprus
through the Research Promotion Foundation (RESTART 2016 - 2020 PROGRAMMES, Excellence Hubs,
Project EXCELLENCE/1216/0481).
 
\appendix
\section{Derivation of expressions for the fluid  fluid flow rate}
\label{ap_A}

In \cite{Wrobel_Arxiv} expressions for the fluid velocity and the average fluid flow rate through the channel cross sections were derived for the slit flow of a generalised Newtonian fluid. In the following we derive respective relations for the flow in an elliptic channel.

Let us consider a fully developed flow of a generalised Newtonian fluid in an elliptic channel of semi-axes $w/2$ and $H/2$ respectively (see Fig. \ref{PKN_geom}). Due to the problem symmetry it is sufficient to consider only one quarter of the ellipse, e.g.:
\[
y \in [0,w/2], \quad z \in [0,H/2].
\]

We assume that the following condition holds:
\begin{equation}
\label{assump_1}
H\gg w.
\end{equation}

For the stationary unidirectional flow in the $x$ direction of an incompressible fluid the general Navier-Stokes system of equations can be reduced to \citep{Perkowska_Phd}:
\begin{equation}
\label{NS_red}
-\frac{\partial p}{\partial x}+\frac{\partial \tau_{yx}}{\partial y}+\frac{\partial \tau_{zx}}{\partial z}=0,
\end{equation}
where the corresponding shear stresses are defined as:
\begin{equation}
\label{tau_def}
\tau_{yx}=\eta_\text{a}\frac{\partial V}{\partial y}, \quad \tau_{zx}=\eta_\text{a}\frac{\partial V}{\partial z},
\end{equation}
with $V(y,z)$ being the velocity profile over the elliptic cross section. Respective boundary conditions read:
\begin{equation}
\label{BCs-ellip}
V\big|_{\partial A}=0, \quad \frac{\partial V}{\partial y}\Big|_{y=0}=\frac{\partial V}{\partial z}\Big|_{z=0}=0,
\end{equation}
where $\partial A$ defines the channel wall.

We employ a transformation to the cylindrical coordinate system $(x,r,\theta)$:
\begin{equation}
\label{cylind_coord}
y=\frac{w}{2}r\cos \theta, \quad z=\frac{H}{2}r\sin \theta,
\end{equation}
where $r \in [0,1]$, $\theta \in[0,\pi/2]$. The following notation for the velocity is adopted now:
\[
V(x,y)=u(r,\theta).
\]
Additionally, assuming that  velocity is constant along the concentric ellipses around the channel longitudinal axis ($x$ axis)  one has:
\begin{equation}
\label{du_0}
\frac{\partial u}{\partial \theta}=0.
\end{equation}
Respective boundary conditions read now:
\begin{equation}
\label{BCs_cyl}
u(1,\theta)=0, \quad \frac{\partial u}{\partial r}\Big |_{r=0}=0.
\end{equation}

\begin{figure}[htb!]
\begin{center}
\includegraphics[scale=0.6]{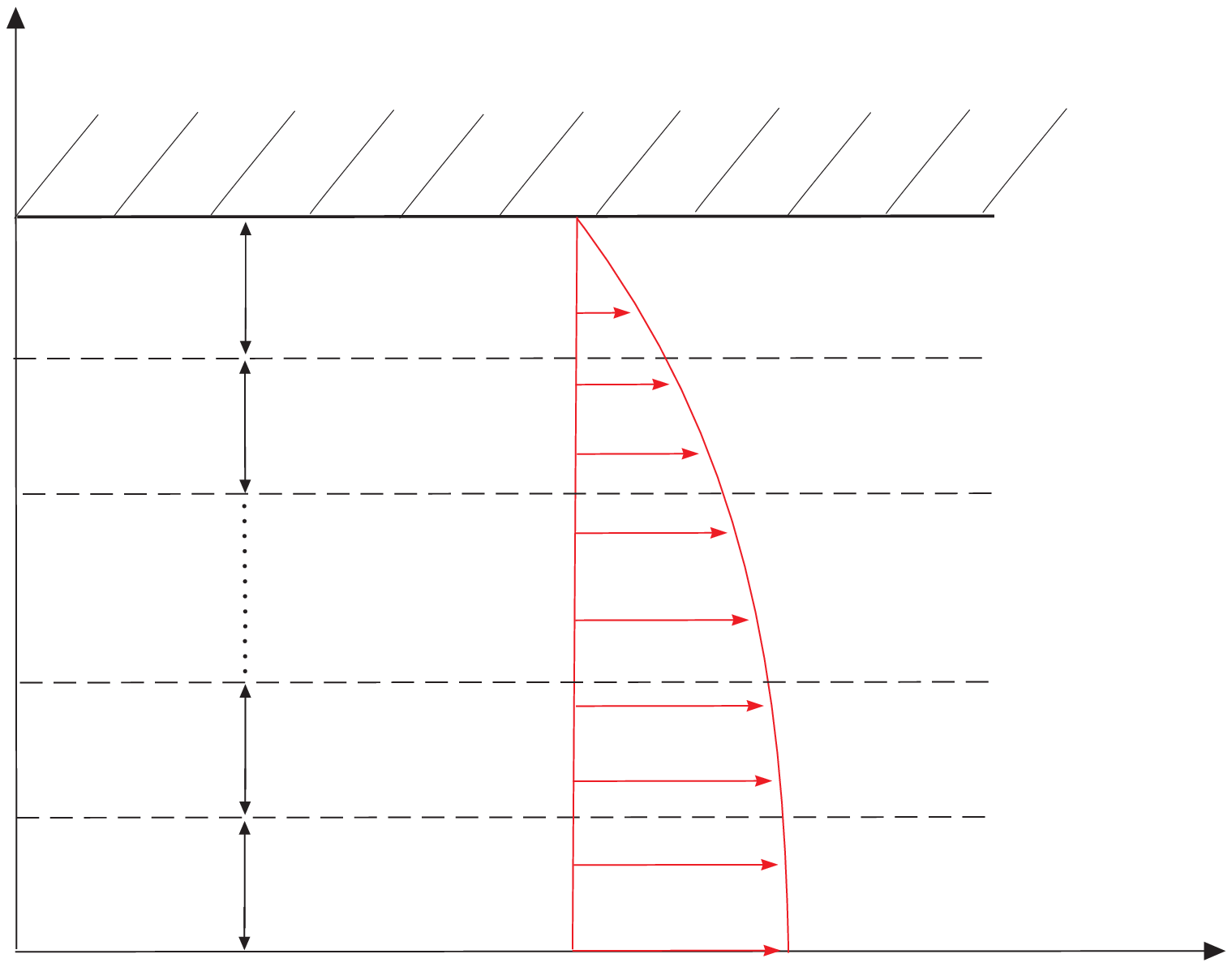}
\put(-13,7){$x$}
\put(-253,190){$y$}
\put(-270,0){$0$}
\put(-273,30){$\delta_1$}
\put(-293,58){$\delta_1+\delta_2$}
\put(-301,97){$\sum_{j=1}^{N-1}\delta_j$}
\put(-300,127){$\sum_{j=1}^{N}\delta_j$}
\put(-271,156){$\frac{w}{2}$}
\put(-220,12){\rotatebox{90}{$\delta_1$}}
\put(-220,41){\rotatebox{90}{$\delta_2$}}
\put(-220,108){\rotatebox{90}{$\delta_{N}$}}
\put(-220,132){\rotatebox{90}{$\delta_{N+1}$}}
\put(-95,75){$V(y,0)$}
\put(-195,35){$V_1, \dot \gamma_1$}
\put(-195,65){$V_2, \dot \gamma_2$}
\put(-195,105){$V_{N-1}, \dot \gamma_{N-1}$}
\put(-195,133){$V_N, \dot \gamma_{N}$}
\caption{The channel cross section and velocity profile for $z=0$. Only one of the symmetrical parts is shown.}
\label{channel}
\end{center}
\end{figure}

Under the above conditions equation \eqref{NS_red} transforms to:
\begin{equation}
\label{NS_cyl}
\begin{aligned}
-\frac{1}{4}\frac{\partial p}{\partial x}+\left(\frac{\sin^2\theta}{H^2}+\frac{\cos^2\theta}{w^2} \right)\frac{\partial}{\partial r}\left(\eta_\text{a}\frac{\partial u}{\partial r}\right)+\left(\frac{1}{H^2}-\frac{1}{w^2}\right)\sin \theta \cos \theta \frac{\partial}{\partial r}\left(\frac{\eta_\text{a}}{r}\frac{\partial u}{\partial r}\right) \\
+\left(\frac{\cos^2\theta}{H^2}+\frac{\sin^2\theta}{w^2} \right)\frac{\eta_\text{a}}{r}\frac{\partial u}{\partial r}=0.
\end{aligned}
\end{equation}
Note that in the cylindrical coordinate system the velocity profile does not change with changing the value of $\theta$. Thus, it is sufficient to solve the equation \eqref{NS_cyl} for a single value of $\theta$. We set $\theta=0$ ($z=0$) for which equation \eqref{NS_cyl} simplifies to:
\begin{equation}
\label{NS_simp}
r\frac{\partial}{\partial r}\left(\eta_\text{a}\frac{\partial u}{\partial r} \right)+\frac{w^2}{H^2}\eta_\text{a}\frac{\partial u}{\partial r}=\frac{w^2}{4}\eta_\text{a}\frac{\partial p}{\partial x}.
\end{equation}
When solving \eqref{NS_simp} with respect to $\eta_\text{a}\frac{\partial u}{\partial r}$ under the boundary condition \eqref{BCs_cyl}$_2$ one arrives at the following relation:
\begin{equation}
\label{sol_1_cyl}
\eta_\text{a}\frac{\partial u}{\partial r}=\frac{1}{4}\frac{w^2H^2}{w^2+H^2}\frac{\partial p}{\partial x}r.
\end{equation}
Form \eqref{assump_1} it follows that:
\begin{equation}
\label{assump_2}
\frac{H^2}{w^2+H^2} \to 1.
\end{equation}
Thus, equation \eqref{sol_1_cyl} can be rewritten as:
\begin{equation}
\label{sol_2_cyl}
\eta_\text{a}\frac{\partial u}{\partial r}=\frac{w^2}{4}\frac{\partial p}{\partial x}r.
\end{equation}
Note that, when analysing the flow in the plane $z=0$ (i.e. in which the PKN fracture width is defined), equation \eqref{sol_2_cyl} is identical to its counterpart obtained in  \cite{Wrobel_Arxiv} for a slit flow. Therefore, for a piecewise rheology of the type \eqref{eta_ap}, results from \cite{Wrobel_Arxiv} that involve thicknesses of respective shear rate layers and velocity profiles are directly transferable. In this way, for $N$ boundary values of the shear rates $\dot \gamma_j$ in representation \eqref{eta_ap} up to $N+1$ shear rate layers appear over each of the cross section ($z=0$) symmetrical parts depending on the magnitudes of $w$ and $\partial p/\partial x$ - see Fig. \ref{channel}. Thicknesses of these layers (in the plane $z=0$) are defined as:
\begin{itemize}
\item{for the Newtonian-type layer in the core of the flow associated with the viscosity $\eta_0$
\begin{equation}
\label{delta_1_ap}
\delta_1=\left( \frac{dp}{dx}\right)^{-1}\eta_0 \dot \gamma_1,
\end{equation}}
\item{for the power-law layers in the range $|\dot \gamma_1|<|\dot \gamma|<|\dot \gamma_N|$
\begin{equation}
\label{delta_j_ap}
\delta_{j+1}=\left(-\frac{dp}{dx}\right)^{-1}C_j\left[(-\dot \gamma_{j+1})^{n_j} -(-\dot \gamma_{j})^{n_j}\right], \quad j=1,...,N-1,
\end{equation}}
\item{for the Newtonian layer adjacent to the channel wall with the viscosity $\eta_\infty$
\begin{equation}
\label{delta_N_ap}
\delta_{N+1}=\frac{w}{2}-\sum_{j=1}^{N}\delta_j.
\end{equation}}
\end{itemize}
Naturally, for certain values of $w$ and $\partial p/\partial x$ some of these layers may not be present or some of them can be reduced by the overall height of the channel (full explanation of this problem can be found in \cite{Wrobel_Arxiv}). 

For a predefined rheological model described by apparent viscosity $\eta_\text{a}$ one can calculate the velocity profile by integrating \eqref{sol_2_cyl} under the boundary condition \eqref{BCs_cyl}$_1$:
\begin{equation}
\label{u_op}
u(r,\theta)=-\frac{w^2}{4}\frac{\partial p}{\partial x}\int_r^1 \frac{\xi}{\eta_\text{a}}\text{d}\xi.
\end{equation}
Consequently, the fluid flow rate through the channel cross section can be obtained as:
\begin{equation}
\label{q_op}
Q=\frac{wH}{4}\int_0^{2\pi}\int_0^1ru\text{d}r\text{d}\theta=\frac{wH\pi}{2}\int_0^1ru\text{d}r.
\end{equation}
In order to compute the normalised fluid flow rate for the PKN problem the scaling factor $\frac{4}{\pi H}$ is to be employed:
\begin{equation}
\label{q_norm}
q=\frac{4}{\pi H}Q=\frac{8}{w}\int_0^{w/2}yV(y,0)\text{d}y.
\end{equation}
When integrating the velocity profile in a piecewise manner over the respective shear rate layers one arrives at the following computational formula for $q$:
\begin{equation}
\label{q_sum}
q=\frac{8}{w}\sum_{j=0}^N \int_{y_j}^{y_{j+1}}yV(y,0)\text{d}y, 
\end{equation}
where:
\[
y_0=0, \quad y_{N+1}=w/2, \quad y_j=\sum_{k=1}^j\delta_k, \quad j=1,...,N.
\]
The component integrals in \eqref{q_sum} are expressed as:
\begin{equation}
\label{q_int_1}
\int_0^{y_1}yV(y,0)\text{d}y=\frac{\delta_1^2}{2}\left(V_1-\frac{1}{4\eta_0}\frac{\text{d}p}{\text{d}x}\delta_1^2 \right),
\end{equation}
\begin{equation}
\label{q_int_j}
\begin{split}
 \int_{y_j}^{y_{j+1}}yV(y,0)\text{d}y={}& \frac{V_{j+1}}{2}\left(y_{j+1}^2-y_j^2\right)-\frac{n_j}{n_j+1}C_j^{-1/n_j}\left(\frac{\text{d}p}{\text{d}x}\right)^{-1} \Bigg\{ \frac{y_{j+1}^2-y_j^2}{2}\left(-\frac{\text{d}p}{\text{d}x}y_{j+1}-D_j\right)^{\frac{n_j+1}{n_j}} \\
& -\left(\frac{\text{d}p}{\text{d}x}\right)^{-2}\Bigg[\frac{n_j}{3n_j+1}\left(-\frac{\text{d}p}{\text{d}x}y_{j+1}-D_j\right)^{\frac{3n_j+1}{n_j}} 
-\frac{n_j}{3n_j+1}\left(-\frac{\text{d}p}{\text{d}x}y_j-D_j\right)^{\frac{3n_j+1}{n_j}} \\
&+ \frac{n_j}{2n_j+1}D_j\left(-\frac{\text{d}p}{\text{d}x}y_{j+1}-D_j\right)^{\frac{2n_j+1}{n_j}}-
\frac{n_j}{2n_j+1}D_j\left(-\frac{\text{d}p}{\text{d}x}y_j-D_j\right)^{\frac{2n_j+1}{n_j}}\Bigg] \Bigg\}, 
\end{split}
\end{equation}
\begin{equation}
\label{q_int_N}
\int_{y_N}^{w/2}yV(y,0)\text{d}y=-\frac{1}{2\eta_\infty}\frac{\text{d}p}{\text{d}x}\left[\frac{w^4}{64}+\frac{y_N^2}{4}\left(y_N^2-\frac{w^2}{2} \right) \right]-\frac{D_N}{\eta_\infty}\left[\frac{w^3}{48}+y_N^2\left(\frac{y_N}{3}-\frac{w}{4}\right)\right],
\end{equation}
where:
\begin{equation}
\label{D_j}
D_j=-y_j^\frac{w^2}{H^2}\left[C_j\left(-\dot \gamma_j \right)^{n_j}+\frac{H^2}{w^2+H^2}\frac{\text{d}p}{\text{d}x}y_j \right],
\end{equation}
\begin{equation}
\label{D_N}
D_N=y_N^\frac{w^2}{H^2}\left[\eta_\infty\dot \gamma_N-\frac{H^2}{w^2+H^2}\frac{\text{d}p}{\text{d}x}y_N \right].
\end{equation}
The interfacial velocities (see Fig. \ref{channel}) are denoted as $V_j$ ($j=1,...,N$).

\begin{remark}
\label{rem_q}
Note that the expression for the fluid flow rate for the classical power-law model \eqref{q_PL} can be recreated from \eqref{q_sum} and \eqref{q_int_j} by setting $y_j=0$, $y_{j+1}=w/2$ and $V_{j+1}=0$.
\end{remark}


\begin{thebibliography}{1}

\bibitem[Adachi $\&$ Detournay, 2002]{Adachi_2002}  Adachi J., Detournay E. (2002) Self-similar solution of a plane-strain fracture driven by a power-law fluid. \textit{International Journal of Numerical and Analytical Methods in Geomechanics}, 26, 579--604

\bibitem[Bao et al., 2017]{Bao_2017} Bao K., Lavrov A., Nilsen H. (2017) Numerical Modeling of Non-Newtonian Fluid Flow in Fractures and Porous Media. \textit{Computational Geosciences}, 21(5-6): 1313--1324

\bibitem[Barbati et al., 2016]{Barbati_2016}  Barbati A.,  Desroches J.,  Robisson A.,  McKinley G. (2016) Complex Fluids and Hydraulic Fracturing. \textit{Annual Review of Chemical and Biomolecular Engineering},  7: 415--453

\bibitem[Bird, 1987]{Bird_1987} Bird R., Armstrong R.,  Hassager  O. (1987) Dynamics of Polymeric Liquids, Wiley, New York, Vol. 1

\bibitem[Garagash, 2006]{Garagash_2006} Garagash D. (2006) Transient solution for a plane-strain fracture driven by a shear-thinning, power-law fluid.\textit{International Journal for Numerical and Analytical Methods in Geomechanics}, 30(14): 1439--1475

\bibitem[Garagash et al., 2019]{Garagash_2019} Garagash I., Osiptsov A., Boronin S. (2019) Dynamic bridging of proppant in a hydraulic fracture. \textit{International Journal of Engineering Science}, 135: 86--101

\bibitem[Gholipour et al., 2018]{Gholipour_2018_1} Gholipour A.,  Ghayesh M.,  Zander A.,  Mahajan R. (2018) Three-dimensional biomechanics of coronary arteries. \textit{International Journal of Engineering Science}, 130: 93 -- 114

\bibitem[Habibpour $\&$ Clark, 2017]{Habibpour_2017} Habibpour M., Clark P. (2017) Drag reduction behavior of hydrolyzed polyacrylamide/xanthan gum mixed polymer solutions. \textit{Petroleum Science}, 14: 412 -- 423

\bibitem[Huang $\&$ Desroches, 2004]{Huang_2004} Huang H., Desroches J. (2004) A PKN hydraulic fracturing model with piecewise fluid rheology. In: ARMA/NARMS 04-560, pp 42-52

\bibitem[Lavrov, 2015]{Lavrov_2015} Lavrov A. (2015) Flow of truncated power-law fluid between parallel walls for hydraulic fracturing applications. \textit{Journal of Non-Newtonian Fluid Mechanics}, 223: 141--146

\bibitem[Moukhtari $\&$ Lecampion, 2018]{Lecampion_2018} Moukhtari F., Lecampion B. (2018) A semi-infinite hydraulic fracture driven by a shear-thinning fluid. \textit{Journal of Fluid Mechanics}, 838: 573--605

\bibitem[Nordgren, 1972]{Nordgren} Nordgren R. (1972) Propagation of a Vertical Hydraulic Fracture.  \textit{Society of Petroleum Engineers Journal},  253: 306-314

\bibitem[Peck et al., 2018]{Peck_2018_1} Peck D., Wrobel M., Perkowska M., Mishuris G. (2018) Fluid velocity based simulation of hydraulic fracture: a penny shaped model - part I: the numerical algorithm. \textit{Meccanica}, 53(15): 3615--3635

\bibitem[Peck et al., 2018a]{Peck_2018_2} Peck D., Wrobel M., Perkowska M., Mishuris G. (2018) Fluid velocity based simulation of hydraulic fracture - a penny shaped model. Part II: new, accurate semi-analytical benchmarks for an impermeable solid.  \textit{Meccanica}, 53(15): 3637--3650

\bibitem[Perkowska et al., 2016]{Perkowska_2016} Perkowska M., Wrobel M., Mishuris G. (2016) Universal hydrofracturing algorithm for shear--thinning fluids: particle velocity based simulation. \textit{Computers and Geotechnics}, 71: 310--337

\bibitem[Perkowska, 2016]{Perkowska_Phd} Perkowska M. (2016) Mathematical and numerical modeling of hydraulic fractures for non-Newtonian fluids. PhD thesis, Aberystwyth University

\bibitem[Wang et al., 2018]{Wang_2018} Wang J., Elsworth D., Denison M. (2018) Propagation, proppant transport and the evolution of transport properties of hydraulic fractures. \textit{Journal of Fluid Mechanics}, 855: 503--534

\bibitem[Wrobel $\&$ Mishuris, 2015]{Wrobel_2015} Wrobel M., Mishuris G. (2015) Hydraulic fracture revisited: Particle velocity based simulation. \textit{International Journal of Engineering Science}, 94: 23--58

\bibitem[Wrobel et al., 2017]{Wrobel_2017} Wrobel M., Mishuris G., Piccolroaz A. (2017) Energy Release Rate in hydraulic fracture: can we neglect an impact of the hydraulically induced shear stress? \textit{International Journal of Engineering Science}, 111: 28--51

\bibitem[Wrobel et al., 2018]{Wrobel_2018} Wrobel M., Mishuris G., Piccolroaz A. (2018) On the impact of tangential traction on the crack surfaces induced by fluid in hydraulic fracture: Response to the letter of A.M. Linkov. Int. J. Eng. Sci. (2018) 127, 217--219. \textit{International Journal of Engineering Science}, 127: 220--224

\bibitem[Wrobel, 2019]{Wrobel_Arxiv} Wrobel M. (2020) An efficient algorithm of solution for the flow of generalized Newtonian fluid in channels of simple geometries. \textit{Rheologica Acta}, DOI: 10.1007/s00397-020-01228-2


\end{thebibliography}
\end{document}